\documentclass[final,3p,times,twocolumn]{elsarticle}
\usepackage[cp850]{inputenc}
\usepackage{amsmath,amssymb,amsfonts,color}
\usepackage{graphicx}
\usepackage{latexsym}
\usepackage{psfrag}
\usepackage{array}
\usepackage{wasysym}
\usepackage{epsfig}
\biboptions{sort&compress}

\newcommand*{\vcenteredhbox}[1]{\begingroup
\setbox0=\hbox{#1}\parbox{\wd0}{\box0}\endgroup}

\newcommand{\textin}[1]{\mbox{\scriptsize{#1}}}

\begin{document}

\begin{frontmatter}
\title{Dripping, jetting and tip streaming}

\author[1]{J. M. Montanero}%\corref{cor1}
\ead{jmm@unex.es}
\author[2]{A. M. Ga\~n\'an-Calvo}
\ead{amgc@ues.es}
%\cortext[cor1]{Corresponding author. Tlf: +34924289610 Fax: +34924189602}
\address[1]{Depto.\ de Ingenier\'{\i}a Mec\'anica, Energ\'etica y de los Materiales and\\
Instituto de Computaci\'on Cient\'{\i}fica Avanzada (ICCAEx),\\ Universidad de Extremadura, E-06006 Badajoz, Spain}
\address[2]{Depto, de Ingenier\'{\i}a Aeroespacial y Mec\'anica de Fluidos,\\
Universidad de Sevilla, E-41092 Sevilla, Spain}

\begin{abstract}
Dripping, jetting and tip streaming have been studied up to a certain point separately by both fluid mechanics and microfluidics communities, the former focusing on fundamental aspects while the latter on applications. Here, we intend to review this field from a global perspective by considering and linking the two sides of the problem. In the first part, we present the theoretical model used to study interfacial flows arising in droplet-based microfluidics, paying attention to three elements commonly present in applications: viscoelasticity, electric fields and surfactants. We review both classical and current results about the stability of jets affected by these elements. Mechanisms leading to the breakup of jets to produce drops are reviewed as well, including some recent advances in this field. We also consider the relatively scarce theoretical studies on the emergence and stability of tip streaming in open systems. In the second part of this review, we focus on axisymmetric microfluidic configurations which can operate on the dripping and jetting modes either in a direct (standard) way or via tip streaming. We present the dimensionless parameters characterizing these configurations, the scaling laws which allow predicting the size of the resulting droplets and bubbles, as well as those delimiting the parameter windows where tip streaming can be found. Special attention is paid to electrospray and flow focusing, two of the techniques more frequently used in continuous drop production microfluidics. We aim to connect experimental observations described in this section of topics with fundamental and general aspects described in the first part of the review. This work closes with some prospects at both fundamental and practical levels.
\end{abstract}


\end{frontmatter}

\tableofcontents

\newpage

\section{Introduction}
\label{sec1}

% Matter fragmentation
A multitude of technological applications demands the fragmentation of a continuous phase (gas, liquid or solid) down to the submillimeter scale in a controlled manner. This fragmentation can be produced by gently deforming, stretching and splitting matter in its fluid form. The resulting drops, bubbles, emulsions or capsules are subsequently solidified (if necessary). In this way, these fluid entities are used as templates for the synthesis of complex micro-objects, like multi-component and non-spherical microparticles \citep{XNSLKSGWGW05}, or large aspect ratio microfibers \citep{BK10}. These micro-objects can be utilized in very diverse technologies, including drug synthesis and delivery, field responsive rheological fluids, tissue engineering scaffolds, food additives, photonic materials, particle-based display technologies, high-performance composite filler materials, etc. For more details about these technologies, the reader is referred to, e.g., the review of \citet{NTWS13} and references therein.

\begin{figure}[hbt]
%\begin{center}
\centering{\resizebox{0.35\textwidth}{!}{\includegraphics{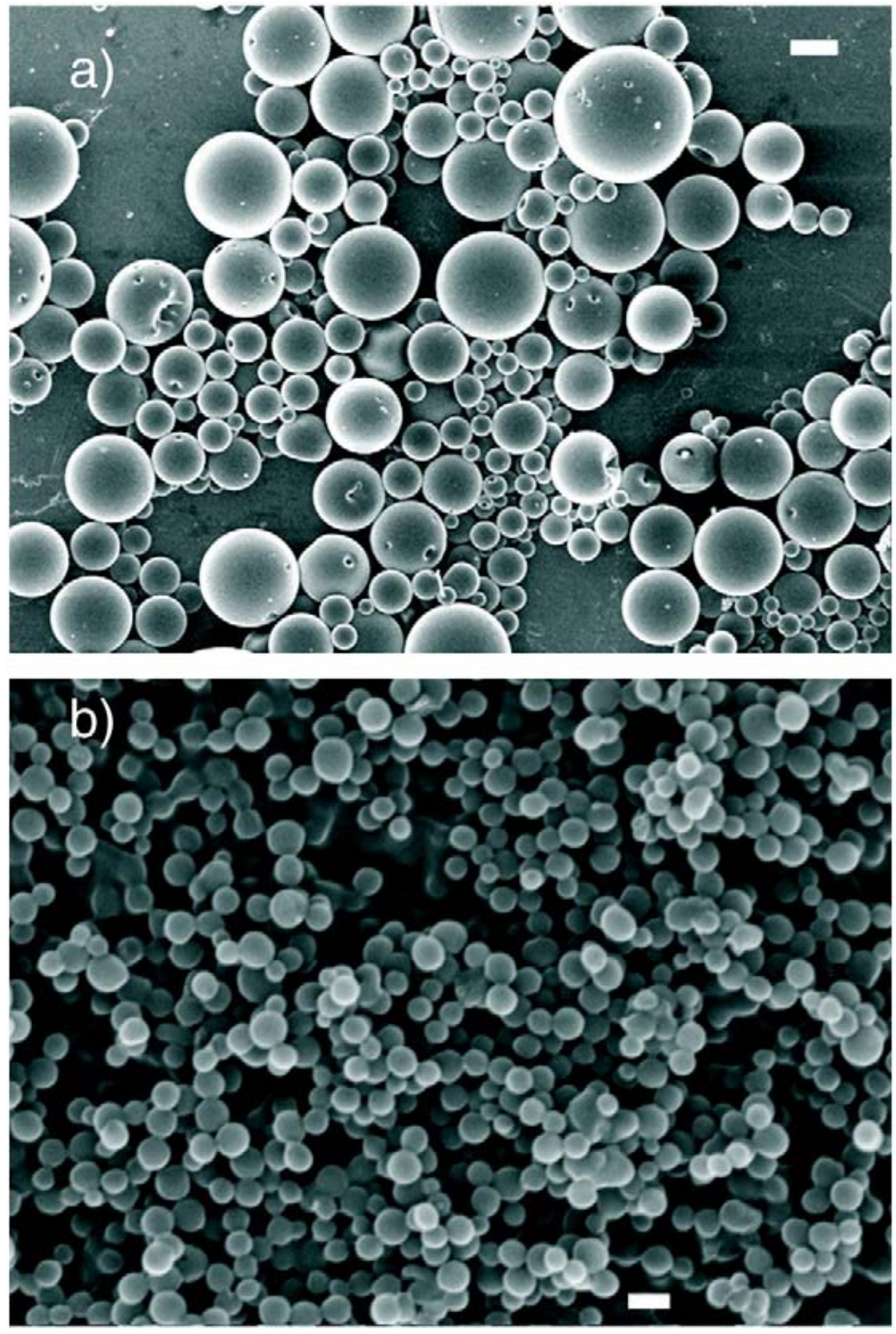}}}
%\end{center}
\caption{SEM images of Gem-loaded PLGA particles produced by (a) turbulent fragmentation followed by solvent extraction, and (b) capillary breakup using Flow Focusing$\circledR$ followed by solvent extraction (bar length: 1 $\mu$m) \citep{GMMF13}.}
\label{monodis}
\end{figure}

% Size and monodispersity
The formation of the above-mentioned fluid entities on the micro and nanometer scales has been extensively investigated over the last thirty years. Driven by their technological relevance, studies have mainly focused on both the size of the fluidic individuals and the monodispersity degree of the population. Experience has repeatedly shown that these two features are somehow antagonistic with the usual atomization technologies (see, e.g., \citep{RG08}): the smaller sizes are reached only at the expense of monodispersity, and {\em vice versa}. Reducing the size of the produced fluid entities requires overcoming the resistance offered by both viscosity and surface tension, which can only be achieved by injecting a significant amount of energy into the process. Only those procedures in which that injection is carefully focused can lead to high monodispersity degrees. Figure \ref{monodis} illustrates how the atomization mechanism reflects in the internal structure of the produced dispersion.

\begin{figure}[hbt]
%\begin{center}
\centering{\resizebox{0.45\textwidth}{!}{\includegraphics{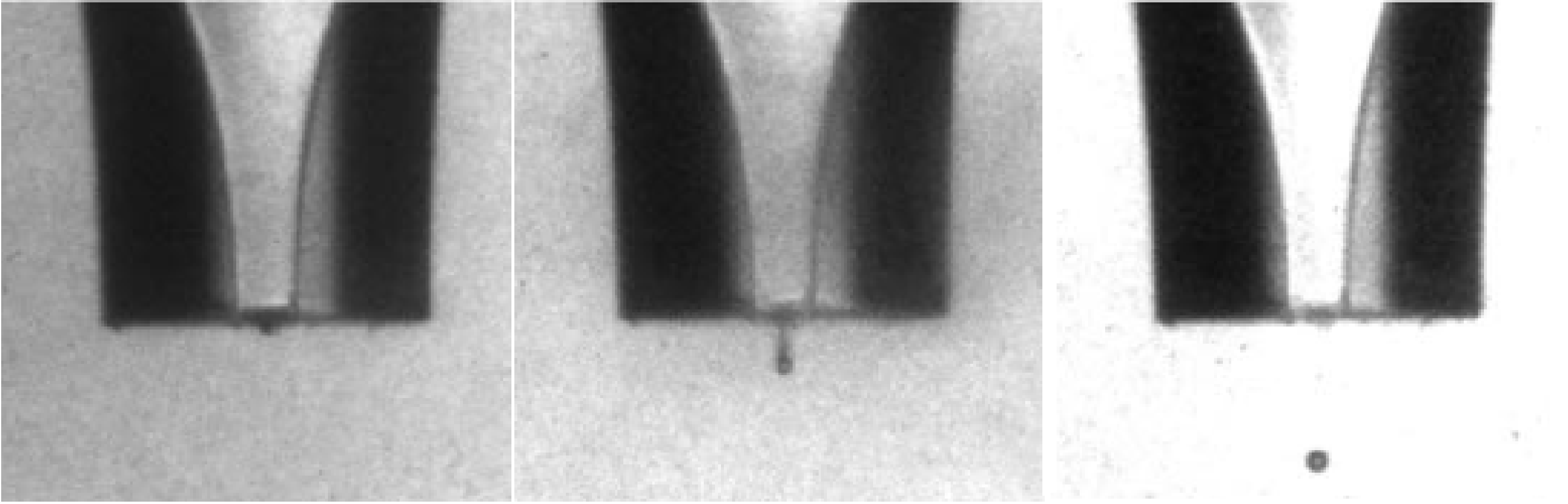}}}
%\end{center}
\caption{Formation of an ultra-small drop upon
application of voltage waveform to a piezoelectric transducer \citep{CB02}. The drop diameter is about 32$\mu$m.}
\label{DoD}
\end{figure}

\begin{figure}[hbt]
%\begin{center}
\centering{\resizebox{0.45\textwidth}{!}{\includegraphics{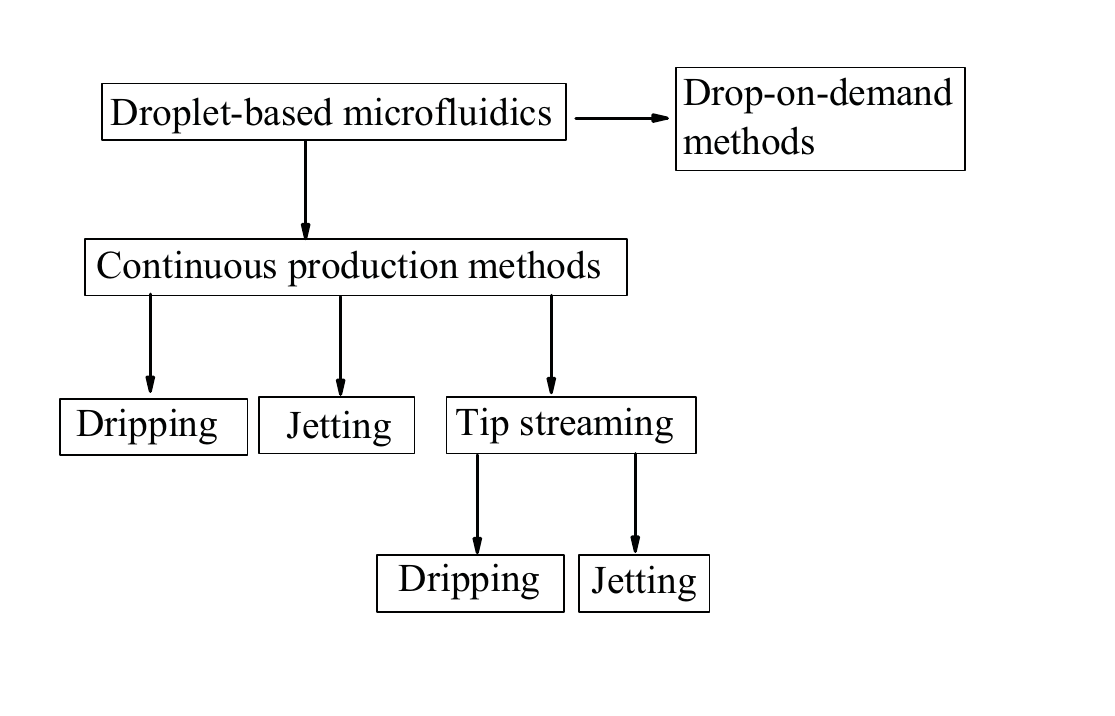}}}
%\end{center}
\caption{Methods for producing droplets/bubbles on the micrometer and nanometer scales. Tip streaming is a route to dripping and jetting with sizes much smaller than that of the microfluidic device.}
\label{modes}
\end{figure}

% Drop-on-demand versus continuous methods
An ample variety of methods can be used to produce droplets/bubbles of different nature and morphology with narrow diameter distributions on the micro and nanometer scales. Among them, we can distinguish drop-on-demand techniques from those in which droplets/bubbles are continuously generated (Fig.\ \ref{modes}) \citep{B02a,W10b}. The thermal and piezoelectric inkjet methods constitute important examples of the first class. In the thermal inkjet method \citep{N85}, a resistor heats the ink until it vaporizes. Then, a bubble grows and collapses quasi-instantaneously producing the ejection of a droplet through the nozzle. This technique requires inks with high vapor pressure, low boiling point, and high kogation stability. In the piezoelectric inkjet method \citep{F84,W10b}, the pressure wave necessary to eject the droplet comes from the contraction of a piezoelectric element. The major limitation of this technique is probably the fact that the ejected inks must have viscosities and surface tensions within relatively narrow ranges. Both the thermal and piezoelectric methods produce drops with diameters similar to that of the ejecting nozzle. The droplet diameter can be adjusted by modulating the electric signal applied to the resistor and piezoelectric element, respectively. Under certain specific conditions, droplets with diameters much smaller than that of nozzle can be formed \citep{CB02} (Fig.\ \ref{DoD}). Drop-on-demand methods were originally devised to recreate a digital image onto paper, plastic or other substrates. This technology has been subsequently extended to many other fields. Among them, the building of functional structures in tissue engineering \citep{BXDC06} has deserved special mention.

% Dripping/bubbling and jetting
In this review, we will pay attention to microfluidic configurations commonly used to {\em continuously} produce drops and bubbles \citep{CA07}. These configurations can operate in the dripping/bubbling and/or the jetting mode. In the dripping/bubbling mode, drops/bubbles are produced right behind the exit of a feeding capillary or ejecting nozzle. On the contrary, a fluid thread long compared with its diameter is formed in the jetting regime. In this case, the surface tension eventually triggers instabilities which yield the breakage of the thread into a collection of droplets/bubbles. 

% Distinction between dripping and jetting regimes
The distinction between the dripping/bubbling and jetting regimes is not always clear. There are many applications in which that distinction is established ambiguously: jetting becomes dripping/bubbling as the precursor fluid thread shortens. \citet{ASPB04} have proposed to define the dripping-to-jetting transition in a leaky faucet as the parameter conditions for which certain measures of the dynamics undergo sharp changes. According to this criterion, the jet length at the dripping-to-jetting transition ranges from a few to hundreds of jet radii as the viscosity increases. The fact that the jet breakup can be regarded as a local phenomenon in terms of the jet's length can also be considered as the defining condition of the jetting regime. In fact, if the axial size of the breakup region were commensurate with the jet length, then the growth of the capillary perturbation would be affected by the discharge orifice, as it is characteristic of the dripping regime.

% The outcome
While the dripping/bubbling mode generally yields higher monodispersity degrees, the generation of droplet streams in the jetting regime is also very attractive because it leads to larger production rates. Typically, dripping/bubbling produces drops with sizes that are commensurate with, or even much greater than that of the nozzle. The diameters of the droplets resulting from the inertio-capillary breakup of jets are around twice that of the precursor jet. As will be shown in Sec.\ \ref{sec5}, this proportion can be significantly altered by viscosity, electric fields, confinement and other factors.

\begin{figure*}[hbt]
%\begin{center}
\centering{\resizebox{0.75\textwidth}{!}{\includegraphics{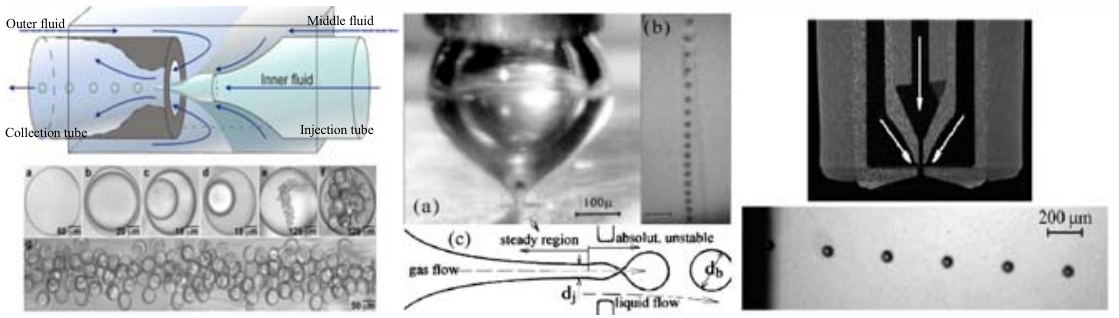}}}
%\end{center}
\caption{Some examples of dripping/bubbling from tip streaming: double emulsions produced with hydrodynamic focusing (left-hand images) \citep{ULLKSW05}, bubbling in flow focusing (central images) \citep{GG01,G04b}, and dripping in gaseous flow focusing (right-hand images) \citep{CMG16}.}
\label{tip2}
\end{figure*}

\begin{figure*}[hbt]
%\begin{center}
\centering{\resizebox{0.65\textwidth}{!}{\includegraphics{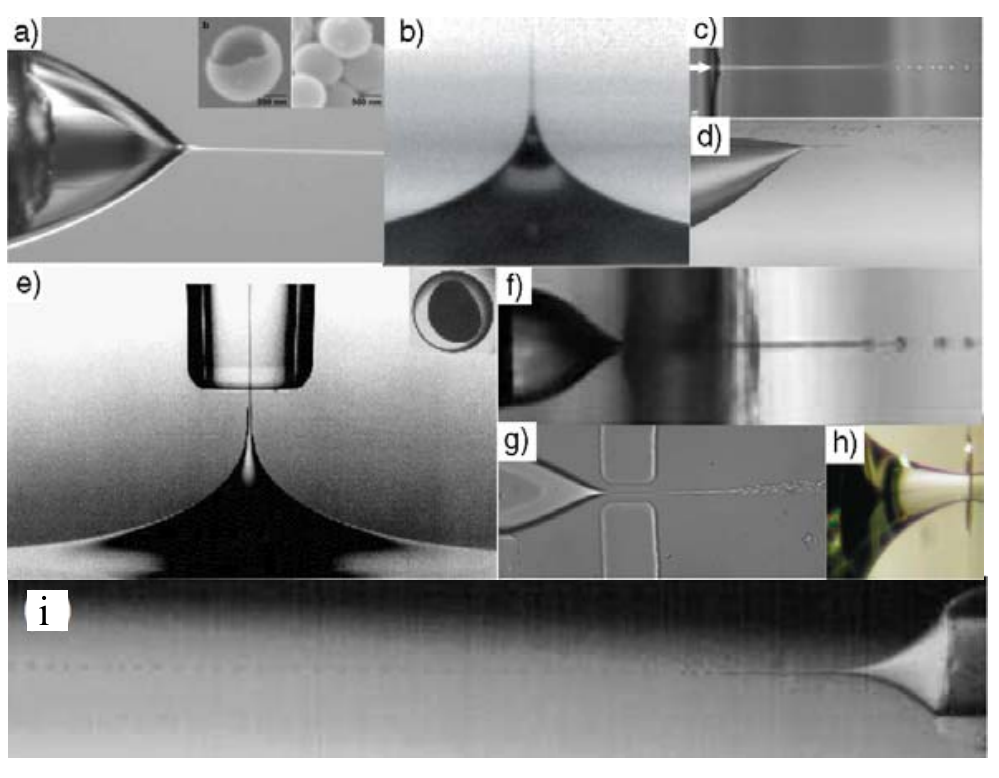}}}
%\end{center}
\caption{Some examples of jetting from tip streaming: (a) double electrospray configuration to produce capsules down to the nanometer scale \citep{BL07}, (b) electrified film of a low-conductivity liquid \citep{CJHB08}, (c) liquid transport due to light scattering \citep{SWCZD07}, (d) tip streaming due to the action of a shear flow in the presence of surfactants \citep{E97}, (e) selective withdrawal to produce microcapsules \citep{CLHMN01}, (f) original flow focusing configuration \citep{SLYY09}, (g) flow focusing in a 2D microfluidic device \citep{ABS03}, (h) double flow focusing arrangement to coat microparticles \citep{GGRHF07} and (i) liquid-liquid coflowing configuration to produce emulsions \citep{CGFG09}.}
\label{tip}
\end{figure*}

% Tip streaming
In some configurations such as electrospray \citep{T64,Z14}, coflowing \citep{SB06}, hydrodynamic focusing \citep{KVBA98} and flow focusing \citep{G98a,ABS03}, there is a narrow parameter window leading to the so-called tip streaming \citep{SB06}. In this regime, the fluid is directed by some external actuation towards the tip of a deformed film, drop or stretched meniscus attached to a feeding capillary. This tip emits small drops/bubbles either directly (Fig.\ \ref{tip2}) or via the breakage of a very thin jet (Fig.\ \ref{tip}). In both cases, the droplets/bubbles are smaller or even much smaller than any characteristic length of the microfluidic device. In almost all the cases, the external actuation mentioned above is gently exerted by interfacial stresses, whether their origin is electrical (Maxwell stresses), hydrodynamic (an outer stream) or any other. Maxwell stresses result from both the accumulation of free electric charges at the interface and the jump of electrical permittivity across this surface \citep{MT69}. Hydrodynamic forces are caused by the suction (decrease of hydrostatic pressure) and/or viscous traction exerted by an outer stream moving faster than the dispersed phase.

% The balance in tip streaming
In tip streaming, normal interfacial stresses can cause variations of the hydrostatic pressure on the inner side of the interface, which gives rise to bulk forces throughout the dispersed phase. Tangential stresses accelerate the fluid layer next to the interface, and feed recirculation cells for low enough viscosities and injected flow rates \citep{GM09}. Both types of stresses may play a critical role depending on the specific configuration considered. Interestingly, normal stresses may play a subdominant role in the final energy budget of some tip streaming realizations. However, they are necessary for this phenomenon to take place. In any case, tip streaming is the result of a delicate balance between the forces driving and opposing the flow. When that balance is tilted in favor of one of those forces, regular or intermittent dripping is obtained. This explains why tip streaming is sometimes an elusive phenomenon, only found under very specific conditions. Despite the important advances in the understanding of tip streaming, there are still many open questions about both the origin of this phenomenon and the instability mechanisms which limit its appearance.

% Tip streaming advantages
Tip streaming has been shown to be very advantageous at the technological level essentially because it allows for the production of droplets with sizes well below that of the feeding channels, avoiding clogging effects. When designing new devices based on tip streaming, one must concentrate efforts on enhancing the stability and robustness of this mode. For instance, the presence of surfactants at the tip weakens the interfacial tension and makes the phenomenon more robust \citep{D93,ETS01,A16}. Modifications of the injection system to eliminate recirculating patterns in the liquid source is another via to stabilize tip streaming \citep{LM05,WDYF85,SDKL06,ARMGV13}.

% Topologies
Most microfluidic devices possess a planar (two-dimensional, 2D) topology. These devices can be manufactured in essentially one single step, either through photolithography and etching in substrates of silicon or via soft lithography in substrates of polymer materials (PDMS) \citep{DMSW98,UCTSQ00}. This property has conferred great popularity among researchers on the 2D configuration. However, the planar topology also presents certain disadvantages. In these devices, an emerging droplet typically touches the walls of the channel, which can damage fragile particles and cause problems associated with the competitive wetting between immiscible liquids \citep{TGWW05}. In addition, PDMS channels swell in strong organic solvents and siloxane-based compounds and tend to deform with intense applied pressures due to their high elasticity. The planar geometry usually requires specific coatings dedicated surfactants. The problems mentioned above are eliminated in an axisymmetric device. The circular cross-section of the outlet channel allows the continuous phase to surround completely and shield the dispersed one at all flow rates. Axisymmetric devices can be fabricated with glass, which is a resistant, smooth and transparent material. Finally, axisymmetry necessarily entails a considerable increase of the droplet production rate with respect to that taking place in the 2D topology. The major drawback of the axisymmetric geometry is probably the fact that microfluidic devices commonly consist of several pieces that must be carefully aligned to obtain the desired outcome. For this reason, fabrication techniques are in most cases 
art-dependent and not scalable.

% This work I
There is an immense body of literature about properties and functionality of microfluidic devices designed for the continuous production of droplets and bubbles. Excellent reviews have summarized the major achievements in this field \citep{SSA04,CT04,W06a,CA07,A16}. Here, we will focus on the axisymmetric configuration, which has been reviewed on many fewer occasions, and normally as part of a work with a broader scope \citep{BL07,GDM11,VKN12,ZW17}. In an attempt to present an original vision, we will group the results according to the production mode (instead of the employed technique), distinguishing the ``simple"\ dripping and jetting regimes from their counterparts obtained via tip streaming (Fig.\ \ref{modes}). We will devote special attention to electrospray and flow focusing, probably the most popular techniques in this area.

% This work II
Microfluidics researchers typically pay attention to the development and experimental characterization of microfluidic techniques. Experimental results are rationalized using dimensional analysis, which looks for scaling relationships among dimensionless governing parameters. These studies are frequently assisted by direct numerical simulations to describe global, qualitative or involved aspects of the problem. On the other hand, in the quest to reveal the physics of those aspects, fluid dynamics researchers focus on rather fundamental questions, reducing reality to models kept as simple as possible, which sometimes have little connection with experiments or technological applications. The present review aims to serve and bridge both the microfluidic and fluid dynamicist communities, indicating and emphasizing the connections between results obtained from both approaches. For this reason, we will contemplate not only practical aspects of the problem but also fundamental issues which may help experimentalists to understand those aspects.

% Organization of the review
This review is organized as follows. In Sec.\ \ref{sec2}, we describe the theoretical models, approximations and assumptions typically used to examine the microfluidic configurations considered throughout this review. In Sec.\ \ref{sec3}, we explain some of the fundamental ideas involved in the stability analysis of those configurations. Sections \ref{sec4}--\ref{sec6} present some important results about the linear stability of capillary jets. Here, we also mention studies on the global stability of tip streaming flows. The results are presented in more detail in subsequent sections, once the corresponding microfluidic configurations have been described. Section \ref{sec7} shows relevant results about the nonlinear breakup of fluid threads. We devote Sec.\ \ref{sec8} to discuss fundamental and general features of tip streaming and present relevant results of tip streaming in open systems. The microfluidic configurations considered in this review are described in detail in Sec.\ \ref{sec9}, where the governing dimensionless numbers are introduced too. Sections \ref{sec10} and \ref{sec11} show how those configurations work in the simple dripping and jetting modes, and in their counterparts from tip streaming. We review the scaling laws predicting the sizes of the produced droplets/bubbles and discuss the instability mechanisms which determine the parameter regions where the different modes operate. We pay special attention to electrospray and flow focusing operating in the steady tip streaming regime in Secs. \ref{sec12} and \ref{sec13}, respectively. The paper closes with some prospects in Sec.\ \ref{sec14}.

\section{Theoretical model}
\label{sec2}

In this section, we present the theoretical model which frames the microfluidic applications described in this review. It includes the three major factors that increase the level of complexity of the problem: viscoelasticity, electric fields and surfactants. We also introduce two approximations frequently considered in this context: the leaky-dielectric model for electrohydrodynamic processes, and the one-dimensional (1D) approximation for slender configurations.

% Laminarity
Liquid-liquid microfluidic devices operate in the laminar regime essentially because of their smallness. However, there are gas-liquid configurations in which turbulence may play a relevant role. In particular, the mixing layer between the high-speed gaseous jet and the surrounding ambient in flow focusing \citep{G98a} becomes unstable and renders the flow turbulent at small distances from the discharge orifice. Turbulent viscosity slows down the gaseous jet, which losses most of its kinetic energy a few nozzle diameters beyond the orifice. This effect influences the amount of energy transferred by the gaseous stream to the liquid through viscous shear stresses beyond the discharge orifice. The spinning \citep{OMCAMOM11} and electrospinning \citep{WYPLLLXLW09} of polymeric solutions assisted by a high-speed gas current constitute good examples of partially turbulent microfluidic realizations.

\subsection{Bulk equations}
\label{sec2.1}

% Continuity equation
Consider the density $\rho^{(j)}({\bf r},t)$ and velocity ${\bf v}^{(j)}({\bf r},t)$ fields for the inner ($j=i$) and outer ($j=o$) fluid phases. These fields verify the continuity equation
\begin{equation}
\label{conn}
\frac{\partial \rho^{(j)}}{\partial t} +{\boldsymbol \nabla} \cdot (\rho^{(j)} {\bf v}^{(j)})=0,
\end{equation}
which in the incompressible regime reduces to ${\boldsymbol \nabla}\cdot {\bf v}^{(j)}=0$. This last equation applies to all liquid-liquid configurations reviewed here, and also to microfluidic devices used for producing bubbles \citep{GG01,G04b,RSMG15}. It can also be safely used to describe gas-liquid flows in which the outer gaseous stream moves with velocities smaller than, say, 100 m/s. This last condition holds for gaseous flow focusing devices \citep{G98a} and other liquid ejections assisted with airflow \citep{G07a,JWLLLZ18} if the applied pressure drop does not exceed around 100 mbar. The comparison between numerical simulations and experiments shows that ${\boldsymbol \nabla}\cdot {\bf v}^{(j)}=0$ constitutes a relatively good approximation even for pressure drops up to 250 mbar \citep{CHGM17}. However, compressibility effects must be accounted for in some specific applications; for instance, in solution blow spinning \citep{OMCAMOM11,DBSK16} or when using gaseous flow focusing in the Serial Femtosecond Crystallography \citep{Cetal11}, where the fluid streams are injected on high-vacuum conditions.

% Momentum equation
In the absence of viscoelasticity and electrical forces, the momentum equation reduces to
\begin{equation}
\label{momV}
\rho^{(j)}\left(\frac{\partial {\bf v}^{(j)}}{\partial t}+{\bf v}^{(j)}\cdot {\boldsymbol \nabla}{\bf v}^{(j)}\right)=-{\boldsymbol \nabla}p^{(j)}+{\boldsymbol \nabla}\cdot {\boldsymbol \tau}^{(j)},
\end{equation}
where $p^{(j)}({\bf r},t)$ is the reduced pressure field,
\begin{equation}
\label{NS}
{\boldsymbol \tau}^{(j)}=2\mu^{(j)}{\boldsymbol {\sf D}}^{(j)}+\lambda_{\mu}^{(j)} (\nabla\cdot{\bf v}) \mathbf I
\end{equation}
the total extra stress tensor, $\mu^{(j)}$ the fluid viscosity, ${\boldsymbol {\sf D}}^{(j)}=1/2[{\boldsymbol \nabla}_s {\bf v}^{(j)}+({\boldsymbol \nabla}{\bf v}^{(j)})^T]$ the deformation rate tensor, $\lambda_{\mu}^{(j)}$ the dilatational coefficient of viscosity, and ${\bf I}$ the identity matrix.

% Energy equation
The energy equation 
\begin{eqnarray}
\label{e}
&&\frac{\partial (\rho^{(j)} e^{(j)})}{\partial t}+{\boldsymbol \nabla}\cdot (\rho^{(j)} \mathbf v^{(j)} e^{(j)})=
\nonumber\\&&-{\boldsymbol \nabla}\cdot (p^{(j)}{\bf v}^{(j)})+{\boldsymbol \nabla}\cdot(\boldsymbol\tau^{(j)}\cdot {\bf v}^{(j)})-{\boldsymbol \nabla}\cdot{\mathbf q}^{(j)}
\end{eqnarray}
and the ideal gas law $p^{(j)}=\rho^{(j)} R_g^{(j)} T^{(j)}$ are considered in the gaseous phases when compressibility effects are accounted for. Here, $e^{(j)}=c_v^{(j)} T^{(j)}+1/2\, v^{(j)2}$ and $T^{(j)}$ are the specific energy and temperature fields in each phase, respectively, while $c_v^{(j)}$ and $R_g^{(j)}$ are the corresponding specific heat coefficients and gas constants, respectively. In addition, ${\bf q}^{(j)}=-\kappa_T^{(j)} {\boldsymbol \nabla} T^{(j)}$ is the heat flux vector and $\kappa_T^{(j)}$ the thermal conductivity.

\subsubsection{Viscoelasticty}

% The Oldroyd model: motivation
Many microfluidic applications involve dilute polymer solutions. These liquids exhibit a constant viscosity (shear thinning can be neglected) over a wide range of shear rates so that the major polymer effects are the increase of the solution viscosity $\mu$ with respect to that of the solvent and elasticity \citep{J09}. For quasi-monodisperse molecular weight distributions, it is frequently assumed that elasticity can be approximately quantified by a single characteristic time $\lambda_s$, related to the slowest relaxation process of the entire molecular chain \citep{CPKOMSVM06}. The Olroyd-B model \citep{O50}, or similar approximations including polymer finite extensibility effects \citep{EV08}, has been frequently used in microfluidics to calculate the total extra stress tensor of this type of non-Newtonian liquids. The Olroyd-B model popularity can be attributed to its relative simplicity and the fact that it can be derived from kinetic theory by assuming that the viscoelastic solution is an ideal system of Hookean dumbbells dissolved in a Newtonian liquid \citep{BAH87}. It can also be obtained following a pure continuum approach by assuming (i) a linear relationship between the polymer stress and a certain state variable, (ii) a linear relaxation law for that variable, and (iii) affine motion (i.e., each material point of the polymer follows the flow) \citep{SPHE19}.

% The Oldroyd model: equation
The total extra stress tensor ${\boldsymbol \tau}^{(j)}$ in the Olroyd-B model verifies the constitutive relationship
\begin{equation}
\label{h10}
\left(1+\lambda_s^{(j)}{\sf G}\right){\boldsymbol \tau}^{(j)}=2\mu^{(j)} (1+\lambda_r^{(j)}{\sf G}){\boldsymbol {\sf D}}^{(j)}\; ,
\end{equation}
where $\lambda_s^{(j)}$ and $\lambda_r^{(j)}=\lambda_s^{(j)}\mu_s^{(j)}/\mu^{(j)}$ are the stress relaxation and retardation time, respectively, $\mu_s^{(j)}$ and $\mu^{(j)}$ are the solvent viscosity and solution viscosity at zero shear rate, respectively, and ${\sf G}$ is the upper convective derivative operator. The Navier-Poisson law (\ref{NS}) for an incompresible fluid is recovered for $\lambda_s^{(j)}=0$.

% The Oldroyd model: limitations
The Olroyd-B model is believed to provide reasonable predictions in capillary extensional flows \citep{BER01} when the stress relaxation time is properly adjusted. For this reason, one expects to obtain reliable results from this or similar approximations for microfluidic configurations such as electrospinning \citep{F02,CJ06}, flow focusing or selective withdrawal \citep{ZF10b}, in which the polymer is subject to a strong extensional flow in the tip of the tapering liquid meniscus. In any case, caution must be taken when other capillary flows are analyzed because the Olroyd-B model can lead to important errors for certain polymer solutions \citep{PAHM18}. For instance, \citet{TLEAD18} have recently found considerable discrepancies between the Olroyd-B and experimental \citep{CEFLM06} self-similar dynamics for the final stages of the thinning of a viscoelastic filament.

\subsubsection{Electric fields}
\label{sec2.1.2}

% Momentum equation
Electric forces drive the liquid motion in important microfluidic configurations, such as electrospray \citep{Z14,T64} and electrospinning \citep{DR95,YKR01,YKR01b,TZY04,TCYR07,RY08,HYR08,BK10,Y11}. In the absence of magnetic fields and permittivity gradients in the bulk, the electric volumetric forces are caused by the net free charge exclusively, and the momentum equation (\ref{momV}) reduces to
\begin{equation}
\label{momVI}
\rho^{(j)}\left(\frac{\partial {\bf v}^{(j)}}{\partial t}+{\bf v}^{(j)}\cdot {\boldsymbol \nabla}{\bf v}^{(j)}\right)=-{\boldsymbol \nabla}p^{(j)}+{\boldsymbol \nabla}\cdot {\boldsymbol \tau}^{(j)}+\rho_e^{(j)} {\bf E}^{(j)}.
\end{equation}
Here, $\rho_e^{(j)}({\bf r},t)$ is the (volumetric) charge density and ${\bf E}^{(j)}$ the electric field given by the Maxwell electrostatic equations
\begin{equation}
{\boldsymbol \nabla} \cdot \mathbf{E}^{(j)}=\frac{\rho_e^{(j)}}{\varepsilon^{(j)}}\; , \quad {\boldsymbol \nabla} \times \mathbf{E}^{(j)} ={\bf 0}\; ,
\label{A0b}
\end{equation}
where $\varepsilon^{(j)}$ is the electrical permittivity.

% Conservation of charges in the bulk
In some microfluidic applications, such as electrospray or electrospinning, ionic species, initially present in the liquid or generated at an upstream electrode, migrate across the bulk with zero net production of positive/negative charges owing to electrochemical reactions. In this case, the conservation equation for the volumetric charge density $\rho_e(\mathbf{r},t)$ becomes \citep{GLHRM18}
\begin{equation}
\label{cond}
\frac{\partial \rho_e^{(j)}}{\partial t}+{\bf v}^{(j)}\cdot {\boldsymbol \nabla}\rho_e^{(j)}={\cal D}_{\rho_e}^{(j)}\nabla^2 \rho_e^{(j)}+{\boldsymbol \nabla}\cdot (-K^{(j)} \mathbf{E}^{(j)})\; ,
\end{equation}
where ${\cal D}_{\rho_e}^{(j)}$ is the thermal diffusion coefficient, and 
\begin{equation}
\label{ecv}
K^{(j)}(\mathbf{r},t)=\, \sum_k \omega_k\, {\cal F}\, e\, z_k^2\, n_k^{(j)}
\end{equation}
is the electrical conductivity. Here, $\omega_k$, $z_k$ and $n_k^{(j)}(\mathbf{r},t)$ are the mobility, valence and number of mols per unit volume of the $k$-species, respectively, while ${\cal F}$ and $e$ are the Faraday constant and elementary charge, respectively. The ejection of micrometer size fluid objects demands intense electric fields. In this sense, it is sensible to neglect the migration of electric charges due to thermal diffusion versus the electric drift under the applied electric field. It is also frequent to assume that the dissolved species are distributed over most part of the bulk with a certain degree of uniformity, and, therefore, the electrical conductivity (\ref{ecv}) takes a constant value in that region (the so-called Ohmic conduction model) \citep{GLHRM18}. When this condition does not hold, an electrokinetic model must be adopted.

% Electrokinetic effects
In an electrokinetic model, the distributions of ions, $n_k^{(j)}$, are calculated throughout the fluid domain by solving the corresponding Nernst-Planck conservation equations \citep{GLHRM18}. The electrical conductivity (\ref{ecv}) is calculated from the spatial distributions of ions and their mobilities. Electrokinetic effects must be taken into account when, for instance, the size of the system is comparable to the Debye layer thickness \citep{GLHRM18}. In this case, the electrical conductivity exhibits a strong spatial dependence, and the Ohmic model fails to describe the transport of free charges across the fluid medium. The predictions provided by an electrokinetic model also differ from those of the Ohmic approximation in the disintegration of microdroplets and the pinching of fluid threads. In these problems, an interface can be created at a rate at least of the order of the inverse of the electric relaxation time, which makes the Ohmic model overestimate the injection of charges from the bulk onto the fresh interface. In these examples, the evolution of solutions consisting of ions of opposite charges and different mobilities can significantly depend on the polarity of the applied electric field, an effect not contemplated in the Ohmic model. This is an area of research which has not as yet properly explored.

% Anisotropy
Electrokinetic effects are not the only cause that invalidates the Ohmic model. Equation (\ref{cond}) implicitly assumes that the conduction of electrical charges in the bulk is isotropic. However, in microfluidic configurations such as electrospinning, the presence of macromolecules significantly stretched along the streamwise direction may limit the validity of that assumption in the critical cone-jet transition region. Electrical conduction along the Debye layer may significantly differ from that in the bulk, which may constitute another noticeable source of anisotropy for large surface-to-volume ratios. 
\subsubsection{Surfactants}

% Soluble surfactants
Soluble surfactants play a fundamental role in many microfluidic applications \citep{A16}. For bulk concentrations below the critical micellar concentration (CMC), soluble surfactants are present as monomers in solution. Above that critical concentration, fluid-like aggregates called micelles form spontaneously. The volumetric concentration of surfactant as monomers, $c^{(j)}({\bf r},t)$, and micelles, $m^{(j)}({\bf r},t)$, are calculated in a fluid dynamical problem from the conservation equations \citep{CMP09}
\begin{equation}
\label{c11}
\frac{\partial c^{(j)}}{\partial t}+{\bf v}^{(j)}\cdot {\boldsymbol \nabla} c^{(j)}={\cal D}_c^{(j)}\nabla^2 c^{(j)}-n{\cal J}_{cm}^{(j)},
\end{equation}
\begin{equation}
\label{c22}
\frac{\partial m^{(j)}}{\partial t}+{\bf v}^{(j)}\cdot {\boldsymbol \nabla} m^{(j)}={\cal D}_m^{(j)}\nabla^2 m^{(j)}+{\cal J}_{cm}^{(j)},
\end{equation}
where ${\cal D}_c^{(j)}$ and ${\cal D}_m^{(j)}$ are the diffusion coefficients for the surfactant as monomers and micelles, respectively, $n$ is the number of monomers that constitute a micelle, while ${\cal J}_{cm}^{(j)}$ stands for the net rate either of formation (${\cal J}_{cm}^{(j)}>0$) or breakup (${\cal J}_{cm}^{(j)}<0$) of micelles per unit volume.

% The sublayer
In many droplet production techniques, the dispersed phase is injected from a reservoir at equilibrium (${\cal J}_{cm}^{(j)}=0$) with uniform monomer and micelle concentrations ($c=$const., $m=$const.). In this case, Eqs.\ (\ref{c11}) and (\ref{c22}) show that those concentrations are convected by the fluid particles so that they remain constant throughout most of the liquid domain. Spatial variations of surfactant concentration can arise in the sublayer next to the interface, which constitutes a source/sink of surfactant molecules during the system evolution. The transfer of surfactant molecules from the bulk to the fresh interface created during the atomization is essentially governed by the adsorption/desorption process and/or diffusion within the sublayer, while bulk diffusion and convection are much less relevant.

\subsection{Interface boundary conditions}
\label{sec2.2}

% The physics of interfaces. Topological reduction
{\it ``God made the bulk; surfaces were invented by the devil" (Wolfgang Pauli)}. 

Due to the large surface-to-volume ratios reached in microfluidics, interfaces play a critical role in the dynamics of the fluid system. In fact, they contain most of the physics of the problem, which must be modeled accurately. Interfaces are barriers preventing the continuous diffusion of free ions under applied electric fields. The accumulation of charges onto those surfaces and the jump of electrical permittivity in that region substantially affect surface forces and their equilibrium. Surface active molecules adsorb at interfaces and form monolayers which locally reduce the interfacial tension and can exhibit rheological properties. Interfacial (Debye, surfactant,\ldots) layers are typically much thinner than the rest of the fluid domain, and, therefore, the resolution of their spatial structure is a difficult task. For this reason, they are topologically reduced to surfaces and introduced into the problem as boundary conditions.

% Balance of stresses
The balance of stresses on the two sides of the interface reflects the complexity of the problem considered. In the absence of electric fields and surfactants, it yields
\begin{equation}
\label{stred}
\mathbf{n}\cdot ||{\boldsymbol \tau}^{(j)}||-\left[||p^{(j)}||+||\rho^{(j)}|| ({\bf g}\cdot{\bf r})\right]\mathbf{n}=\sigma ({\boldsymbol \nabla}\cdot {\bf n}) {\bf n},
\end{equation}
where ${\bf n}$ is the unit outward normal vector, $||A||$ denotes the difference $A^{(o)}-A^{(i)}$ between the values taken by the quantity $A$ on the two sides of the interface, ${\bf g}$ is the gravitational acceleration, and $\sigma$ is the the interfacial tension. As will be shown below, Eq.\ (\ref{stred}) is completed by additional stresses when electric fields and surfactants are present.

% Kinematic compability
Neither of the two phases can cross the interface separating immiscible fluids, which leads to the kinematic compatibility boundary condition
\begin{equation}
\frac{\partial f}{\partial t}+ \mathbf{v}^{(j)}\cdot \nabla f=0\; .
\end{equation}
The equation $f(\mathbf{r}_s,t)=0$ determines the interface position $\mathbf{r}_s$. Alternatively, one can also define the distance $F(\theta,z,t)$ of an interface element from the axis $z$ of a cylindrical coordinate system $(r,\theta,z)$. This function obeys the equation
\begin{equation}
\frac{\partial F}{\partial t}-v^{(j)}_r+\frac{v^{(j)}_{\theta}}{F}\frac{\partial F}{\partial \theta}+v^{(j)}_z\frac{\partial F}{\partial z}= 0\; ,
\end{equation}
where $v_r^{(j)}$, $v_{\theta}^{(j)}$ and $v_z^{(j)}$ stand for the radial, angular and axial components of the velocity field, respectively. This last formulation allows for the implementation of interface-tracking techniques \citep{RK98} and boundary fitted methods \citep{TW82a} to numerically integrate the hydrodynamic equations. It also facilitates imposing the anchorage of triple contact lines in the numerical simulation.

\subsubsection{Electric fields}

% Electric boundary conditions. Conservation of charges
When electric fields are present, they must be calculated considering the interface boundary conditions
\begin{equation}
\label{maxs}
||\varepsilon^{(j)} {\bf E}^{(j)}||\cdot {\bf n}=\sigma_e, \quad ||\varepsilon^{(j)} {\bf E}^{(j)}||\times {\bf n}={\bf 0},
\end{equation}
where $\sigma_e$ is the surface charge density. The conservation equation for this quantity reads
\begin{equation}
\label{sigmae}
\frac{\partial \sigma_e}{\partial t}+{\boldsymbol \nabla}_s(\sigma_e {\bf v}_s)+\sigma_e{\bf n}\cdot ({\boldsymbol \nabla}_s\cdot {\bf n}){\bf v}=-||K^{(j)} {\bf E}^{(j)}||\cdot {\bf n},
\end{equation}
where ${\boldsymbol \nabla}_s={\boldsymbol {\sf I}}_s {\boldsymbol \nabla}$ is the surface gradient operator,
${\boldsymbol {\sf I}}_s={\boldsymbol {\sf I}}-{\bf n}{\bf n})$ the tensor that projects any vector onto the interface, ${\boldsymbol {\sf I}}$ the identity tensor, and ${\bf v}_s={\boldsymbol {\sf I}}_s{\bf v}^{(j)}$ the surface velocity. In the above equation, both conduction and diffusion {\em along} the interface have been neglected. In configurations such as liquid bridges between two electrodes, the variation of the charge density due to surface compression/dilatation and convection is typically neglected, which allows to decouple the calculation of the electric field from that of the velocity field \citep{PET01,BS02,MP20}. However, this approximation is not valid in tip streaming configurations such as electrospray or electrospinning, in which surface charge convection plays an important role.

% Maxwell stresses
When electric fields are applied, the balance of stresses at the interface (\ref{stred}) is completed by adding the Maxwell stresses
\begin{equation}
{\boldsymbol \tau}_M^{(j)}=\varepsilon^{(j)} \left({\bf E}^{(j)}{\bf E}^{(j)}-\frac{1}{2}{\boldsymbol {\sf I}}E^{(j)2}\right)
\end{equation}
to the left-hand side of that equation ($||{\boldsymbol \tau}^{(j)}||\to ||{\boldsymbol \tau}^{(j)}||+||{\boldsymbol \tau}_M^{(j)}||$).

% Applications of the full equations
The equations presented above and in Sec.\ \ref{sec2.1.2} allow one to describe electrohydrodynamic phenomena with net free charge both in the bulks and the interface. Typical examples of these phenomena are some flows driven by AC electric fields \citep{YGC05,CMC08,MC09,CCGC10}, the initial phase of ejections from charged liquid surfaces \citep{GT94,GLRM16}, the oscillation of liquid menisci with periodic emissions of charged liquid droplets \citep{JK99,HLIH15}, the free surface pinching in charged liquid jets \citep{LGPH15}, or the disintegration of very small drops \citep{PBHD16}. \citet{CMCP11a} presented an electrokinetic model to describe the breakup of a jet loaded with electrically charged surfactants. This model involves non-zero volumetric and surface charge densities, and, therefore, requires integrating the corresponding conservation equations (\ref{cond}) and (\ref{sigmae}) for the bulk and interface, respectively.

% Coarse-grained simulations
When the flow conditions are such that interfaces move slowly in comparison with the diffusion velocity of charges under the action of electric fields, charges accumulate onto those interfaces creating a layer where molecular diffusion is halted by electric drift (the so-called Debye layer). As mentioned above, the resolution of the Debye layer structure becomes computationally unaffordable in many cases due to the disparity between the layer thickness (the Debye length) and the system size. This problem has been obviated in coarse-grained simulations where molecular diffusion is not considered, the Debye layer is not resolved, and the electrical conductivity is assumed to be constant \citep{FLHMA13,GLRM16,DJE18}.

\subsubsection{The leaky-dielectric model}

% The Taylor-Melcher leaky-dielectric model
The Taylor-Melcher leaky-dielectric model \citep{T66,MT69,S97,SY15} has become the most popular alternative to overcome the obstacle mentioned above. The fundamental approximation of this model is to assume that all the net free charge accumulates at the interface within a Debye layer much thinner than the system size. This implies that (i) the charge distribution can be described in terms of the surface charge density $\sigma_e$ exclusively ($\rho_e=0$), which accounts for the net free charge contained in the Debye layer, and (ii) the electrical conductivity can be regarded as a constant throughout the liquid domain (the Ohmic conduction model). Therefore, Eq.\ (\ref{cond}) is no longer necessary, and electric forces in the bulk can be neglected. One can probably state that \citet{MT69} defined through their pioneering work the field of electrohydrodynamics, where the interaction between low-conductivity liquids and strong electric fields continues to yield new and intriguing phenomena \citep{V19}. 

\begin{figure}[hbt]
%\begin{center}
\centering{\resizebox{0.45\textwidth}{!}{\includegraphics{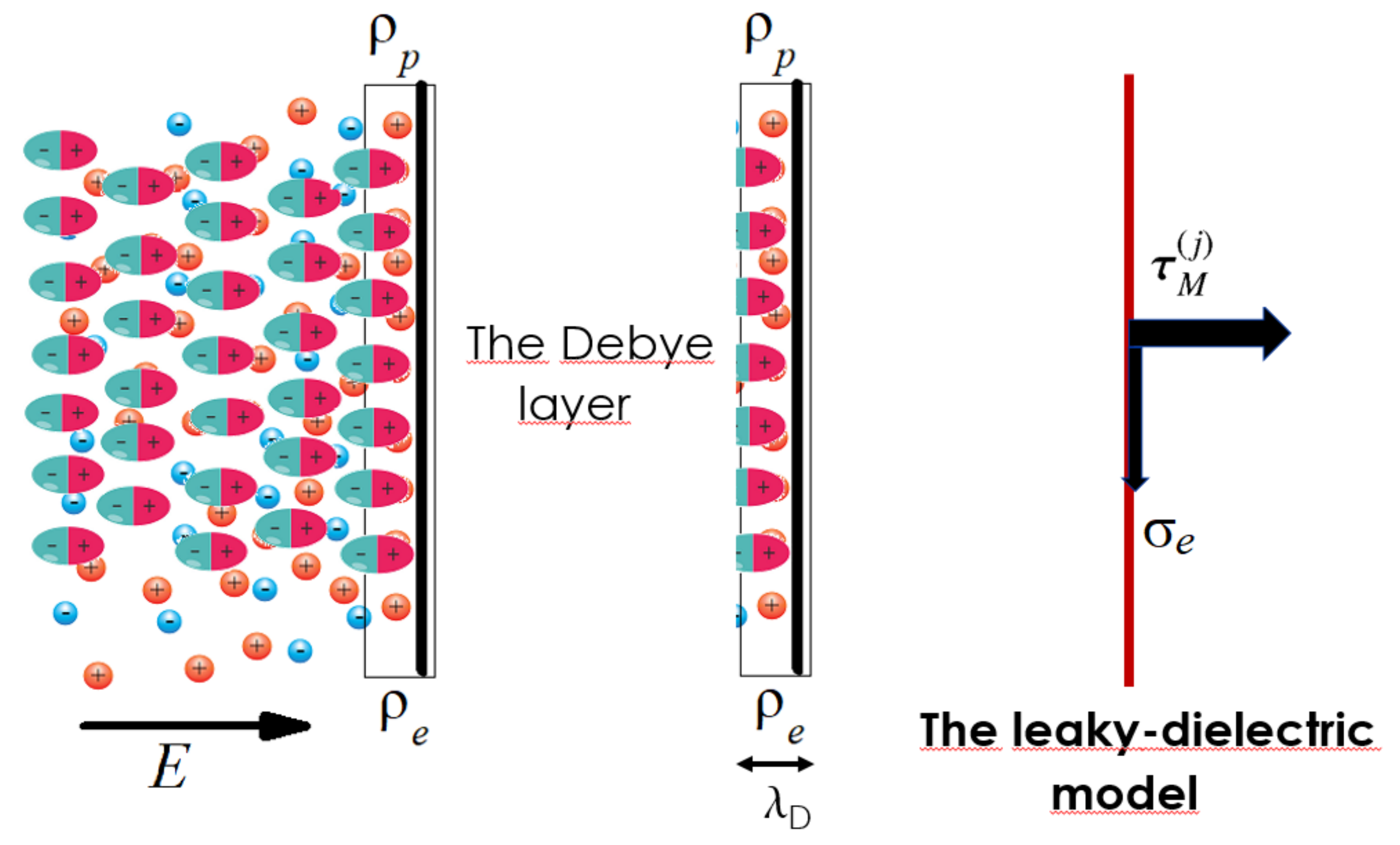}}}
%\end{center}
\caption{The simplification in the leaky-dielectric model. The Debye layer of thickness $\lambda_D$ is reduced to a infinitely thin surface with a free charge density $\sigma_e$ and subject to Maxwell stresses ${\boldsymbol \tau}_M^{(j)}$. These stresses result from both the net free charge accumulated in the Debye layer ($\rho_e$) and the interfacial charge polarization ($\rho_p$). The net free charge is neglected in the bulk.}
\label{debye}
\end{figure}

% Applications
The leaky-dielectric model has proved to be a useful tool to examine the dynamical behavior of poorly conducting droplets in poorly conducting baths. In particular, it provides accurate predictions for the steady cone-jet mode of electrospray \citep{G97a,G04a,F07,H10,HLGVMP12,PRHGM18,B18,GM18,JGS20} and electrospinning \citep{CJ06,RY08}, two techniques reviewed in this paper. It has been used to describe AC electrospray phenomena \citep{YLWC04,DPR11}, and has been extended to simulate ionic liquid menisci undergoing evaporation of ions \citep{H08c,CML19}. 

% Anisotropy
The leaky-dielectric model is not exempt from severe limitations. For instance, its extension to include anisotropic and/or inhomogeneous conductivity can violate the conservation of volumetric charge ($\nabla \cdot \mathbf{j}^{(j)}=0$, $\mathbf{j}^{(j)}$ is the current density), which is automatically satisfied for constant scalar conductivity ($\nabla \cdot \mathbf{j}^{(j)}=\nabla \cdot (K^{(j)} \mathbf{E}^{(j)}) =K^{(j)}\nabla \cdot \mathbf{E}^{(j)}=\rho_e^{(j)}/\varepsilon^{(j)}=0$). This implies that if the electrical conductivity is linked to, e.g., the state of dissolved polymers in a viscoelastic solution, then the volumetric charge density in the bulk must be calculated even if electric forces are neglected there.

\subsubsection{Surfactants}

% Surfactant surface equations
Surfactants present in solution as monomers adsorb onto the interface. At equilibrium, the Langmuir adsorption isotherm relates the volumetric concentration $c^{(j)}$ and surface distribution $\Gamma$ of surfactant if $c^{(j)}<\text{CMC}$. For $c^{(j)}>\text{CMC}$, the surface concentration $\Gamma$ saturates to an approximately constant value $\Gamma_c$.

In a non-equilibrium state, $\Gamma$ verifies the conservation equation \citep{CMP09}
\begin{equation}
\frac{\partial \Gamma}{\partial t}+{\boldsymbol \nabla}_s(\Gamma {\bf v}_s)+\Gamma{\bf n}\cdot({\boldsymbol \nabla}_s\cdot {\bf n}){\bf v}={\cal D}_s\nabla_s^2 \Gamma+{\cal J}^{(i)}_{\Gamma c}+{\cal J}^{(o)}_{\Gamma c},
\end{equation}
where ${\cal D}_s$ is the surface diffusion coefficient, and ${\cal J}_{\Gamma c}^{(j)}$ denotes the net flux of surfactant from the bulk to the interface due to the adsorption/desorption process. To derive this equation, one supposes that the micelles do not adsorb directly onto the interface, but that they completely dissociate into bulk monomers prior to the adsorption process \citep{CMP09}.

The adsorption/desorption net flux must equal the diffusive flux in the sublayer next to the interface, i.e.
\begin{equation}
{\cal D}_c^{(j)}{\boldsymbol \nabla}c^{(j)}\cdot {\bf n}=\pm {\cal J}^{(j)}_{\Gamma c}, \quad {\cal D}_m^{(j)}{\boldsymbol \nabla}m^{(j)}\cdot {\bf n}=0,
\end{equation}
where the sign $+$ and $-$ applies to $j=o$ and $i$, respectively.

Due to the small values taken by surface diffusion coefficient ${\cal D}_s$ for strong surfactants \citep{T97}, the convection of these molecules over the interface is much more important than the diffusion mechanism in most microfluidic applications (the surface P\'eclet number is much greater than unity), and the latter is neglected.

% The two limits
In many configurations, the hydrodynamic characteristic time is much smaller than that characterizing the adsorption/desorption process, which implies that the surfactant can be regarded as insoluble. This considerably simplifies the analysis because it eliminates both Eqs.\ (\ref{c11}) and (\ref{c22}) and the (normally unknown) quantities $\{{\cal D}_c^{(j)}$, ${\cal D}_m^{(j)}$, ${\cal J}_{cm}^{(j)}$, ${\cal J}^{(j)}_{\Gamma c}\}$ from the problem.

The opposite limit to insolubility is that in which the adsorption/desorption kinetics is sufficiently fast for the volumetric concentration at the interface, $c_s$, to be in local equilibrium with the surface concentration $\Gamma$ \citep{JS07}. In this case, $c_s$ evolves according to the Langmuir equation
\begin{equation}
\frac{\Gamma}{\Gamma_{\infty}}=\frac{\hat{\beta}/\hat{\alpha}\, c_s}{1+\hat{\beta}/\hat{\alpha}\, c_s},
\end{equation}
where $\Gamma_{\infty}$ is the maximum packing density, and $\hat{\beta}$ and $\hat{\alpha}$ are the kinetic constants for adsorption and desorption, respectively. In this limit, the transport of surfactant molecules between the bulk and the interface is limited by diffusion in the sublayer next to that surface.

% Isotherm
The dependence of the surface tension $\sigma$ upon the surfactant surface concentration $\Gamma$ is frequently calculate from the Langmuir equation of state \citep{T97}
\begin{equation}
\label{es}
\sigma=\sigma_0+R_g T \Gamma_{\infty}\ \text{ln}\left(1-\frac{\Gamma}{\Gamma_{\infty}}\right),
\end{equation}
where $\sigma_0$ is the surface tension of the clean interface, $R_g$ the gas constant, and $T$ the temperature. Experiments show that $\sigma(\Gamma)$ reaches a plateau at $\Gamma\simeq \Gamma_c$. This effect is not captured by Eq.\ (\ref{es}), which must be replaced by an appropriate equation of state if the volumetric concentration $c^{(j)}$ is expected to reach values close to the CMC.

% Balance of stresses
Fresh interfaces between two immiscible fluids are constantly formed in droplet emulsification produced by microfluidic devices. If the multiphase system contains surfactants with adsorption times larger than or comparable to the droplet formation time, the interface may be subjected to a dynamic interfacial tension different from that measured at equilibrium [Eq.\ (\ref{es})]. The balance of stresses at the interface, Eq.\ (\ref{stred}), involves now the local value of the surface tension (the so-called solutocapillarity effect). In addition, this boundary condition is completed by adding the term ${\boldsymbol \tau}_{st}+\tau_{sn} {\bf n}$ to the right-hand side of Eq.\ (\ref{stred}). Here, ${\boldsymbol \tau}_{st}$ and $\tau_{sn}$ are the surface stresses tangential and normal to the interface, respectively, both associated with the existence of a surfactant monolayer at that surface.

% Surface stresses
The tangential component of the surface stress includes both the Marangoni stress due to the surface tension (surfactant concentration) gradient and the superficial viscous stress associated with the variation of the surface velocity ${\bf v}_s$. Marangoni and superficial viscous stresses tend to eliminate inhomogeneities of surfactant concentration and surface velocity, respectively.

% Newtonain interface
The superficial viscous stress obeys different constitutive relationships depending on the surfactant molecule nature. For a Newtonian interface \citep{S60}, the surface stress can be calculated as \citep{LH98}
\begin{equation}
{\boldsymbol \tau}_{st}={\boldsymbol \nabla}_s\sigma+{\boldsymbol \nabla}_s\left[\left(\kappa^s-\mu^s\right) \left({\boldsymbol \nabla}_s\cdot {\bf v}_s\right)\right]+2{\boldsymbol \nabla}_s\cdot (\mu^s{\boldsymbol {\sf D}}_s),
\end{equation}
where ${\boldsymbol {\sf D}}_s=1/2\, [{\boldsymbol \nabla}_s {\bf v}_s\cdot {\boldsymbol {\sf I}}_s+{\boldsymbol {\sf I}}_s\cdot ({\boldsymbol \nabla}_s {\bf v}_s)^T]$ is the surface rate of deformation tensor, $\kappa^s$ the dilatational surface viscosity, and $\mu^s$ the shear surface viscosity. These two surfactant properties depend on the surfactant surface concentration $\Gamma$ \citep{LSB19}. Adsorbed surfactant monolayers at fluid surfaces usually exhibit rheological properties too. In fact, surface viscosities frequently depend on the timescale and amplitude of the deformation owing to surface relaxations and nonlinear responses \citep{L14}.

% Importance of surface viscosities
The lack of precise information about the values taken by the surface viscosities, as well as the mathematical complexity of the calculation of the surface viscous stresses, has motivated that most of the experimental and theoretical works in microfluidics do not take into account those stresses. However, they may considerably affect the dynamics of interfaces even for surface viscosities much smaller than the bulk ones \citep{PMHVV17}. This may occur for two reasons: (i) surface viscous stresses may significantly alter the transport of surfactants over the interface, which may have important consequences in the resulting solutocapillarity effect and Marangoni stresses; and (ii) their relevance increases as the surface-to-volume ratio increases, as happens, for instance, during the interface breakup \citep{PMHVV17}.

% Surface viscosity and foams
It is believed that foam and emulsion stability can be caused by the surface shear viscosity of the surfactant used to stabilize them. In fact, surface viscosity can significantly increase the drainage time during the coalescence of two bubbles/droplets \citep{OJ19}. However, there is no clear evidence that soluble and small-molecule surfactants have measurable surface shear viscosities \citep{ZNMLDMTS14}. This raises doubts about the role played by surface shear rheology in the stability of foams and emulsions treated with soluble surfactants. In fact, surfactants can stabilize emulsions through Marangoni stresses too. The surfactant depletion in the center of the gap between two approaching interfaces produces surface tension gradients. The resulting Marangoni stresses resist the outwards radial flow in the gap, thus preventing coalescence \citep{EBW91}. Surface diffusion of surfactant hinders this mechanism as the size of the coalescending droplets decreases. A similar effect is produced by the surfactant solubility when the adsorption-desorption time is comparable to that of the gap drainage.

\subsection{Solid boundary conditions}

The formulation of the problem is completed by imposing the noslip condition and zero diffusion flux of surfactants at the solid surfaces. In addition, triple contact lines must be pinned when they meet edges delimiting solid elements of microfluidic devices.

% Anchorage boundary condition
The triple contact line anchorage condition must also be imposed when studying the linear stability of capillary systems interacting with real surfaces, i.e., those exhibiting contact angle hysteresis. As discussed by \citet{D79}, the contact angle of the unperturbed state takes a value in the interval delimited by the receding and advancing contact angles, for which the contact line velocity vanishes. Because linear perturbations produce only infinitesimal variations around that angle, the triple contact line remains fixed during the evolution of those perturbations. In the nonlinear regime, the dynamic contact angle can take values outside the interval mentioned above. In this case, the triple contact line slips over the solid surface. The dynamic contact angle depends on the triple contact line speed, although it loses its sensitivity to that quantity as the latter increases in value \citep{D79}.

% Motion of the triple contact line
When the triple contact line moves, the noslip boundary condition inevitably leads to a singularity at that line. For this reason, one usually adopts the so-called slip model \citep{G78,D80,YD87},
\begin{equation}
\label{slip}
{\bf v}^{(j)}\cdot {\bf e_{\parallel}}={\sf s}^{(j)}\, ({\bf e_{\parallel}}\cdot \boldsymbol{\tau}^{(j)}\cdot {\bf e_{\perp}}),
\end{equation}
at the solid surface, where ${\bf e_{\parallel}}$ stands for any of the two unit vectors on the solid surface, ${\bf e_{\perp}}$ is the outward unit vector perpendicular to that surface, and ${\sf s}^{(j)}$ is the slip coefficient.

\subsection{The 1D approximation}
\label{sec2.4}

% The 1D model
The theoretical model described above can be greatly simplified when the inner fluid (typically a jet) adopts a slender shape along the streamwise direction $z$. In this case, the inner velocity profile is approximately parallel and uniform \citep{E93}. If one also considers the leaky-dielectric approximation, and neglects the dynamical effects of the outer medium, the 1D model for steady flow becomes \citep{E97,G99b,F02,F03,CJ06,WWKB20}
\begin{equation}
Q=\pi F^2 w,
\end{equation}
\begin{equation}
I=\pi F^2 K^{(i)} E_z+2\pi F w\sigma_e,
\end{equation}
\begin{eqnarray}
\label{m1d}
\rho^{(i)} w w'&=&\rho^{(i)} g+\frac{T'}{\pi F^2}-\left(\frac{\sigma }{F}\right)'+\frac{\sigma_e\sigma_e^{'}}{\varepsilon^{(o)}}\nonumber\\
&+&(\varepsilon^{(i)}-\varepsilon^{(o)})E_z E_z'+\frac{2\sigma_e E_z}{F},
\nonumber\\ &+&\frac{2\sigma'}{F}+\frac{(9\mu^S F w')'}{2F^2}+\frac{(\kappa^S F w')'}{2F^2},
\end{eqnarray}
where $Q$ and $I$ are the flow rate and electric current transported by the jet, respectively, $F(z)$ and $w(z)$ are the interface contour and jet velocity, respectively, $E_z(z)$ is the axial component of the electric field, and $T=\pi F^2 (\tau_{zz}-\tau_{rr})$ is the tensile force in the jet. The prime denotes the derivative with respect to the $z$ coordinate. The elements $\tau_{zz}$ and $\tau_{rr}$ of the total extra stress tensor are given by the corresponding constitutive relationship: Navier-Poisson law \citep{G99a}, Olroyd-B model \citep{CJ06}, FENE-P model \citep{G17b}, etc. The last three terms in Eq.\ (\ref{m1d}) correspond to the Marangoni stress and the surface viscous stresses caused by a surfactant monolayer \citep{WWKB20}. The shear and dilatational viscosity terms have the same form, and, therefore, the relevant parameter to this order becomes $9\mu^S+\kappa^S$. Something similar occurs in the analysis of films, in which the two surface viscosities are indistinguishable from each other because they enter the problem through the single parameter $\mu^S+\kappa^S$ \citep{HFY20}.

% Applications
The 1D approximation provides useful predictions for many microfluidic configurations considered in this review, such as jets emitted at large enough flow rates in the gravitational \citep{JEBHR13}, electrospray \citep{PRHGM18}, electrospinning \citep{CJ06}, coflowing \citep{GSC14} and flow focusing \citep{GFM11} configurations (in the last two cases, the dynamical effects of the outer medium are to be taken into account). However, important 2D effects can be erroneously neglected in the inception of the emitted jet and in the later stages of a viscoelastic filament thinning \citep{TLEAD18}.

% Twofold interpretation
Equation (\ref{m1d}) admits a twofold interpretation. In an Eulerian frame of reference, its spatial integration leads to the balance between the driving and resistant forces acting on the whole liquid thread. On the other hand, if one focuses on a liquid slice moving throughout the fluid domain, then the terms of Eq.\ (\ref{m1d}) yields the kinetic energy supplied to or withdrawn from that slice between $z$ and $z+dz$, which allows one to approximately compute the so-called ``energy budget"\ for the flow.

\subsection{Searching for scaling laws}

Many microfluidic applications described in this review involve complex phenomena, for which theoretical analyses based on first principles may not lead to practical results. In this case, it is very useful to search for scaling laws that unify the description of similar experiments and allow one to identify the physical mechanisms governing the associated applications.

Little can be said, in general terms, about the above-mentioned purpose beyond what \citet{B03a} explained in his remarkable text on dimensional analysis, self-similarity, intermediate asymptotics and scaling laws. In particular, the author devotes one of the chapters to the most general class of problems that exhibit scaling laws: those showing incomplete similarity. These problems are characterized by the existence of a canonical set $\{\chi_1,\chi_2,\ldots,\chi_N\}$ of $N$ non-dimensional parameters governing the analyzed variable $a_0$ when the latter is written in a {\em physically meaningful} non-dimensional form, $\chi_0=a_0/\widetilde{a}_0$, where $\widetilde{a}_0=\Pi_{m=1}^M a_i^{\beta_m}$ is the characteristic scale of $a_0$ expressed in terms of $M$ significant dimensional parameters $\{a_1,a_2,\ldots,a_M\}$ and $\beta_m$ are rational exponents.

% Fitting the variables
The sought scaling law typically involves $P$ fitting dimensionless parameters $\{\alpha_1,\alpha_2,\ldots,\alpha_P\}$, i.e. 
\begin{equation}
\chi_0=f\left(\chi_1,\chi_2,\ldots,\chi_N;\alpha_1,\alpha_2,\ldots,\alpha_P\right).
\end{equation}
To determine the values of those parameters, one may calculate the  probability density function PDF$(\alpha_1,\alpha_2,\ldots,\alpha_P)$ of the logarithmic errors
\begin{eqnarray}
\epsilon^{(k)}&=&\log f(\chi_1^{(k)},\chi_2^{(k)},\ldots,\chi_N^{(k)};\alpha_1,\alpha_2,\ldots,\alpha_P)\nonumber\\&&-\log\chi_0^{(k)}\quad (k=1,2,\ldots,K)
\end{eqnarray}
for different values of the set $\{\alpha_1,\alpha_2,\ldots,\alpha_P\}$. Here, $\{\chi_0^{(k)},\chi_1^{(k)},\chi_2^{(k)},\ldots,\chi_N^{(k)}\}$ represent the values of the corresponding dimensionless variables measured in the $k$th experimental realization or numerical simulation. Then, the normal distribution with zero average is fitted to the resulting PDF. The optimum values of $\{\alpha_1,\alpha_2,\ldots,\alpha_P\}$ are those leading to the normal distribution with minimum variance.

In most problems, this general guidance is hampered by the lack of sufficient experimental data or the limited physical knowledge of the problem. Nevertheless, these limitations do not pose insurmountable barriers in many areas of physics. This is the case of the particular field considered in this review, characterized by the existence of {\it motion} and fluid {\it interfaces}. The former always demands a source of energy, while the latter enables the concurrence of different types of surface energies. A finite amount or a continuous flow of energy can be supplied to the system depending on whether its motion is incipient or steady, respectively. In most cases, that motion is limited by either dissipative or mass (potential) forces. Typically, the balance between driving and resistant forces at a certain critical situation allows completing the set of equations that determine the exponents of the scaling laws. Most of the scaling laws reviewed here have been derived following these general ideas.

\section{Stability analysis}
\label{sec3}

In this section, we review some of the general ideas which underlie the stability analysis of the configurations considered in this work. Results related to those configurations are reviewed in Sec.\ \ref{sec4}.

\subsection{Local stability analysis}
\label{sec3.1}

% Linear stability analysis
The direct numerical simulation of the 3D model described in Sec.\ \ref{sec2} constitutes a difficult task, even if simplifications like the leaky-dielectric model or the surfactant insolubility condition are taken into account. Nevertheless, relevant information can be extracted by conducting the linear stability analysis of the base flow. In this analysis, one avoids the time integration of the model by splitting the calculation into two parts: the steady base flow and its linear modes. These modes describe the base flow response to small-amplitude perturbations, which determines the system stability in most cases.

% Local linear stability analysis
The problem becomes analytically or semi-analytically tractable when the stability analysis is conducted locally. In this analysis, one assumes that the characteristic length of the perturbations is much smaller than the hydrodynamic length of the base flow in the streamwise direction (the symmetry axis). In this way, one supposes that this flow is locally homogeneous in that direction. This is commonly referred to as the WKBJ approximation and allows one to examine the stability of slowly spatially varying base flows. In this approximation, the evolution of perturbations in the linear regime at a given flow station can be described as the superposition of the normal modes
\begin{equation}
\Phi^{(j)}(r,\theta,z;t)=\phi^{(j)}(r) e^{i(m\theta+kz-\omega t)},
\end{equation}
where $\Phi^{(j)}$ represents any variable of the problem, $(r,\theta,z)$ is a cylindrical coordinate system whose $z$-axis is the base flow symmetry axis (the streamwise direction), $\omega=\omega_r+i\omega_i$ and $k=k_r+ik_i$ are the perturbation eigenfrequency and wavenumber, respectively, while $m$ is the azimuthal mode number.

% The dispersion relationship
The fulfillment of the hydrodynamic equations and boundary conditions determines the spatial structure of the linear mode, $\phi^{(j)}(r)$, and, more importantly, leads to the (dimensionless) dispersion relationship
\begin{equation}
\label{disper0}
{\cal D}_m(k,\omega;\{{\cal P}_i\})=0,
\end{equation}
which gives the eigenfrequency $\omega$ of a mode with azimuthal and axial wavenumber $m$ and $k$, respectively, as a function of the parameters $\{{\cal P}_i\}$ ($i=1,2,\ldots,N$) characterizing the problem. The dispersion relationship is applied throughout the flow considering the local values of the parameters $\{{\cal P}_i\}$ of the problem. The total growth of the perturbation results from the integration of the Eulerian growth rate $\omega_i(z)$ along Lagrangian paths, taking into account the variation of $k$ and $\{{\cal P}_i\}$ along those paths (see, e.g., \citep{JEBHR13}).

The relative simplicity of the calculation of the dispersion relationship has favored the application of the local stability analysis to a plethora of problems, many of them with little connection with experiments or applications. 

\subsubsection{Temporal and spatial stability analyses}

\begin{figure}[hbt]
%\begin{center}
\centering{\resizebox{0.15\textwidth}{!}{\includegraphics{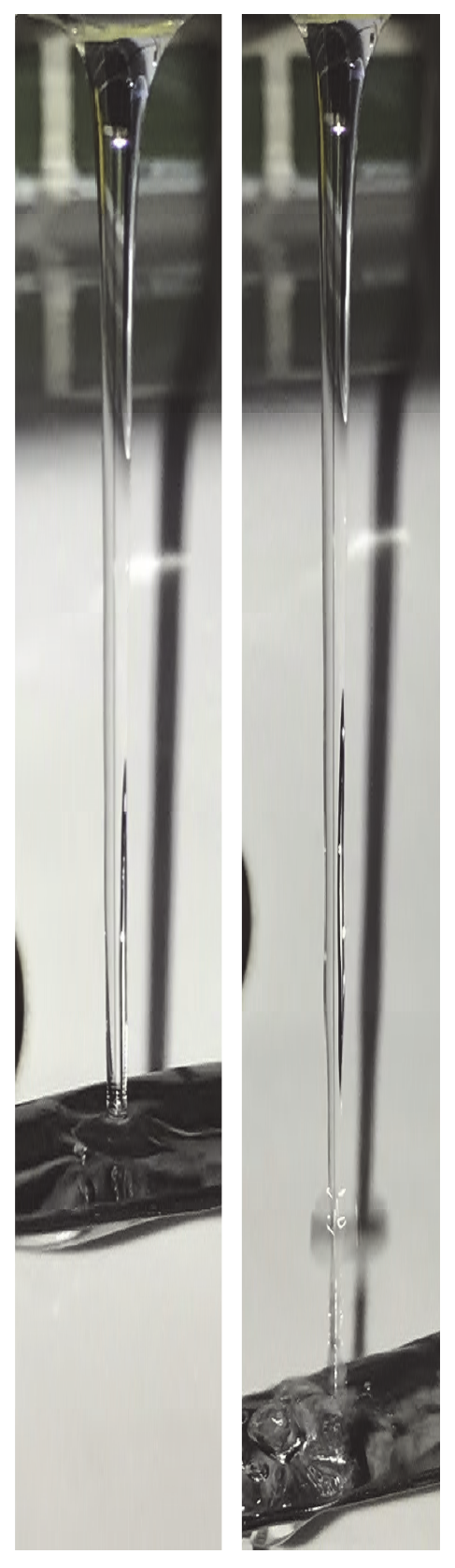}}}
%\end{center}
\caption{(Left) Gravitational-capillary water jet intercepted at a distance from the tap smaller than the breakup length. The jet is strictly stable. One may observe the classical small wavelength ripples formed right upstream of the interception point. (Right) The jet is intercepted beyond the breakup region, which does not alter its convectively unstable nature.}
\label{chorros}
\end{figure}

% Stable jets. Intact region
Most of the jets produced in microfluidic applications eventually break up, and, therefore, they are unstable in a strict sense. In this context, the adjective {\em stable} means {\em convectively unstable}, as will be explained in the next section. In many cases, the shape of a stable (convectively unstable) jet is nearly indistinguishable from that corresponding to the base flow (unperturbed sate) except close to the breakup region (Fig.\ \ref{chorros}). The fluid domain where perturbations are hardly noticeable is frequently called the {\em intact region}. There are certain situations in which strictly stable jets are formed; for instance, when a jet hits a downstream steady boundary condition that precludes the growth of perturbations (Fig.\ \ref{chorros}).

% Temporal stability analysis
If a jet is convectively unstable, the temporal stability analysis allows one to predict the most important aspects of the breakup process. In this analysis, the growth rate $\omega_i$ is calculated as a function of the real wavenumber $k$ and the parameters $\{{\cal P}_i\}$ characterizing the problem. One is typically interested in whether a certain factor (electric field, viscoelasticity, surfactant, \ldots) has a stabilizing or destabilizing effect. In the former case, the growth rates, the range of unstable wavenumbers and the most unstable wavenumber generally decrease, while the opposite occurs when destabilization takes place.

% Most unstable jet, breakup length, droplet diameter
In the temporal stability analysis of a capillary jet, the (dimensional) growth rate $\omega_i^{\textin{max}}$ and wavelength $\lambda^{\textin{max}}$ of the most unstable mode are probably the most interesting quantities. They allow one to estimate the jet breakup length $l_b$ and droplet diameter $d_d$ as
\begin{equation}
\label{esti}
l_b\sim V_j/\omega_i^{\textin{max}}, \quad \pi R_j^2 \lambda^{\textin{max}}\simeq \pi d_d^3/6,
\end{equation}
where $V_j$ and $R_j$ are the jet's mean velocity and radius, respectively. In the first expression, one implicitly assumes that the perturbation responsible for the breakup is born next to the jet inception region, and that this perturbation is convected by the jet, i.e. the capillary velocity is much smaller than that of the jet. In the second expression, we take into account that the volume distribution after the jet breakup is essentially decided before nonlinear effects come into play.

Equation (\ref{esti})-left has been used to calculate the breakup length of gravitational \citep{JEBHR13} and, more recently, electrified \citep{IYXS18} jets. Equation (\ref{esti})-right is the expression most commonly used to estimate the droplet diameter in the jetting regime. \citet{CGFG09} have proposed an alternative way to derive that expression, and have shown how to correct it to calculate the droplet diameter following the breakup of widening jets in the coflowing configuration. 

% Umemura
The temporal stability approach may suggest that the breakup length should depend on the details of the ejection procedure and geometry, which are expected to play a relevant role in the excitation of the dominant capillary mode. However, both experimental and numerical results for different ``smooth"\ ejectors indicate that the breakup length in well-controlled experimental realizations essentially depends on the liquid properties and operating parameters, which raises questions about the idea that the perturbation origin is located in the ejector. \citet{GCHWKDGHLCBM19} have calculated the natural breakup length in terms of the transient growth of perturbations coming from the surface energy excess at the breakup \citep{U16}. This quasi-periodic source of energy may regularly feed the perturbations leading to each breakup event, which would explain the rather deterministic manner in which unforced capillary jets spontaneously break up. We will explain these results in more detail in Sec.\ \ref{sec11}.

% Presence of the nozzle: spatial analysis
The fact that perturbations in the temporal analysis are characterized by a real wavenumber $k$, implies that they grow at the same rate both in the vicinity of the nozzle and downstream. This unrealistic assumption is eliminated in the spatial stability analysis, where the complex waver number is calculated as a function of the real eigenfrequency. \citet{KRT73} claimed that there are spatial modes with growth rates larger than the dominant temporal one, although they are not observed in the experiments probably because their wavelengths are too long to become established in a finite jet. However, and as explained by \citet{E97}, the modes in question violate a radiation condition, and hence do not exist with proper boundary conditions at infinity. The spatial and temporal stability analyses are equivalent if the speed of the jet is much larger than that of the small-amplitude capillary waves \citep{KRT73}.

\subsubsection{The convective-to-absolute instability transition analysis}
\label{sec3.1.2}

% Convective and absolute instabilities
The breakup mode adopted by a fluid thread can be predicted in terms of the so-called convective-to-absolute instability transition, a concept widely used in instabilities of shear flows and wakes \citep{HM90a}. In convectively unstable jets, capillary waves are swept away downstream by the current, which keeps a considerable portion of the jet free of perturbations. Conversely, growing perturbations travel both downstream and upstream along absolutely unstable jets, precluding their formation.

\begin{figure}
%\begin{center}
\centering{\resizebox{0.475\textwidth}{!}{\includegraphics{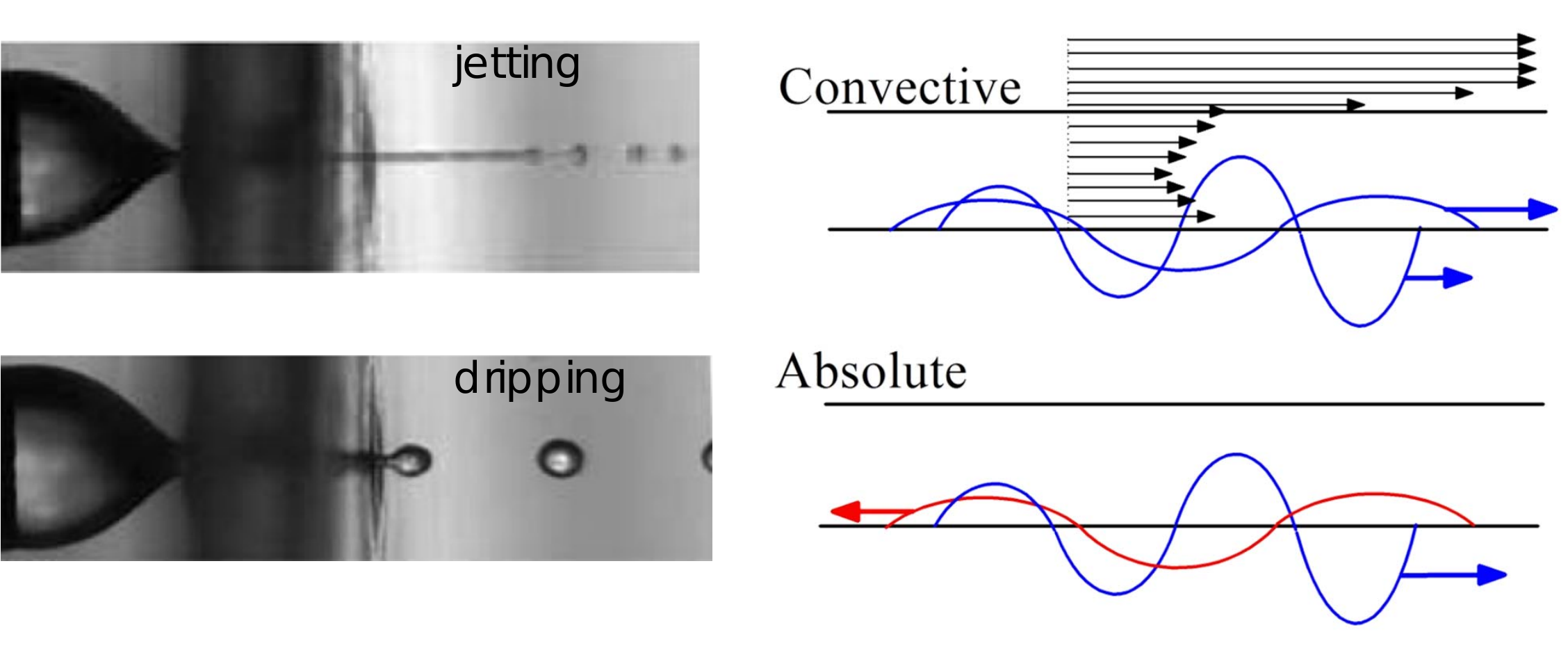}}}
%\end{center}
\caption{Relationship between the jetting-to-dripping and convective-to-absolute instability transitions. The images on the left side show the transition from jetting to dripping in gaseous flow focusing for small applied pressure drops \citep{SLYY09}. The sketches on the right side represent the evolution of growing capillary waves in a convectively and absolutely unstable jet.}
\label{conabs0}
\end{figure}

% The convective-to-absolute instability transition
Under certain conditions, the jetting-to-dripping transition of liquid \citep{L03b,GCUA07} and gaseous \citep{GHG06} jets has been successfully linked to the convective-to-absolute instability transition for axisymmetric ($m=0$) perturbations (Fig.\ \ref{conabs0}). However, and as will be explained below, we have to appeal to other instability mechanisms to explain many jetting-to-dripping transitions observed in microfluidic applications. In fact, the correspondence between convective instability and jetting is not always clear even in relatively simple cases. For example, inclined jets can suffer from self-sustained oscillations when they are convectively unstable throughout the entire fluid domain \citep{Y97a}. There can be significant discrepancies between the conditions leading to absolute instability and dripping in both plane liquid sheets \citep{L99} and round jets \citep{L97}. However, and despite of its limitations, the convective-to-absolute instability transition has proved to provide useful information on the relatively scarce occasions in which it has been applied.

\begin{figure}
%\begin{center}
\centering{\resizebox{0.35\textwidth}{!}{\includegraphics{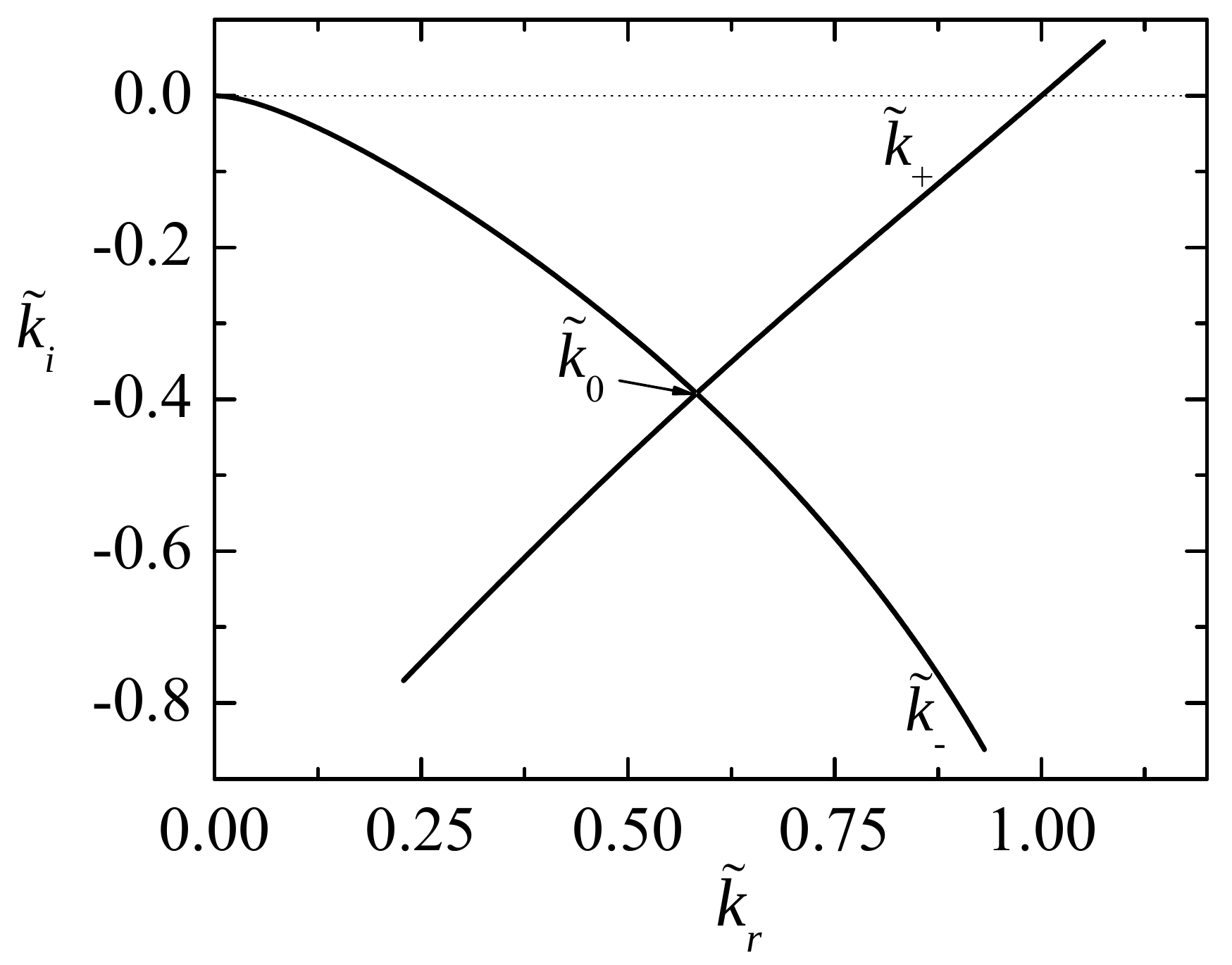}}}
%\end{center}
\caption{Spatial branches $\widetilde{k}_{\pm}(\omega^{'*})$ for a jet moving in a vacuum \citep{MG08a}. These branches originated from the solution $\widetilde{k}_0=0.5828-0.3921 i$ and $\omega^{'*}_{0}=0.5757$ of the dispersion relation. As can be observed, the branch $\widetilde{k}_{+}$ crosses the $\widetilde{k}_r$ axis at $\widetilde{k}_r=1$ (Rayleigh stability limit). Here, $\widetilde{k}=k R_j$ and $\omega^{'*}=\omega' R_j/V_j$.}
\label{brigg}
\end{figure}

% Calculation of the convective-to-absolute instability transition
The critical conditions leading to the convective-to-absolute instability transition are determined by the spatio-temporal analysis of the dispersion relationship (\ref{disper0}). In this analysis, one explores the response of the system to perturbations characterized by a {\em complex} axial wavenumber $k$ observed by a fixed observer anchored at the nozzle. The dispersion relationship is typically derived using the frame of reference travelling with the jet. To change the frame of reference from a traveling observer to a fixed one, one just needs to replace the wave frequency $\omega$ by $\omega'-V_j k$ in the dispersion relation (\ref{disper0}). For fixed values of the control parameters $\{{\cal P}_i\}$ ($i=2,\ldots,N$), one calculates the critical value ${\cal P}_{1c}$ for which Brigg's pinch condition \citep{B64,HM90a} is satisfied. This condition establishes that there must be at least one pinching of a $k^+$ and a $k^-$ spatial branch with $\omega'_i=0$, where the $k^+$ is the path of ${\cal D}_m=0$ in the complex $k$ plane which moves into the $k_i>0$ half-plane as $\omega'_i$ increases, while the $k^-$ branch always remains in the $k_i<0$ half-plane as $\omega'_i$ increases (Fig. \ref{brigg}).

% Equivalence between Brigg's and van Sarloos' criteria
\citet{S87} proposed an alternative criterion for determining the convective-to-absolute instability transition based on the analysis of the propagation front velocity. Specifically, the system becomes absolutely unstable when the rear front velocity of a localized initial distortion becomes zero. This has been used by many researchers in capillary flows because is very intuitive and immediately understandable in physical terms. \citet{MG08b} showed the equivalence between this and the classical saddle-point criterion \citep{B64,HM90a}.

\subsection{Global stability analysis}
\label{sec3.1.3}

% Instability of the source and the jet.
One of the central problems in droplet-based microfluidics is to determine the parameter conditions leading to the dripping-to-jetting transition for the varied experimental configurations. In general, the existence of the jetting regime demands (i) enough mechanical energy to overcome the viscosity force and to create a large liquid-fluid interface, and (ii) the stability of the base flow sustaining the liquid ejection. This last condition involves the stability of both the fluid source and the emitted jet.

% The slender meniscus
In many applications, the fluid source is a slender meniscus hanging on a feeding capillary. This fluid configuration can be seen as a simple upstream extension of the emitted jet, where the velocity field is quasi-parallel and the liquid velocity is smaller than that of the jet. Then, the system's stability essentially reduces to that of the jet, and the jetting-to-dripping transition can be explained in terms of the convective-to-absolute instability transition described above.

% Tip streaming
The above consideration does not apply to tip streaming. In this case, the source is a fluidic structure (in most cases, a cone-like meniscus) fundamentally different from the emitted jet, which can exhibit complex flow patterns including boundary layers, stagnation points and recirculation cells. In tip streaming, the jet's stability becomes a necessary but not sufficient condition for jetting. The determination of that sufficient condition requires the stability analysis of the entire base flow.

% Global stability
Tip streaming is not the only phenomenon which invalidates the local stability analysis. As explained above, the local spatio-temporal stability analysis is valid as long as the base flow explored by the perturbations is quasi-parallel and quasi-homogeneous in the streamwise direction (the WKBJ approximation). There are many applications where the hydrodynamic length characterizing the base flow is of the order of, or even much smaller than, that of the dominant perturbation. In this case, an accurate stability analysis of the steady base flow also requires the calculation of the so-called global modes.

% Global modes
Global modes are patterns of motion depending in an inhomogeneous way on two or three spatial directions, and in which the entire system oscillates harmonically with the same (complex) frequency $\omega$ and a fixed phase relation \citep{C05b,T11}. This implies that space and time variables are separable when describing the system response to small-amplitude perturbations. The global modes are calculated from the ansatz
\begin{equation}
\Phi^{(j)}({\bf r};t)=\phi^{(j)}(r,z)\, e^{m\theta-i\omega t},
\label{EVP}
\end{equation}
where $\Phi^{(j)}$ represents any variable of the problem and $m$ is the azimuthal mode number. The global modes (\ref{EVP}) are calculated as the eigenfunctions of the linearized Navier-Stokes operator as applied to a given configuration (base flow). The base flow is linearly and {\em asymptotically} stable if the spectrum of eigenvalues is in the stable complex half-plane. In this case, any initial small-amplitude perturbation will decay exponentially on time for $t\to \infty$ (as long as the linear approximation applies). Global stability analysis has been rarely used in microfluidics, although one can expect its application to spread in the coming years \citep{SB05a,TLS12,RSG13,GSC14,CHGM17,PRHGM18,AFG18,BHGM19}.

\begin{figure}[hbt]
%\begin{center}
\centering{\resizebox{0.4\textwidth}{!}{\includegraphics{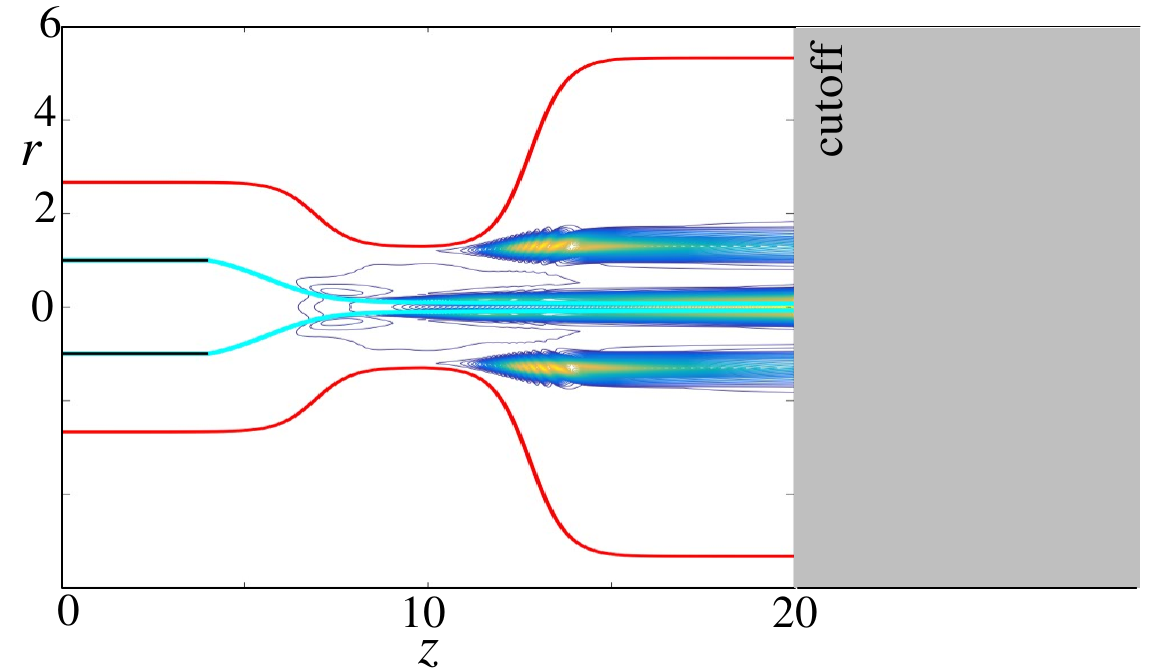}}}
%\end{center}
\caption{Energy per unit volume $e\equiv p^{(j)}+1/2\, \rho^{(j)}\, (|v_r^{(j)}|^{2}+|v_z^{(j)}|^{2})$ of the dominant eigenmode at marginal stability of a water stream focused by an air current \citep{CHGM17}. The scalar fields $e(r,z)$ in the liquid and gas domains have been normalized with their corresponding maximum values (the maximum value in the liquid domain is around 132 that of the gas stream). Higher (lower) values of $e$ correspond to the color yellow (blue).}
\label{energywater}
\end{figure}

% Finite system
As mentioned in Sec.\ \ref{sec3}, capillary jets are almost always unstable because they eventually break up into droplets or suffer other types of convective instabilities. Therefore, the global stability analysis of a base flow unlimited in the downstream direction should show almost always the existence of unstable convective modes independently from the operating conditions. In practice, we set a boundary (cutoff) in the downstream direction to define the finite fluid domain considered in the analysis (Fig.\ \ref{energywater}). ``Soft"\ boundary conditions, such as the so-called outflow or traction-free boundary condition, can be applied at that cutoff. In viscous systems, such as the liquid-liquid coflowing configuration \citep{GSC14,AFG18}, global modes of convective nature can be subdominant, and the outlet boundary conditions are practically irrelevant provided that the cutoff length is much larger than the injector diameter \citep{TLS12,GSC14,AFG18}. For this reason, perturbations can be forced to vanish at the outlet and even so the results are accurate \citep{SB05a,TLS12,AFG18}. In any case, the analyzed fluid domain must contain an ejected fluid thread much longer than its diameter, and one needs to verify that the cutoff arbitrarily imposed in the analysis does not significantly affect the eigenvalues for a sufficiently large interval of jet lengths. When all the global modes of that finite system decay on time, the flow is assumed to operate in the jetting regime. On the contrary, the growth of axisymmetric ($m=0$) modes is supposed to cause self-sustained oscillations when non-linear terms saturate the perturbation and dripping otherwise. In addition, the instability of nonaxisymmetric ($m\geq 1$) modes is assumed to produce the whipping (bending) of the emitted jet \citep{BRCHGM20}.

\subsection{Short term response}

% Short term response
It is frequently believed that global stability is a sufficient condition for the linear stability of the base flow. However, this is not necessarily true. If the linearized Navier-Stokes operator is non-normal, the short-term dynamics of the system can be the result of a ``constructive interference"\ of stable global modes, which can lead to a bifurcation before those modes are damped out \citep{C05b,S07,LCC02}. In other words, the superposition of decaying small-amplitude perturbations introduced into a microfluidic configuration can destabilize the flow before those perturbations disappear, which prevents the system from reaching the jetting regime.

In the case described above, global stability becomes a necessary but not sufficient condition for jetting, and the stability analysis must be completed with direct numerical simulations of the system to examine its response to initial perturbations (the initial value problem). To speed up the calculation, the nonlinear terms can be dropped when integrating the hydrodynamic equations \citep{CHGM17}. This allows one to see whether the superposition of linear global modes makes the resulting perturbation grow within the small-amplitude response regime. Of course, non-linear terms must be taken into account to study the subsequent evolution of that perturbation. In any case, this is a complex problem because the outcome can significantly depend on the type and location of the initial perturbation, something difficult to determine in an experiment.

\begin{figure}
%\begin{center}
\centering{\resizebox{0.475\textwidth}{!}{\includegraphics{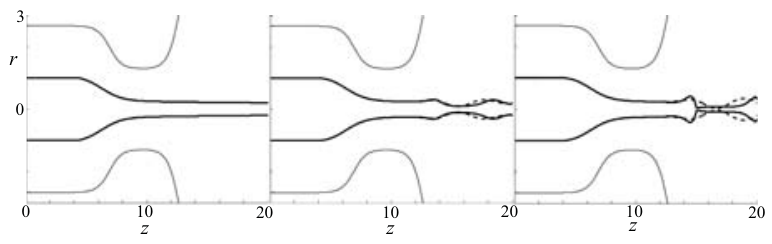}}}
%\end{center}
\caption{Free surface deformation calculated with the linearized (dash line) and nonlinear (solid line) hydrodynamic equations at three instants for an asymptotically stable base flow of
gaseous flow focusing \citep{CHGM17}.}
\label{ipv}
\end{figure}

% Example: flow focusing
The gaseous flow focusing configuration constitutes a good example of the situation described above. \citet{CHGM17} has recently shown that there is a transient growth of linear perturbations before the asymptotic exponential regime is reached. This growth leads to dripping for small applied pressure drops. Figure \ref{ipv} shows the free surface deformation at three instants for an asymptotically stable base flow. The superposition of decaying linear global modes triggered by the perturbation gives rise to the free surface pinch-off within the numerical domain.

% Line of argument
In this section, we have described some general ideas and methodological aspects of the stability analyses applied to the configurations considered in this work. In Secs.\ \ref{sec4}-\ref{sec7}, we will show results for the convective-to-absolute instability (jetting-to-dripping) transition, and then we will describe the linear capillary instability and nonlinear breakup in the jetting regime. In this way, we first try to determine the conditions leading to jetting, and then we describe the evolution of the system in that regime. The results obtained from the global stability analysis of the microfluidic configurations considered in this review are presented in Sec.\ \ref{sec11}, once those configurations have been described in detail.

\section{Results of spatio-temporal and global stability analyses}
\label{sec4}

\subsection{Convective-to-absolute instability transition}

% Basic result
In the absence of viscosity effects, the convective-to-absolute instability transition of axisymmetric ($m=0$) capillary perturbations growing along a fluid jet takes place for $V_j\simeq \overline{V}_{\sigma}$ \citep{LG86a,MG08a}, where $V_j$ is the jet velocity, $\overline{V}_{\sigma}=[\sigma/(\overline{\rho} R_j)]^{1/2}$ is the inertio-capillary speed, and $\overline{\rho}$ is an effective density of the jet-environment ensemble. This result has a straightforward interpretation: for $V_j\gtrsim \overline{V}_{\sigma}$ the liquid sweeps away the growing capillary waves that swim against the current at a speed of the order of $\overline{V}_{\sigma}$.

% Liquid jets
The effective density $\overline{\rho}$ for a liquid jet surrounded by a gaseous ambient is essentially that of the jet. When the jet is directly extruded from a nozzle by the action of the upstream pressure, the condition $V_j\gtrsim \overline{V}_{\sigma}$ is generally satisfied because it coincides with $\rho_j V_j^2\gtrsim \sigma/R_j$, which is a necessary condition for the jet extrusion. Therefore, absolute instability does not generally constitute an obstacle for the jet formation in this simple application.

% Gaseous jets
The effective density $\overline{\rho}$ for gaseous jets moving in liquid baths is much smaller than that for the inverse case. Therefore, higher jet velocities are demanded to enable convective instability in gaseous jets, which partially explains why is so difficult to produce them. In fact, direct injection of a gas into a quiescent pool of liquid has produced long jets only in the supersonic regime \citep{CR97}. Long gaseous threads can be formed for lower injection velocities with the help of surfactants or mixtures reducing the surface tension \citep{GHG06,CHLG11}, and in the presence of a solid substrate/core \citep{XBC07,HGM13}.

% Some results for the convective-to-absolute instability transition
As explained in Sec.\ \ref{sec3}, the parameter surface corresponding to the convective-to-absolute instability transition for a specific configuration can be accurately determined by conducting the spatio-temporal analysis of the dispersion relationship derived from the linear stability analysis. This is a relatively complex calculation for capillary systems, which may explain why its use has not sufficiently spread in this context \citep{LR98}. \citet{LG86a} studied the absolute instability of a jet in a mechanically inert ambient, while \citet{LL89} took into account the effect of the surrounding medium. Subsequently, more complex configurations have been considered by several authors, including the effect of a small inner-to-outer density ratio \citep{GHG06,G07b}, a viscous coflowing current \citep{GGRHF07,UFGW08,ECG16}, confinement in various geometries \citep{GCUA07,GCA08}, jet swirling \citep{H08b}, a twofold interface \citep{CMPR06,HMFG10}, viscoelasticity \citep{CMPR06,HMFG10,AU19}, and gravity \citep{AU19} among others. In many cases, the dispersion relationship can be derived analytically, while in others the linearized equations are spatially discretized.

\begin{figure}
%\begin{center}
\centering{\resizebox{0.4\textwidth}{!}{\includegraphics{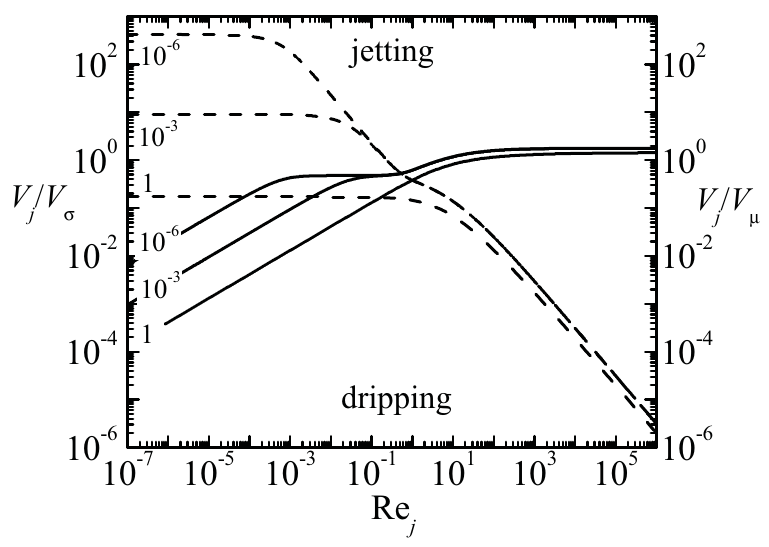}}}
%\end{center}
\caption{$V_j/V_{\sigma}$ (solid lines) and $V_j/V_{\mu}$ (dashed lines) corresponding to the convective-to-absolute instability transition for a liquid jet moving with a uniform velocity in a fluid bath coflowing at the same \citep{MG08a}. The labels indicate the values of $\rho$. In all the cases, $\rho=\mu$.}
\label{conabs}
\end{figure}

% Figure Leib-Goldstein
Figure \ref{conabs} shows the curves $V_j/V_{\sigma}$ corresponding to the convective-to-absolute instability transition for a liquid jet moving as solid body in a bath coflowing with the jet at the same velocity \citep{LG86a,MG08a}. Here, $V_{\sigma}=[\sigma/(\rho_j R_j)]^{1/2}$ is the inertio-capillary velocity defined in terms of the jet density $\rho_j$. The results were calculated for different values of the density and viscosity ratios, $\rho=\rho_o/\rho_j$ and $\mu=\mu_o/\mu_j$, where $\rho_o$ and $\mu_o$ stand for the outer bath density and viscosity, respectively, while $\mu_j$ is the jet viscosity. The velocity ratio $V_j/V_{\sigma}$ is represented as a function the jet Reynolds number
\begin{equation}
\text{Re}_j=\frac{\rho_j V_j R_j}{\mu_j}.
\end{equation}
The figure also shows the ratio of $V_j$ to the capillary-viscous velocity $V_{\mu}=\sigma/\mu_j$, which is the relevant characteristic quantity in the Stokes limit $\text{Re}_j\to 0$. The velocity ratios $V_j/V_{\sigma}$, $V_j/V_{\mu}$ and $V_{\mu}/V_{\sigma}$ are essentially the jet's Weber, Capillary and Ohnesorge numbers, respectively:
\begin{equation}
\label{oh}
\text{We}_j=\left(\frac{V_j}{V_{\sigma}}\right)^2, \quad \text{Ca}_j=\frac{V_j}{V_{\mu}}, \quad \text{Oh}_j=\frac{V_{\sigma}}{V_{\mu}}.
\end{equation}

% Figure Leib-Goldstein. Analysis. Unconditional jetting
As can be observed in Fig.\ \ref{conabs}, the critical ratios $V_j/V_{\sigma}$ and $V_j/V_{\mu}$ reach constant values in the inviscid and viscous limits, respectively. In the former case, these values are similar for the liquid-gas and liquid-liquid systems. In the latter case, this threshold does not depend on the jet's radius, which means that infinitely thin jets can be formed (the so-called {\it unconditional jetting}) \citep{G08a} provided that their velocities are larger than $K(\mu) V_{\mu}$, where the function $K(\mu)$ is expected to take values of order unity. The viscous limit must be taken with caution, because the effects of the outer medium cannot be neglected in that case even for very small density and viscosity ratios \citep{G07b,RAMMG16}.

% Unconditional jetting. Generalization
\citet{G08a} realized that, in the Stokes limit, the convective-to-absolute instability transition does not depend on the jet and outer medium velocity profiles, but only on the interface speed $V_s$. In addition, the function $K(\mu)$ approximately scales as $\mu^{1/2}$. These results suggest expressing the instability transition in terms of the modified Capillary number
\begin{equation}
\label{modca}
\overline{\text{Ca}}\equiv \mu^{1/2}\, \mu_j V_s/\sigma.
\end{equation}
In a viscosity-dominated flow, a jet is convectively unstable if the interface velocity $V_s$ is such as $\overline{\text{Ca}}\gtrsim \overline{\text{Ca}}^*$, where $\overline{\text{Ca}}^*$ depends on the viscosity ratio $\mu$ and lies in the interval $0.14\lesssim \overline{\text{Ca}}^*\lesssim 0.4$ \citep{G08a}.

% Instability sources
The comparison with experimental data has shown that the analysis of the convective-to-absolute instability transition constitutes an accurate predictive tool provided that both the properties of the fluids involved and the base flow are correctly accounted for \citep{GHG06,GR06,GCUA07,G08a,SLYY09,VMHG10}. However, it must be pointed out again that this analysis considers only the stability of the emitted jet. In general, and as mentioned above, the system's stability condition is twofold: the fluid source must be stable and the emitted jet must be convectively unstable.

\begin{figure}
%\begin{center}
\centering{\resizebox{0.32\textwidth}{!}{\includegraphics{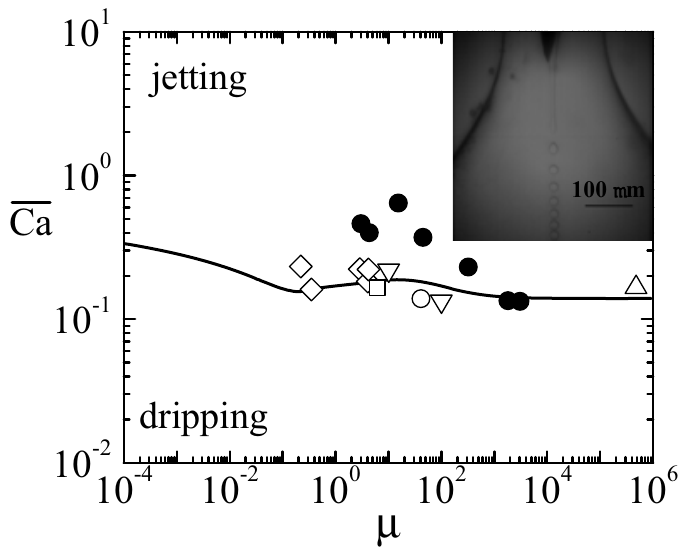}}}
%\end{center}
\caption{Modified Capillary number $\overline{\text{Ca}}\equiv \mu^{1/2}\, \mu_j\ V_s/\sigma$ below which a jet becomes absolutely unstable \citep{ARMGV13}. The line indicates the convective-to-absolute instability transition when the flow is dominated by viscosity \citep{G08a}. Experimental values for the jetting-to-dripping transition (symbols). The data were taken from \citet{ARMGV13} ($\bullet$), \citet{ABS03} ($\square$), \citet{AM06b} ($\circ$), \citet{UFSW07} ($\triangledown$), \citet{GCUA07} and \citet{HGG08} ($\diamond$), and \citet{GHG06} ($\triangle$). The inset shows a liquid-liquid viscosity-dominated jetting realization in a flow focusing device \citep{ARMGV13}.}
\label{unconditional}
\end{figure}

\begin{figure}
%\begin{center}
\centering{\resizebox{0.34\textwidth}{!}{\includegraphics{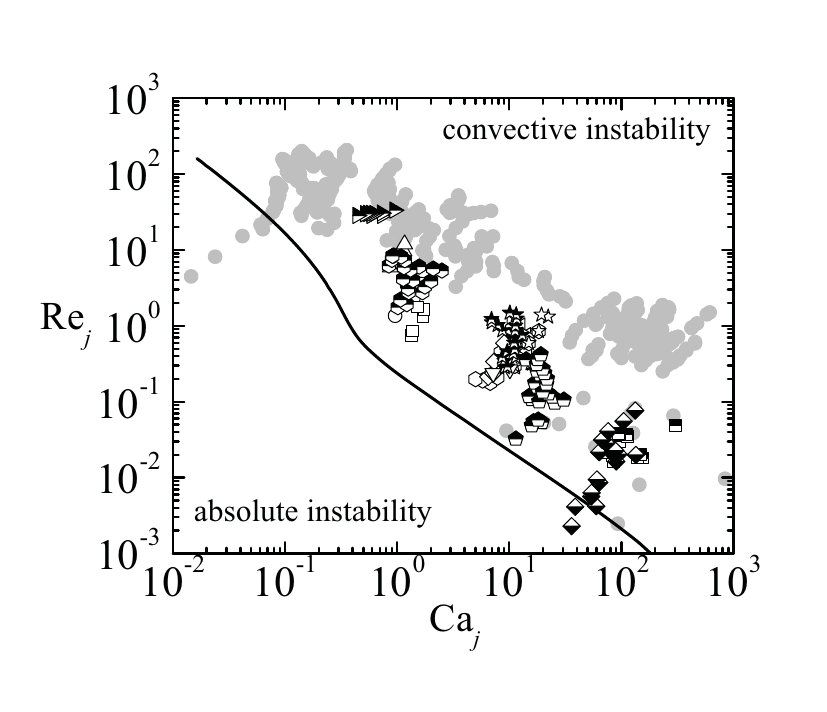}}}
%\end{center}
\caption{Capillary Ca$_j$ and Reynolds Re$_j$ numbers of jets produced with electrospray (open and grey symbols) \citep{PRHGM18} and gaseous flow focusing (half black symbols) \citep{ARMGV13}. The open and half black symbols correspond to steady jetting experiments where the flow rate was decreased down to the stability limit. The line corresponds to the convective-absolute instability transition for a non-electrified jet moving in a vacuum \citep{LG86a}.}
\label{caelec}
\end{figure}

% Results for unconditional jetting and electrospray
Figure \ref{unconditional} shows the critical Capillary number $\overline{\text{Ca}}^*$ below which a jet becomes absolutely unstable when the flow is dominated by viscosity \citep{G08a}. The figure also shows the experimental values of that parameter measured in different microfluidic configurations at the jetting-to-dripping transition. The experimental value significantly exceeded the theoretical prediction in some cases, which probably means that the jetting-to-dripping transition was caused by the liquid source instability. Figure \ref{caelec} also shows that convective instability is a necessary but not a sufficient condition to produce the cone-jet mode of electrospraying \citep{PRHGM18} or the steady jetting of gaseous flow focusing \citep{RAMMG16}. The convective-to-absolute instability transition curve calculated by \citet{LG86a} overestimates the critical Reynolds number for very large and very small values of this parameter probably due to the existence of an inner boundary layer \citep{GM09} and a gaseous environment \citep{ARMGV13} in flow focusing, as well as by electric field effects in electrospray.

% Whipping
The results described above refer to the convective-to-absolute instability transition for axisymmetric perturbations. As will be explained in more detail in Sec.\ \ref{sec6}, the mismatch between the jet and outer bath velocities, as well as the existence of free electric charges accumulated at the interface and subjected to strong electric fields, may make nonaxisymmetric perturbations grow despite the stabilizing effect of the surface tension (the so-called whipping instability). Experience shows that the capillary jets emitted in almost all microfluidics applications are either stable or at most convectively unstable when it comes to nonaxisymmetric perturbations. The lateral jet oscillations frequently observed in configurations like electrospray or flow focusing do not generally propagate upstream, and, therefore, they do not alter the tapering meniscus stability. Absolute whipping has been observed only in liquid jets focused by high-speed gaseous currents inside converging nozzles \citep{AFMG12}. In fact, it seems that the gaseous radial flow in front of the discharge orifice of the original flow focusing configuration \citep{G98a} constitutes a barrier for whipping perturbations. The conditions leading to absolute whipping have been barely studied \citep{HFMG10}.

\subsection{Global stability}
\label{sec4.2}

% Introduction
Global stability analyses have been frequently conducted to study problems such as wakes behind solid obstacles and detached single-phase flows \citep{C05b,T11}. These studies are more scarce in the context of capillary systems. Here, we mention those related with the microfluidic configurations considered in this review. We will come back to these studies in Sec.\ \ref{sec11}.

% Gravitational 
\citet{L97} examined the global modes in falling capillary jets and discussed the relationship between global instability and the jetting-to-dripping transition. \citet{SB05a} carried out a global stability analysis of a gravitational jet using the long-wave (1D) approximation (see Sec.\ \ref{sec2.4}). The results for marginal stability and critical frequencies were in excellent agreement with direct numerical simulations. \citet{RSG13} showed the stabilizing effect of the axial curvature from the 1D model too. 

% Coflowing
\citet{GSC14} studied with the slender-body approximation the global stability of the tip streaming flow produced by a coflowing injector (see Sec.\ \ref{sec9}). The global stability of the axisymmetric flow produced in the coflowing configuration has recently been examined by \citet{AFG18}. For high external flow rates, the predictions almost coincide with those of the local convective-to-absolute instability transition theory \citep{HGG08}. However, the flow is slightly more stable than predicted by the local analysis for small external flow rate and/or a high degree of confinement.

% Electrospray and electrospinning
As will be explained in Sec.\ \ref{sec9}, in the steady cone-jet mode of electrospray very thin jets are produced by tip streaming when strong electric fields are applied to low-conductivity droplets. Theoretical studies on the steady cone-jet mode of electrospray typically use the leaky-dielectric model \citep{MT69,S97} (see Sec.\ \ref{sec2}), i.e. they assume that the liquid exhibits a uniform electrical conductivity (the Ohmic conduction model), and the net free charge accumulates onto the free surface so that the bulk electrostatic mass force is negligible as compared to the superficial one resulting from the Maxwell stresses \citep{G97a,H03,H10,F07,HLGVMP12,G10}. \citet{DC16} studied the global linear stability of the electrospray cone-jet mode with the 1D approximation. They subsequently extended this analysis to electrospinning by incorporating rheological effects described by the Olroyd-B and XPP models \citep{DC17}. \citet{PRHGM18} have calculated the minimum flow rate of the cone-jet mode of electrospray from the global stability of the solution to the 2D leaky-dielectric model, showing good agreement with experiments. \citet{BHGM19} have recently extended this analysis to weakly viscoelastic liquids. The cone-jet mode of electrospray can be stabilized with a coflowing high-speed gas current \citep{CWHBCG19}. The experimental minimum flow rates reasonably agree with the global stability predictions in this case as well \citep{CWHBCG19}.

% Flow focusing
In flow focusing (see Sec.\ \ref{sec9}), tip streaming is achieved with purely hydrodynamic means by making the focused fluid cross the discharge orifice together with an outer (focusing) gas/liquid current. \citet{CHGM17} examined the global stability of the gaseous flow focusing axisymmetric configuration. They found good agreement with experimental values of the minimum liquid flow rate for sufficiently large gas velocities. 

% Short-term instability
As explained in Sec.\ \ref{sec3}, linear {\em asymptotic} global stability does not necessarily imply linear stability. If the linearized Navier-Stokes operator is non-normal, then the perturbation energy may increase during the system's short-term response, and cause the solution bifurcation in asymptotically stable systems. In fact, convective instabilities commonly arising in problems with inflow and outflow conditions are not typically dominated by long-term modal behavior. For instance, asymptotically stable gravitational jets eventually break up due to the growth of non-normal modes \citep{LCC02}. \citet{CHGM17} have shown that flow focusing stability can be explained in terms of the system's short-term response for small gas velocities.

\section{Capillary instabilities}
\label{sec5}

% End-pinching and capillary instability
In the jetting regime, the dispersed phase forms a cylindrical thread long compared with its diameter, which breaks up downstream into a collection of droplets/bubbles. This breakup can be due to the so-called end-pinching mechanism \citep{SBL86}, the Rayleigh capillary instability \citep{R79a}, or a combination of both. In all the cases, the instability is triggered by interfacial energy release.

\subsection{End-pinching instability}

\begin{figure}
\centering{\resizebox{0.45\textwidth}{!}{\includegraphics{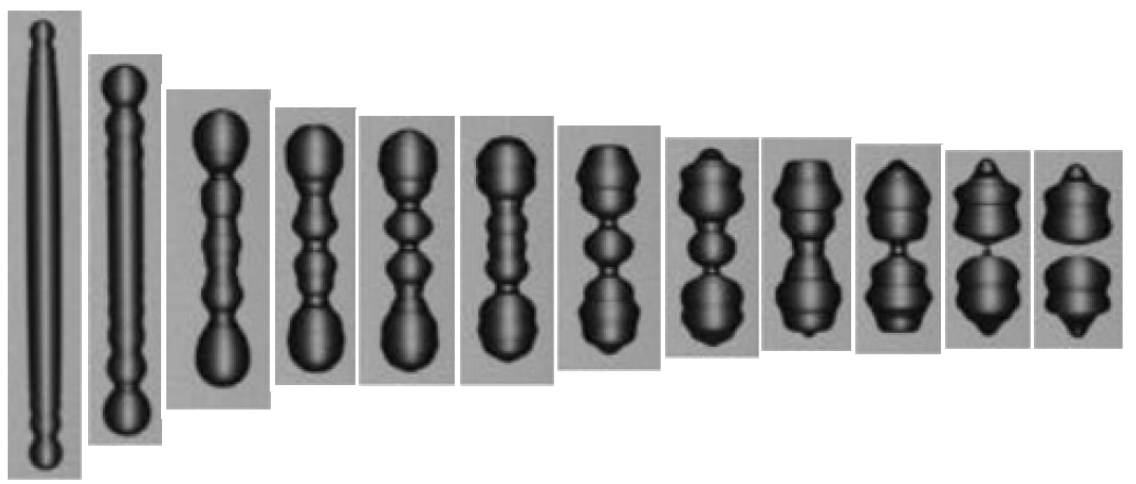}}}
\caption{End-pinching in a liquid thread \citep{WCNCCBWFS19}. The thread aspect ratio is close to the critical value below which the thread recoils without breaking up.}
\label{epinch}
\end{figure}

% End-pinching.
Liquid threads are produced in technological and natural processes such as DOD ink jet printing, crop spraying and atomization coating, or the fragmentation taking place in fountains and many types of sprays \citep{EV08}. For sufficiently large values of the capillary Reynolds number and the thread aspect ratio, the free surface pinches off at the ends of the thread \citep{B58,SBL86,DJWTL13,CCH12,WCNCCBWFS19,AKHB19}, which results in a set of droplets (Fig. \ref{epinch}). This is the so-called end-pinching mechanism. Although the breakup is also driven by surface tension, this process and the Rayleigh capillary instability are clearly distinct.

\begin{figure}
\centering{\resizebox{0.45\textwidth}{!}{\includegraphics{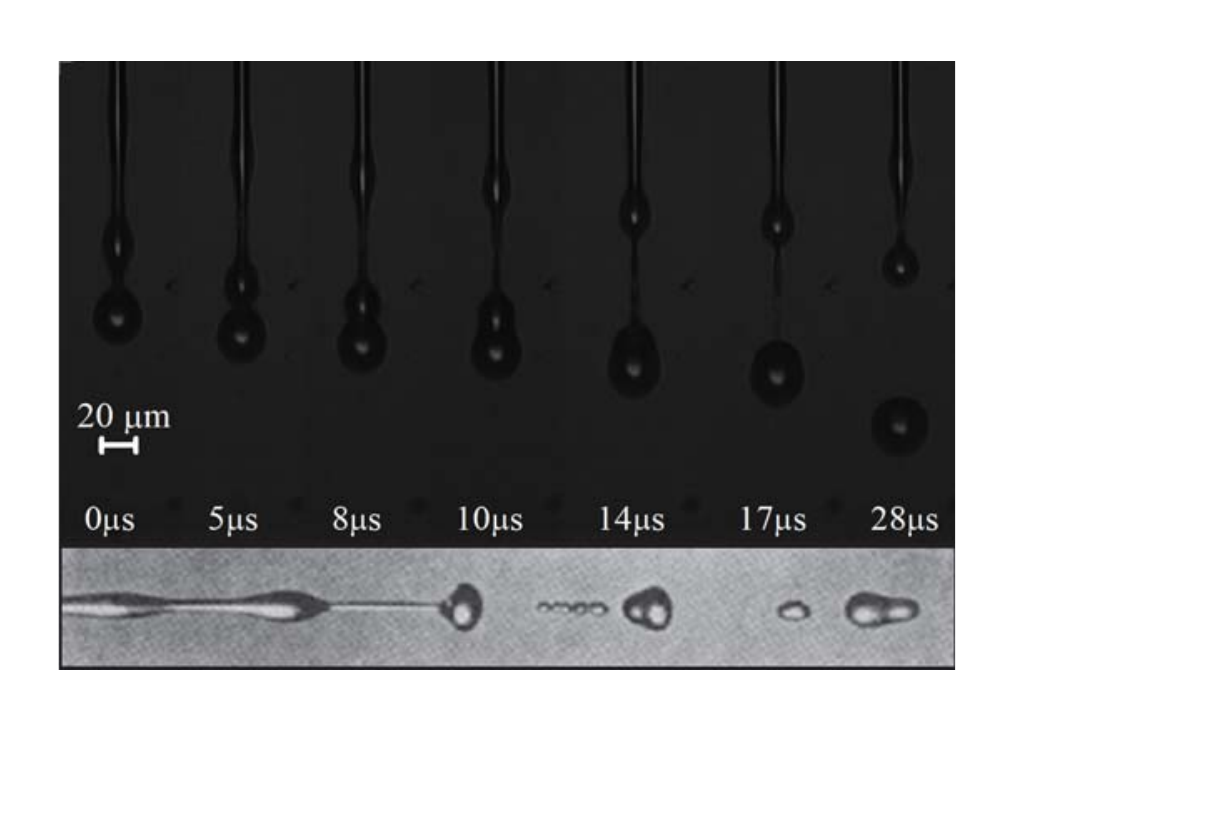}}}
\caption{ (Upper images) End-pinching in a jet emitted close to the minimum flow rate stability limit of electrospray \citep{PRHGM18}. (Lower image) The Rayleigh instability in a millimeter jet \citep{R91a}. }
\label{pinching}
\end{figure}

The end-pinching phenomenon also occurs in a jet when it moves at speeds close to that of the jetting-to-dripping transition. In this case, the jet Weber number We$_j$ takes values around unity, and the liquid inertia hardly overcomes the resistant force exerted by surface tension. In the end-pinching breakup of a jet, a bulb forms at the end of the jet. This bulb moves slower than the fluid in the thread located just behind it. For this reason, the fluid gets into the bulb and inflates it. The neck located between the bulb and the thread stretches due to the capillary pressure and becomes thinner and thinner until a droplet separates from the jet.  Two bulbs can form simultaneously in a jet. When the jet is accelerated under the action of an external force, the rear bulb may catch the lead one, giving rise to the coalescence between them (Fig.\ \ref{pinching}).

\subsection{Rayleigh instability}

% The Rayleigh capillary instability
The Rayleigh capillary instability can be explained as follows. When the interface of a cylindrical fluid thread is perturbed with an infinitesimal sinusoidal deformation, the surface tension produces an axial pressure gradient in the thread. If the perturbation wavelength is larger than the thread perimeter, the pressure decreases/increases in the bulging/narrowing region, which favors the growth of that perturbation. The capillary wave breaks the interface, which gives rise to a quasi-monodisperse collection of droplets. Owing to mass conservation, the droplet diameter can be calculated as $d_d\simeq 3/2\, (R_j \lambda^{\textin{max}})^{1/3}$ [Eq.\ (\ref{esti})-right], where $\lambda^{\textin{max}}$ is the wavelength of the perturbation with the maximum growth rate, and, therefore, responsible for the jet breakup.

\begin{figure}
%\begin{center}
\centering{\resizebox{0.34\textwidth}{!}{\includegraphics{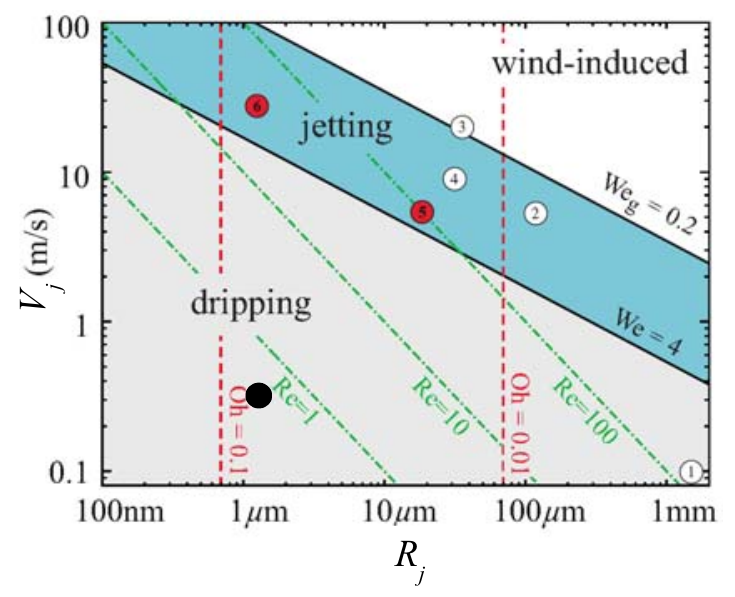}}}
%\end{center}
\caption{Breakup regimes for jets of pure water moving in still air \citep{HGSVBL10}. The encircled numbers (1)-(4) refer to droplet formation studies performed by \citet{APB00}, \citet{KLAS03}, \citet{GG09} and \citet{PL77}, respectively, while the encircled numbers (5) and (6) refer to the diminutive Rayleigh jets produced by \citet{HGSVBL10}. The black circle corresponds to a 500-cSt silicone oil jet with $R_j=1.5$ $\mu$m and $V_j=0.37$ m/s \citep{RAMMG16}. In this figure, Re, We and Oh correspond to the symbolds $\text{Re}_j$, $\text{We}_j$ and $\text{Oh}_j$ defined in this review, respectively.}
\label{Hoeve}
\end{figure}

% Conditions for Rayleigh capillary instability
The phenomenon described above becomes the dominant breakup mechanism for well-established jetting realizations. In the absence of external actuation and viscosity forces, a jet moving in a still ambient breaks up owing to the Rayleigh capillary instability if \citep{CR96}
\begin{equation}
\text{We}_j\gtrsim 4 \quad \text{and}\quad \text{We}_g\equiv \rho\text{We}_j\lesssim 0.2,
\end{equation}
where $\rho$ is the ratio of the ambient density to that of the jet. The first and last condition eliminates the possibility of dripping and whipping (wind-induced instability), respectively. Figure \ref{Hoeve} shows the different breakup regimes for jets of pure water moving in still air.

An outer coflowing stream reduces the value of the Weber number above which Rayleigh capillary instability can be observed. In fact, this breakup mechanism can be found in viscous jets coflowing with high-speed gaseous streams for Weber numbers as low as 0.1 \citep{RAMMG16}.

\subsection{Rayleigh stability analysis}

% Temporal stability
As explained in Sec.\ \ref{sec3}, the temporal linear stability analysis of the jetting mode leads to the dispersion relationship which determines the continuum spectrum of eigenfrequencies characterizing the axisymmetric normal modes (Fourier components) as a function of their (real) wavenumbers \citep{R92b}. The decomposition of initial (spontaneous) perturbations into linear eigenmodes has proved to be useful for studying the short-term evolution of those perturbations. The local stability analysis (WKBJ approximation) assumes that the base flow is quasi-parallel and homogeneous in the streamwise direction over distances of the order of or larger than the perturbation wavelength. This approach has been and still is routinely applied to analyze the influence of all types of effects, on many occasions with very little connection with experiments due to the failure of the locality approximation. Nevertheless, it allows one to identify the stabilizing and destabilizing mechanisms that affect the Rayleigh instability.

% Rayleigh stability analysis
Rayleigh's linear stability analysis \citep{R79a} was based on three major simplifications: the jet moves uniformly, its viscosity has little influence, and the dynamical effects of the environment can be neglected. It predicts that the growth rates are imaginary numbers, which indicates that disturbances are convected downstream with the base flow velocity $V_j$. The maximum growth rate $\omega_i^{\textin{max}}$ is of the order of the inverse of the inertio-capillary time
\begin{equation}
\label{tic}
\hat{t}_{ic}=(\rho_j R_j^3/\sigma)^{1/2},
\end{equation}
which means that the jet breakup time scales as $\hat{t}_{ic}$. Therefore, the distance covered by the dominant perturbation before breakup is of the order of $V_j \hat{t}_{ic}$. The wavelength of the dominant perturbation is $\lambda_R^{\textin{max}}\simeq 2.9\pi R_j$. These predictions have shown to be fairly accurate in many experiments with capillary jets \citep{EV08,GG09}.

% Extensions
Rayleigh's basic analysis has finely been tuned to account for various additional effects, such as the existence of a shear boundary layer on the inner side of the interface \citep{GHM14}, viscous damping \citep{T35}, a non-negligible fluid environment \citep{EV08}, surfactants \citep{TL02}, viscoelasticity \citep{GG82}, electric effects \citep{T69,HBCMS00,LRG05}, etc. These factors can be grouped into two categories: destabilizing and stabilizing. Typically, destabilizing factors increase the perturbation growth rates over the whole spectrum, the most unstable wavenumber and the range of unstable wavenumbers, while stabilizing factors produce the opposite effects. This has important practical implications because, according to Eqs.\ (\ref{esti}), destabilizing factors reduce not only the jet breakup length but also the size of the resulting droplets.

% This section
There are numerous studies on the temporal stability of jets that consider different effects and combinations of them. In this section, we will review those related to the microfluidic configurations considered in this work.

\subsection{Boundary layer and viscosity effects}

% Realistic velocity profiles
The mismatch between the velocities of a liquid jet and the surrounding gaseous medium significantly alters both the liquid and gas velocity profiles for large enough Weber numbers. The consideration of realistic velocity distributions unveils unstable non-axisymmetric modes that do not appear in the stability analysis of inviscid parallel and uniform streams. Those distributions are essential to explain atomization experiments at large enough Weber numbers \citep{GPG01}.

\begin{figure}[hbt]
%\begin{center}
\vcenteredhbox{\resizebox{0.28\textwidth}{!}{\includegraphics{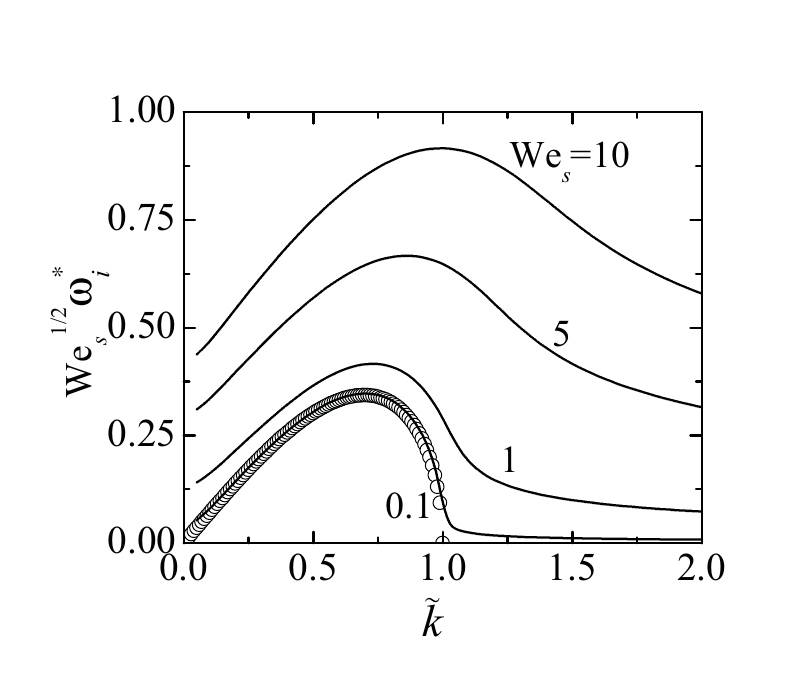}}}\vcenteredhbox{\resizebox{0.2\textwidth}{!}{\includegraphics{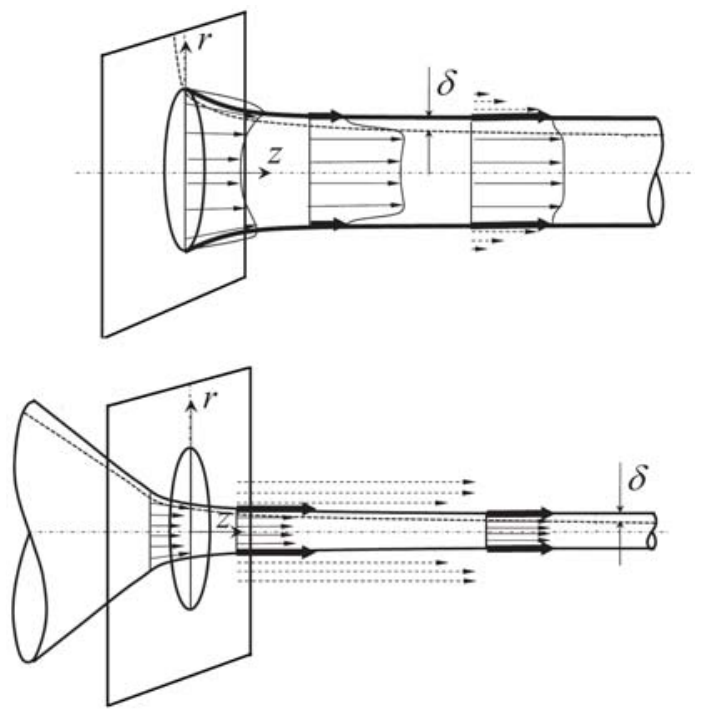}}}
%\end{center}
\caption{Temporal growth rate $\omega^{*}_i$ as a function of the (real) wavenumber $\widetilde{k}$ for $\text{We}_s$=0.1, 1, 5, and 10 (and the additional constraint $\delta=\text{Re}_j^{-1/2}$) \citep{GHM14}. The circles correspond to Rayleigh's dispersion relation. Here, $\omega^*_i=\omega_i R_j/V_j$ and $\widetilde{k}=k R_j$. The sketches on the right-side of the figure show two situations (discharge in a still ambient and flow focusing) where these results apply \citep{GHM14}.}
\label{bl}
\end{figure}

% Boundary layer
\citet{GHM14} have analytically shown that a shear boundary layer on the inner side of the interface can be regarded as a destabilizing factor for distances from the boundary layer origin of the order of the jet's radius. For longer distances, the boundary layer does not affect the jet's stability, and Rayleigh's dispersion relation is recovered. Figure \ref{bl} shows the growth rate as a function of the wavenumber for different values of the modified Weber number We$_s=(1-V_s/V_j)^2\text{We}_j$. The limit We$_s\to 0$ ($V_s/V_j\to 1$) corresponds to a vanishing boundary layer. The growth rate converges to that of the Rayleigh mode in that limit. These results suggest that the inner boundary layer growing in, for instance, gaseous flow focusing can decrease the droplet size and the distance between two consecutive droplets. 

\begin{figure}[hbt]
%\begin{center}
\centering{\resizebox{0.3\textwidth}{!}{\includegraphics{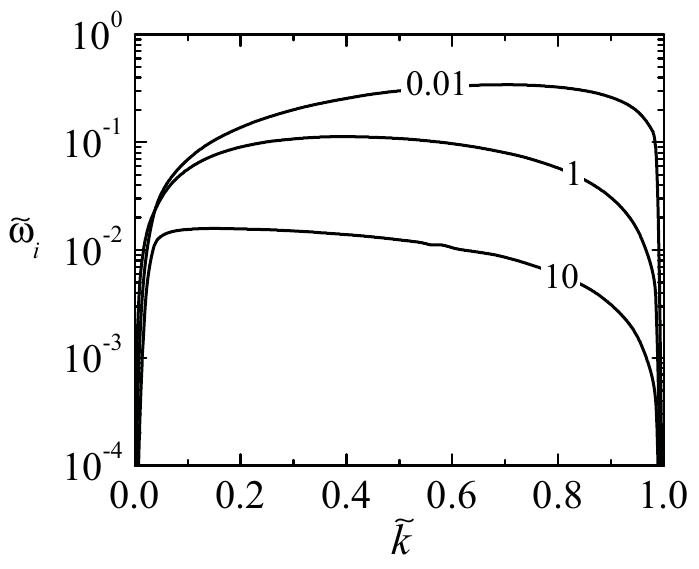}}}
%\end{center}
\caption{Temporal growth rate $\widetilde{\omega}_i$ as a function of the wavenumber $\widetilde{k}$ for $\rho=\mu=10^{-4}$ and different values of Oh$_j$ as indicated by the labels \citep{MG08a}. Here, $\widetilde{\omega}_i=\omega_i \hat{t}_{ic}$ and $\widetilde{k}=k R_j$.}
\label{viscosity}
\end{figure}

% Viscosity
For large enough viscosities, radial diffusion of momentum ensures a quasi-uniform jet velocity profile right behind the nozzle. Jets moving with that velocity profile can be described from the Lagrangian frame of reference solidly moving with the jet. In this case, the velocity $V_j$ no longer enters the problem, and viscous effects are quantified by the Ohnesorge number Oh$_j=\mu_j(\rho_j \sigma R_j)^{-1/2}$ [Eq.\ (\ref{oh})]. Figure \ref{viscosity} shows the influence of this parameter on the temporal growth rate. As can be seen, both the perturbation growth rate and the most unstable wavenumber decrease as the Ohnesorge number increases, which shows the stabilizing effect of viscosity \citep{T35}. The maximum value $k=R_j^{-1}$ of unstable wavenumbers is not affected by this property. For viscous liquid jets, the growth factor scales with the inverse of the viscous-capillary time
\begin{equation}
\label{tvc}
\hat{t}_{vc}=\mu_jR_j/\sigma.
\end{equation}
These results indicate that viscosity can significantly increase the droplet size and the distance between two consecutive droplets produced in jetting realizations.

\begin{figure}[hbt]
%\begin{center}
\centering{\resizebox{0.3\textwidth}{!}{\includegraphics{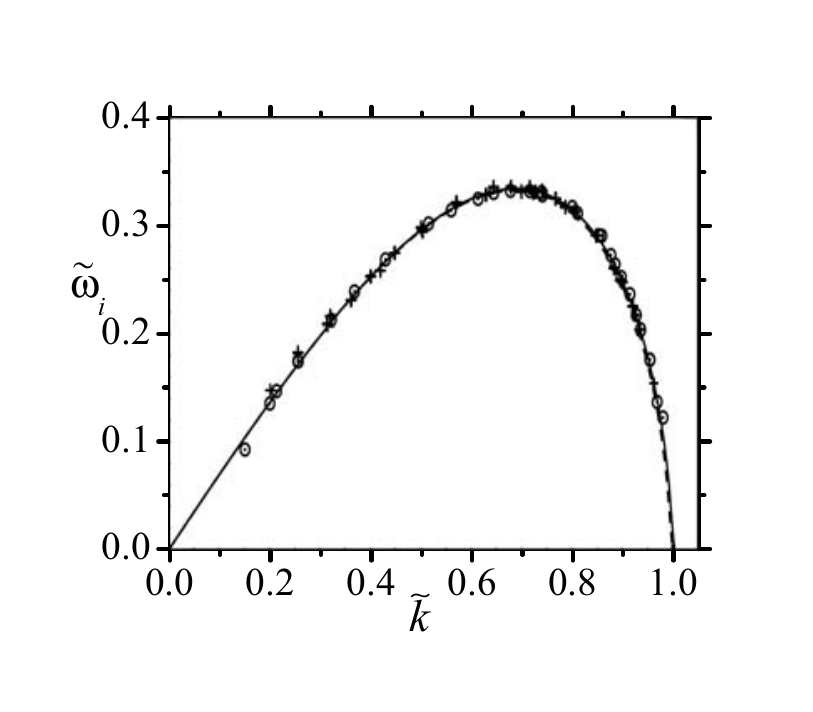}}}
%\end{center}
\caption{Temporal growth rate $\widetilde{\omega}_i$ as a function of the wavenumber $\widetilde{k}$ \citep{GG09}. The crosses and circles correspond to experimental results measured with the breakup time and amplitude-evolution method, respectively. The solid line is the theoretical predictions for Oh$_j=0.021$ and We$_j=59$, while the dashed line was calculated without the surrounding air effect. Here, $\widetilde{\omega}_i=\omega_i \hat{t}_{ic}$ and $\widetilde{k}=k R_j$.}
\label{gonzalez}
\end{figure}

% Gonzalez's validation
When a liquid jet is ejected into a still gaseous ambient, the latter has little influence on the jet's stability under axisymmetric perturbations even for moderately large Weber numbers. \citet{GG09} accurately measured the temporal growth rates of moderately viscous capillary jets emitted in a still atmosphere. The experimental data exhibited remarkable agreement with theoretical predictions calculated by translating spatial analysis \citep{LG86a} results into the temporal variables (Fig.\ \ref{gonzalez}).

\subsection{Confinement effects}

\begin{figure}[hbt]
%\begin{center}
\centering{\resizebox{0.37\textwidth}{!}{\includegraphics{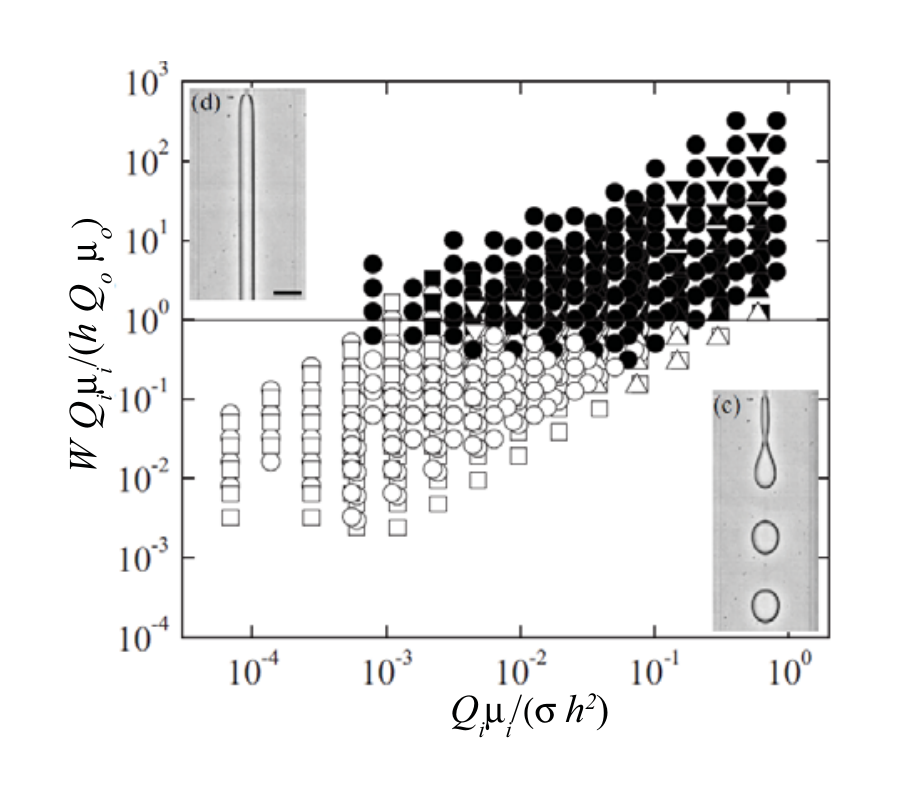}}}
%\end{center}
\caption{Stable jets (open symbols) and unstable jets or dripping (solid symbols) produced in a coflowing liquid-liquid configuration inside a rectangular channel of width $W$ and height $h$ \citep{HAFSW09}. $Q_i$ and $Q_o$ are the inner and outer flow rates, while $\mu_i$ and $\mu_o$ stand for the inner and outer viscosities, respectively.}
\label{confinement}
\end{figure}

% Confinement
The stability of confined capillary jets is of great interest at both fundamental and practical levels, especially in channels with noncircular cross-sections, which are commonly used in microfluidics \citep{SBPH12}. \citet{CM08} showed experimentally that the presence of external walls delays the breakup of liquid threads coflowing with an outer stream, and significantly increases the wavelength of the most unstable mode. In fact, capillary instability can be completely suppressed in a channel with a rectangular cross-section (Fig.\ \ref{confinement}) \citep{HAFSW09}.

% Theoretical works. Cylinders
Most theoretical works on the stability of confined capillary jets have considered the circular geometry \citep{H71,GCUA07,GCA08,KKRBK12}. \citet{H71} studied analytically the stability of the pressure-driven jet in a tube for infinitely small wavenumbers. He found that the flow was unstable under axisymmetric perturbations in all the cases considered. The same conclusion can be derived from the spatio-temporal stability analysis of \citet{GCUA07} and \citet{GCA08}. Using the approach of \citet{H71}, \citet{KKRBK12} found certain parameter conditions for which a jet confined in a circular tube becomes stable.

% Theoretical works. Rectangular shapes
\citet{JAM12} studied numerically the growth of small-amplitude varicose perturbations in jets flowing between parallel surfaces. Confinement did not completely stabilize the jet in any of the cases considered. A recent temporal stability analysis \citep{CHM19b} suggests that a jet sliding over the channel wall is unconditionally stable, while a detached liquid thread can be stable only for very small distances between the interface and the wall, and sufficiently high viscosities or small enough interfacial tensions.

% Summary
Overall, one can state that confinement inhibits capillary instabilities in microfluidic devices. Breakup takes place essentially by quasi-static mechanisms \citep{A16}, which facilitates the control of the size and morphology of the produced fluidic entities.

\subsection{Compound jets}

% Compound jets
Compound jets consist of an inner fluid surrounded by a liquid shell, both moving coaxially at the same velocity \citep{SM85,CMPR00,HMFG10}. Their breakup results in compound drops whose morphology can be controlled by adjusting the diameters and velocities of the inner and outer jets \citep{G98a,BSGM05}. The linear stability analysis of compound jets reveals the existence of the stretching and squeezing unstable modes, which correspond to an in-phase and out-of-phase deformation of the inner and outer interfaces, respectively \citep{SM85,CMPR00,LSPR07,HMFG10}. The stretching and squeezing modes are mainly driven by capillary forces at the inner and outer interfaces, respectively. The stretching mode dominates the breakup process because its maximum growth rate is larger than that of the squeezing mode \citep{SM85,CMPR00,HMFG10}. The growth rates of the stretching mode match those calculated by \citet{T35} when the radius of the outer jet takes sufficiently large values \citep{CMPR00}.

\subsection{Viscoelasticity effects}

% Viscoelasticity. Olroyd-B model. Goren's results
Viscoelasticity is frequently quantified using the Olroyd-B model \citep{BAH87}, which provides reasonably accurate predictions while keeping certain simplicity \citep{BER01} for Boger fluids. These fluids are dilute polymer solutions in viscous and low-viscosity solvents that hardly exhibit shear-thinning \citep{J09}. Non-Newtonian effects in an Olroyd-B liquid are quantified by the stress relaxation and retardation times $\lambda_s$ and $\lambda_r$, respectively (see Sec.\ \ref{sec2}).

% Relaxed jets
Viscoelasticity has a destabilizing effect on relaxed jets: the axisymmetric capillary mode grows in a viscoelastic jet faster than in a Newtonian one with the same Ohnesorge number owing to the apparent shear thinning associated with elasticity \citep{M65,GYPS69,BLD00,FJ03,YYF16}. This destabilizing effect translates into an increase of the minimum Weber number leading to convective instability \citep{MG08b}. These results have relatively little consequences at the practical level because viscoelastic stress is relaxed in the base flow of few applications. 

\begin{figure}[hbt]
%\begin{center}
\centering{\resizebox{0.3\textwidth}{!}{\includegraphics{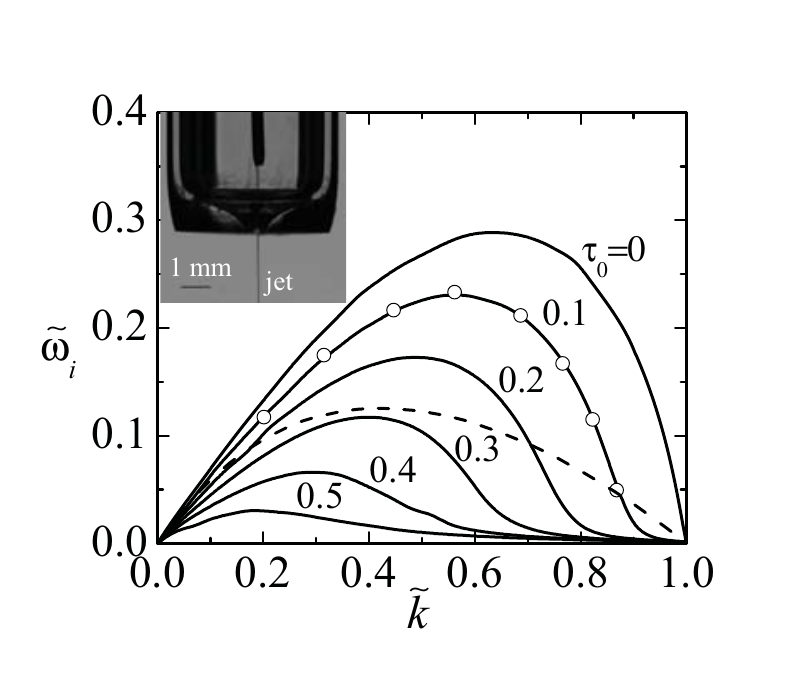}}}
%\end{center}
\caption{Temporal growth rate $\widetilde{\omega}_i$ as a function of the wavenumber $\widetilde{k}$ for the stress relaxation time $\widetilde{\lambda}_s=100$, the retardation time $\widetilde{\lambda}_r=10$, the Ohnesorge number (defined in terms of the Newtonian viscosity) Oh$_j=0.849$, and different values of the unrelaxed axial stress $\tau_0$ \citep{MHGM15}. The circles correspond to the solution obtained by \citet{GG82} for $\tau_0=0.1$. The dashed line is the solution for a Newtonian jet ($\widetilde{\lambda}_s=\widetilde{\lambda}_r=\tau_0=0$). All the quantities have been made dimensionless with $R_j$, the inertio-capillary time (\ref{tic}) and capillary pressure $\sigma/R_j$, i.e. $\widetilde{k}=k R_j$, $\widetilde{\omega}_i=\omega_i \hat{t}_{ic}$, $\widetilde{\lambda}_s=\lambda_s/\hat{t}_{ic}$ and $\widetilde{\lambda}_r=\lambda_r/\hat{t}_{ic}$. The inset shows a microjet of a polyacrylamide aqueous solution at a concentration of 1000 ppm produced with gaseous flow focusing \citep{PMVG16,PVCM17}. }
\label{goren}
\end{figure}

% Unrelaxed jets
The stretching suffered by the polymers in some microfluidic devices \citep{CJ06,ESMRD06,PMVG16,PVCM17} produces an unrelaxed axial stress in the emitted jet. Macromolecules are arranged along the axis of the thin emitted viscoelastic jets, which hinders their relaxation to the coiling state once the external elongational stresses have disappeared. The unrelaxed stress significantly reduces the growth rate of the varicose mode \citep{GG82,MHGM15} (Fig.\ \ref{goren}). This effect explains the stability of long viscoelastic threads extruded in, for instance, electrospinning \citep{RY08} or flow focusing \citep{PVCM17}. Micrometer filaments with lengths up to 1 cm and Weber numbers of the order of $10^{-4}$ were formed in front of the discharge orifice of a gaseous flow focusing device \citep{PMVG16}, while extremely long threads were produced behind that orifice with relatively low polymer concentrations \citep{PVCM17}. It must be noted that the unrelaxed axial stress increases the speed at which capillary waves move over the jet's surface \citep{GG56a}, which favors absolute instability despite the reduction of the growth rates \citep{MHGM15}.

\subsection{Electrified jets}

% Electric fields. New quantities
The stability analysis of electrified jets involves both the relative electrical permittivity and dimensionless electric relaxation time,
\begin{equation}
\label{tk}
\varepsilon=\frac{\varepsilon_j}{\varepsilon_o}, \quad \tau_{ej}=\frac{\varepsilon_j}{K_j \hat{t}_{ic}},
\end{equation}
where $\varepsilon_j$ and $\varepsilon_o$ are the electrical pemittivity of the liquid jet and outer medium, respectively, and $K_j$ is the jet electrical conductivity. The perfect conductor and dielectric limits correspond to $\tau_{ej}\to 0$ and $\tau_{ej}\to \infty$, respectively. In the absence of an externally applied electric field, electric effects are accounted for by the Taylor number
\begin{equation}
\Gamma_e=\frac{R_j \varepsilon_o E_{on}^2}{\sigma},
\end{equation}
which measures the electric normal stress $\varepsilon_o E_{on}^2$ in terms of the capillary pressure $\sigma/R_j$. Here, $E_{on}$ stands for the normal electric field on the outer side of the interface and produced by the surface charge density $\sigma_e$. In the absence of an inner electric field ($\tau_{ej}\to 0$), $E_{on}=\varepsilon_o \sigma_e$. If the jet is subjected to the action of an external electric field (like in electrospray), this last effect is quantified through the dimensionless electric strength
\begin{equation}
\label{strength}
{\cal E}_j=\frac{R_j \varepsilon_o E^2}{4\pi \sigma},
\end{equation}
where $E$ is the axial/radial electric field intensity.

\begin{figure}[hbt]
%\begin{center}
\centering{\resizebox{0.3\textwidth}{!}{\includegraphics{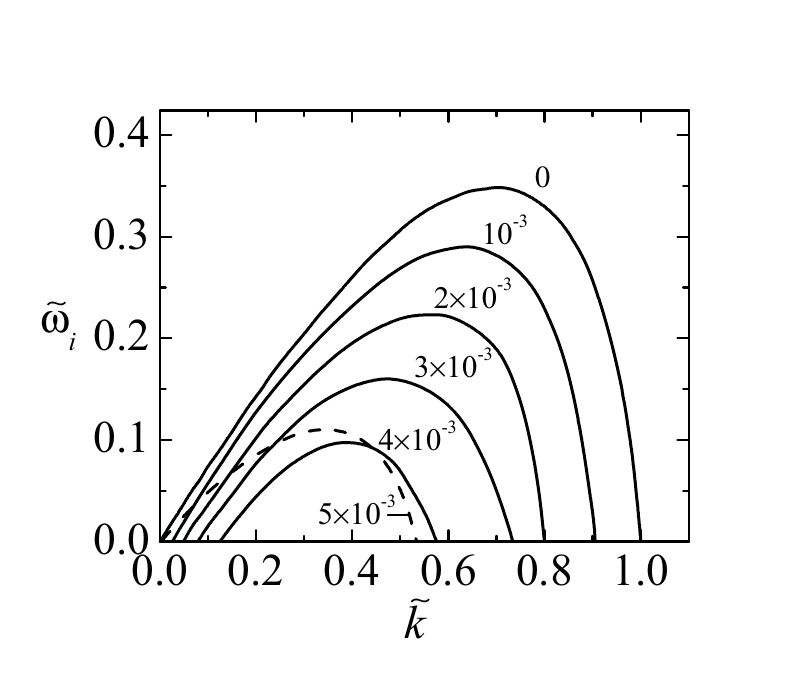}}}
%\end{center}
\caption{Temporal growth rate $\widetilde{\omega}_i$ as a function of the wavenumber $\widetilde{k}$ for a conductor ($\tau_{ej}=0$) inviscid jet with $\varepsilon=78$ (solid lines) in an axial electric field \citep{S71c}. The dashed line shows the results for the dielectric case ($\tau_{ej}=\infty$). The labels indicate the values of the electric strength ${\cal E}_j$. Here, $\widetilde{\omega}_i=\omega_i \hat{t}_{ic}$ and $\widetilde{k}=k R_j$.}
\label{elect}
\end{figure}

% Results
Linear stability analyses consistently show that both the surface charge and the externally applied DC radial electric field destabilize the jet by increasing the maximum growth rate and most unstable wavenumber \citep{HC71,HBCMS00}. These factors also increase the minimum Weber number for convective instability \citep{GLHRM18}. These effects can be explained in terms of an increase of the effective surface tension produced by the electric normal stress. The opposite occurs when axial electric fields are applied on both conductor \citep{S70,S71b,LYY06} and dielectric \citep{S71b,LYY06} liquids (Fig.\ \ref{elect}). In this case, the electric field stabilizes the jet. As will be explained in Sec.\ \ref{sec12}, these results have been used to calculate the breakup length in electrospray \citep{IYXS18}.

\begin{figure}[hbt]
%\begin{center}
\centering{\resizebox{0.3\textwidth}{!}{\includegraphics{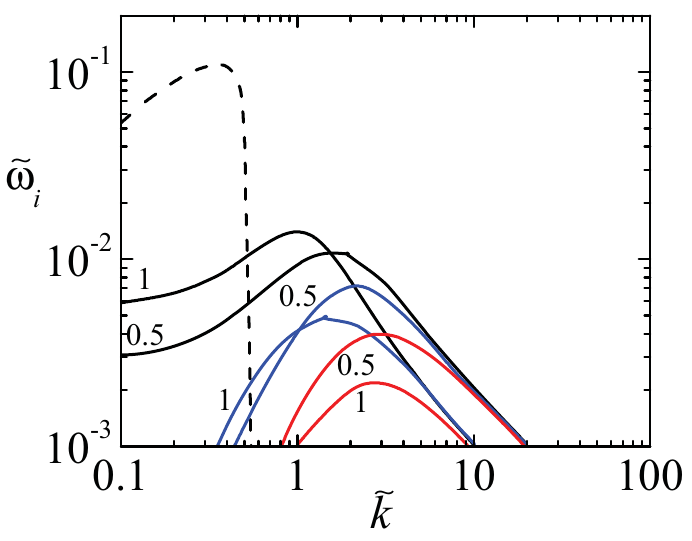}}}
% \end{center}
\caption{Temporal growth rate $\widetilde{\omega}_i$ as a function of the wavenumber $\widetilde{k}$ for an inviscid jet with $\varepsilon=78$ in an axial electric field of intensity ${\cal E}_j=5\times 10^{-3}$ \citep{S71c}. The labels indicate the values of the electric relaxation time $\tau_{ej}$. The dashed line corresponds to the dielectric case ($\tau_{ej}=\infty$). The black, blue and red lines correspond to the modes $m=0$, 1 and 2, respectively. Here, $\widetilde{\omega}_i=\omega_i \hat{t}_{ic}$ and $\widetilde{k}=k R_j$.}
\label{elect2}
\end{figure}

% Leaky-dielectric in an axial electric field
The leaky-dielectric (low-conductivity) jet subject to an axial electric field is the most interesting configuration for the study of the electrospray cone-jet mode. In a leaky-dielectric jet, the surface charge evolves trying to screen the external electric field to eliminate the inner one. \citet{S71c} showed that this surface charge relaxation makes disturbances grow in an oscillatory manner, contrary to what happens in either perfect dielectrics or perfect conductors. The interaction between the electrical and viscous stresses at the electrohydrodynamic boundary layer next to the interface of nearly-inviscid threads can destabilize both asymmetric and axisymmetric perturbations (Fig.\ \ref{elect2}). \citet{M94b,M96} derived the dispersion relationships corresponding to the low- and high-viscosity limits. However, the dependency of the temporal growth rates on the parameters of the problem has not been explored for arbitrary viscosity values. The temporal stability analysis of a charged, leaky-dielectric, Olroyd-B jet under an axial electric field shows that both the electric field \citep{XYQF17} and the axial polymeric stress of the base state \citep{CJ08} inhibit the axisymmetric mode.

\subsection{Surfactant effects}

% Surfactants
When the jet interface is loaded with a monolayer of a surface-active molecule (surfactant), both the outwards surfactant convection and the dilatation of the interface element make the surfactant surface density decrease in the necking region. This surfactant depletion increases the surface tension in that region, which causes Marangoni convection in the opposite direction. Marangoni convection gives rise to additional energy dissipation, which reduces the growth rates of the linear capillary oscillations.

\begin{figure}[hbt]
%\begin{center}
\centering{\resizebox{0.35\textwidth}{!}{\includegraphics{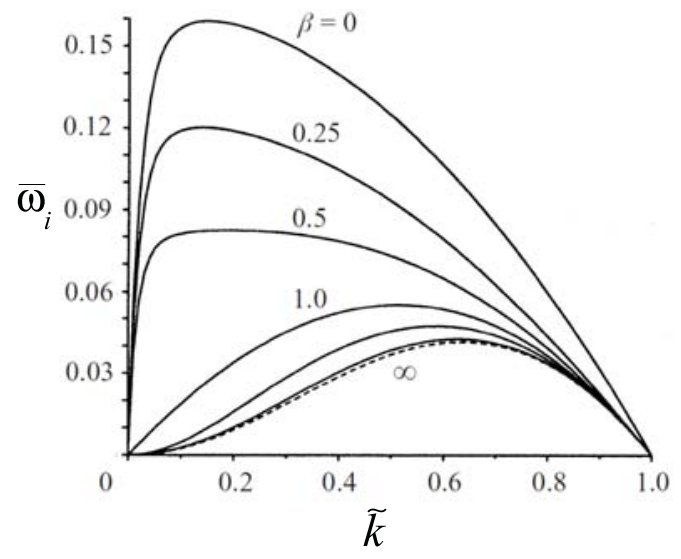}}}
%\end{center}
\caption{Temporal growth rate $\overline{\omega}_i$ as a function of the wavenumber $\widetilde{k}$ for Oh$_j=10$ and different values of the surfactant elasticity $\beta$ \citep{TL02}. Here, $\overline{\omega}_i=\omega_i \hat{t}_{vc}$ and $\widetilde{k}=k R_j$.}
\label{surfactant}
\end{figure}

% Insoluble
In many microfluidic applications, the characteristic time of surfactant adsorption/desorption is much greater than the droplet production time, and surfactant solubility can be neglected at least in part of the process. As explained in Sec.\ \ref{sec2}, this approximation reduces significantly the dimension of the parameter space, which makes the problem more tractable \citep{KP01,TL02}. Figure \ref{surfactant} shows the temporal growth rate as a function of the wavenumber for a liquid thread loaded with an insoluble surfactant characterized by its elasticity $\beta=-(\Gamma_0/\sigma_0)\, d\sigma/d\Gamma|_{\Gamma=\Gamma_0}$ ($\Gamma_0$ and $\sigma_0=\sigma(\Gamma_0)$ are the initial surfactant surface density and surface tension, respectively). The growth rate decreases as the surfactant elasticity (strength) increases \citep{TL02}. As can be seen, $\widetilde{k}^{\textin{max}}$ does not depend monotonically on $\beta$.

% Soluble
The opposite limit to the insolubility case considered above is that in which the surfactant transport between the interface and the bulk fluids is diffusion-limited (see Sec.\ \ref{sec2}). In this case, the stabilizing effects of surfactants are similar to those described for the insoluble one \citep{HPM99}.

\subsection{Relationship with the convective-absolute instability transition}

% Relationship with c/a instability transition
One may think that the destabilizing factors described above necessarily increase the critical Weber numbers leading to the convective-to-absolute instability transition, while the stabilizing elements produce the opposite effect. This is so in most cases (see, e.g., Refs.\ \cite{MG08b,LGLYY13,GLHRM18}). However, there are two notable exceptions to this rule. The existence of a boundary layer on the inner side of the interface makes that surface ``slip"\ over the jet's inviscid core. This favors the motion of capillary waves in the current direction, which delays the jetting-to-dripping transition \citep{VGMCH13,GHM14}. The opposite situation arises in unrelaxed viscoelastic jets \citep{MHGM15}. The speed at which growing waves travel over these jets increases with the elastic axial stress \citep{GG56a}. One may expect that if this stress, measured in terms of the dynamic (convective) pressure, exceeds a certain threshold, then the jet will fail to sweep downstream those waves (absolute instability). This explains the simultaneous reduction of the growth rates (Fig.\ \ref{goren}) and the increase of the critical Weber numbers due to the unrelaxed stress \citep{MHGM15}.

\subsection{Modulated capillary instability}

\begin{figure}[hbt]
%\begin{center}
\centering{\resizebox{0.5\textwidth}{!}{\includegraphics{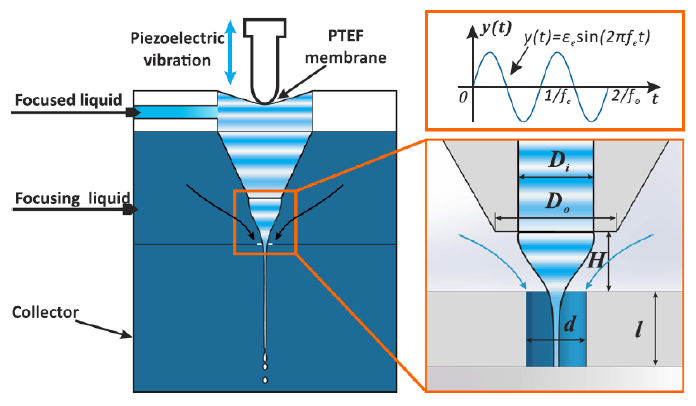}}}
%\end{center}
\caption{Sketch of the active flow focusing principle. The mechanical vibration acts on the focused liquid \citep{YQMZXS19}.}
\label{modulated}
\end{figure}

% Modulated capillary instability.
In many jetting applications, there is a window of operational conditions within which the dependence of the perturbation growth rate on the wavelength exhibits a sharp maximum. This constitutes a ``natural wave filter"\ for the dominant perturbation, which favors the production of monodisperse collections of droplets. When this situation does not occur, external stimuli can be applied to trigger and control the jet breakup, which allows one not only to select the droplet diameter but also to narrow the size distribution. For instance, in continuous inkjet systems, the jet is broken into drops inside a chamber through a pulse produced by a piezoelectric crystal. The resulting droplets are electrically charged, expelled from a printhead nozzle, and positioned on the substrate by appropriately setting their speed and charge. A similar idea has been applied, for instance, to the axisymmetric gaseous \citep{MSLXD18} and liquid-liquid \citep{YQMZXS19} flow focusing (Fig.\ \ref{modulated}), or to the bubble formation in forced co-axial air-water jets \citep{RBSM19}. Pulsed stimulation can produce the detachment of a single drop or a group of drops in the middle of the jet \citep{GGGCC19}. The position and number of the drops can be controlled by adequately selecting the pulse parameters.

\section{Whipping instabilities}
\label{sec6}

% Aerodynamic whipping
As mentioned in the previous section, the existence of an outer fluid medium triggers the transition from the varicose to the whipping (bending, kink) mode for large enough Weber numbers (defined in terms of the jet velocity relative to that of the outer medium) \citep{LH00}. In this case, surface tension has a stabilizing effect, and the destabilizing factor is purely aerodynamic: a perturbation at the interface causes the outer fluid to accelerate as it passes a crest, lowering the pressure at that point and encouraging the crest to increase in size (as in wind-generated ripples on a liquid free surface). An accurate calculation of the perturbation growth rate requires the consideration of the gas viscosity even for small values of this parameter \citep{GP05}. In fact, the existence of an outer boundary layer significantly affects the whipping of a jet discharging into a still gaseous atmosphere.

% Electrical whipping
Electrically charged jets in a hydrodynamically passive dielectric medium may also develop the whipping instability for sufficiently large values of the Taylor number $\Gamma_e$ \citep{HSRB01b,EV08,YDLD14}. The whipping instability in a perfectly conductor jet can be explained as follows: if a small portion of the jet moves slightly off its axis, the charge re-distributes instantaneously along the jet surface accumulating in the ridges and valleys of the deformed interface. This occurs in such a way that the electrical forces push that portion farther away from the axis \citep{HSRB01b}. In fact, while the Rayleigh axisymmetric instability reduces the liquid surface per unit volume, the whipping instability does the contrary. This lowers the surface charge density, separates the electric charges, and reduces the potential energy associated with Coulombic repulsion. For a perfectly conductor, apolar and inviscid cylinder in the absence of an externally applied electric field, whipping instability arises at the so-called Rayleigh limit $\Gamma_e=3/2$ \citep{R81}.

\begin{figure}[hbt]
%\begin{center}
\centering{\resizebox{0.35\textwidth}{!}{\includegraphics{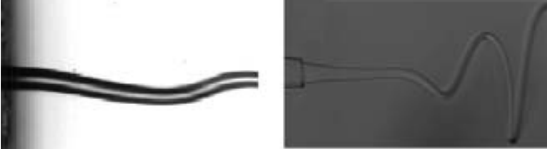}}}
% \end{center}
\caption{Whipping instability in a viscous microjet produced by gaseous flow focusing \citep{AFMG12} (left) and liquid-liquid electrospray \citep{GRGPF14} (right).}
\label{whipping}
\end{figure}

% Common features
Both aerodynamic and electrical whipping (Fig.\ \ref{whipping}) is enhanced by the jet's viscosity, which damps out the varicose mode in favor of bending perturbations. As mentioned in Sec.\ \ref{sec4}, whipping almost always exhibits a convective character, and, therefore, it is observed far beyond the tapering liquid source. In fact, the convective-to-absolute instability transition for the lateral mode $m=1$ has been found in very few works \citep{HMFG10,LGLYY13}.

% Effects
Whipping arises in a number of microfluidic configurations, including electrospray/electrospinning \citep{HSRB01b,SHBR01} and flow focusing \citep{AFMG12}. The violent slashes characterizing the whipping regime are responsible for wider droplet size distributions with smaller average values. Whipping instability precludes the gentle deposition of the emitted jet or the droplets resulting from its breakup. It can be controlled by making the surrounding fluid coflow with the jet, as occurs in electro-coflowing \citep{GRGPF14} and electro-flow focusing \citep{G07a} for ambient pressure mass spectrometry \citep{FS14a}, or in electrohydrodynamic direct-writing \citep{HBDPLYX13} assisted with airflow \citep{JWLLLZ18}. External fields applied to the fiber extruded in near-field electrospinning allow controlling the fiber oscillation to build 3D structures \citep{LRC20}. Whipping usually has a positive effect in the production of polymeric fibers because the slashes mentioned above considerably reduce the diameter of the precursor jet before its solidification. Nevertheless, it can be suppressed by combining low and uniform electrical fields with the mechanical pulling exerted by the collector \citep{KR12}.

\section{Jet breakup}
\label{sec7}

\subsection{Newtonian liquids}

\subsubsection{Satellite droplets}

% Nonlinear effects
The linear (local or global) stability analysis describes only the early stage of the interface deformation during the jet breakup. Non-linear contributions to the hydrodynamic equations invalidate the predictions derived from this analysis for the late phase of the thread breakup. This occurs even in the Stokes limit (when the nonlinear convective term can be neglected) due to the nonlinearity of the capillary pressure.

% Satellite droplet formation
Consider the Lagrangian frame of reference solidly moving with the jet. The growth of the unstable sinusoidal linear mode produces a neck and two bulges on the two sides of the neck. The liquid evacuates the neck towards the bulges driven by the capillary pressure, accelerating in the direction of motion. For $\text{Oh}_j\ll 1$, the acceleration takes its maximum value in a section located between the neck and the bulge, which makes the central neck symmetrically split into two ones. Each neck migrates towards the closest bulge until pinching the interface. This migration is responsible for the formation of a satellite droplet between the two parent drops \citep{CS97}. Satellite droplets are an undesired effect in most microfluidic applications.

% Some classical and recent works on satellite droplets
The formation of satellite droplets in Newtonian liquid jets has been studied since the early seventies. \citet{GY70} compared their experiments with the weakly nonlinear analysis of \citet{Y68}. \citet{CM80} determined the conditions to suppress the satellite droplets by forcing the precursor liquid jet. Direct numerical simulations of the Navier-Stokes equations were conducted by \citet{ML90} and \citet{AM95}. They found very good agreement between the computed droplet radii and the experimental values obtained by \citet{RJ70} and \citet{L75}. As will be explained below, subsequent works have considered the effect of electric fields \citep{SH97,LGP99,LG04,CHB07} and surfactants \citep{CMP02,DSXCS06,MB06,CMP09,PMHVV17,MS18,KWTB18} on the satellite droplet formation in jets and similar configurations (liquid bridges, pendant drops, \ldots). To  study these and other effects, accurate adaptive solvers \citep{P09a}, boundary fitted methods \citep{HM16}, and elliptic mesh generation techniques \citep{KWTB18} have been developed.

\subsubsection{The interface pinch-off}

% Pinching. Scaling theories
The pinching of a Newtonian liquid free surface constitutes a formidable problem that offers a unique opportunity to observe the behavior of fluids with arbitrarily small length and time scales. Theoretical and experimental studies have focused on the evolution of the free surface minimum radius, $R_{\textin{min}}$, to determine which forces are relevant in the vicinity of the pinching region.

% Inviscid
For small viscous effects, the thinning of the liquid thread passes through an inertio-capillary regime characterized by the power law
\begin{equation}
\label{ebbb}
R_{\textin{min}}=A \left(\frac{\sigma}{\rho}\right)^{1/3} \tau^{2/3},
\end{equation}
where $\tau$ is the time to the pinching \citep{KM83,E93}. Different values of $A$ have been determined experimentally \citep{SPADBK12,HDBKB17}. Recent experiments and numerical simulations with very low viscosity liquids have found that the dimensionless prefactor $A$ exhibits a complex, nonmonotonic behavior over many orders of magnitude in $\tau$ \citep{DHHVRKEB18}. In those experiments and simulations, $A$ never fully reached its asymptotic value $A\simeq 0.717$.

\begin{figure}[h]
\centering{\resizebox{0.325\textwidth}{!}{\includegraphics{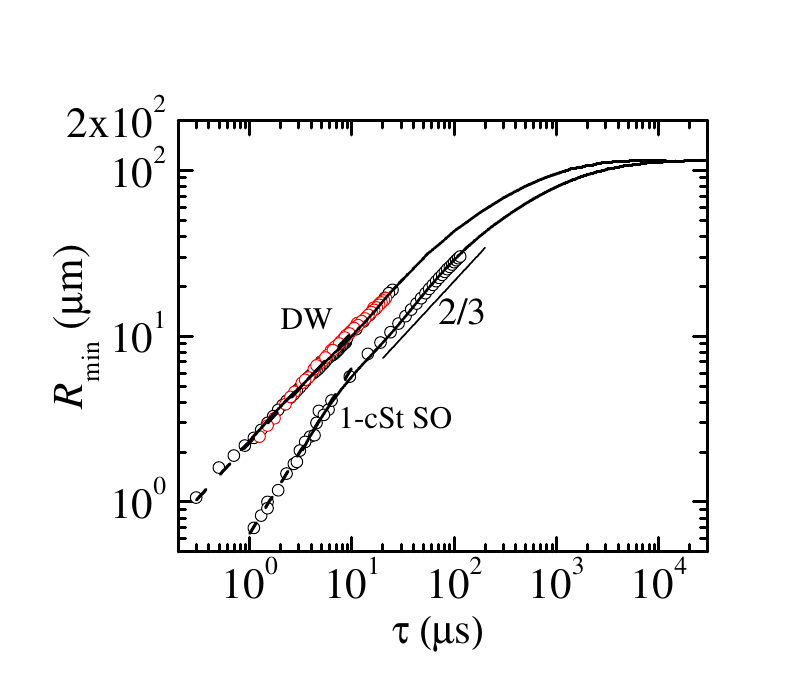}}}
\caption{Free surface minimum radius $R_{\textin{min}}$ as a function of the time to the pinching $\tau$ for deionized water (DW) and 1-cSt silicone oil (SO) \citep{RPVHM19}. The black and red symbols are the experimental data obtained from the breakup of a pendant drop hanging on a nozzle 115 and 205 $\mu$m in radius, respectively. The solid and dashed lines are the full numerical solution of the Navier-Stokes equations and the universal laws (\ref{ebbb}) with $A=0.55$ for DW and (\ref{e0}) for SO.}
\label{both2}
\end{figure}

% Viscous and universal
When viscous stresses are dominant, capillary pressure drives the flow against them during an intermediate phase, where the minimum radius verifies the equation \citep{P95}
\begin{equation}
\label{e1}
R_{\textin{min}}(\tau)=0.0709\ \sigma/\mu_j\ \tau.
\end{equation}
For lengths and times to the pinching of the order of or smaller than the intrinsic characteristic length $\ell_0=\mu_j/(\sigma\rho_j)$ and time $\tau_0=\mu_j^3/(\sigma^2\rho_j)$, the system is expected to reach an inertio-viscous-capillary regime in which all three forces are commensurate with each other. In this regime, the free surface minimum radius obeys the universal law \citep{E93}
\begin{equation}
\label{e0}
R_{\textin{min}}(\tau)=0.0304\ \sigma/\mu_j\ \tau.
\end{equation}
Figure \ref{both2} shows both the universality and validity of (\ref{ebbb}) and (\ref{e0}) \citep{RPVHM19}.

% Thermal fluctuations
The universal inertio-viscous-capillary regime is limited by the appearance of noticeable thermal fluctuations when the thermal length scale $\ell_T=k_B T/\sigma$ ($k_B$ is the Boltzmann constant) is reached, which is of the order of 1 nm at room temperature $T$ \citep{ML00,E02}. Thermal fluctuations make nanojets adopt double-cone neck shapes before breakup, which almost eliminates the formation of satellite droplets.

% Beyond scaling theories. Newtonian in a passive ambient.
A complex scenario of intermediate transitions between the regimes (\ref{ebbb})--(\ref{e0}) may take place before the liquid thread adopts the final inertial-viscous-capillary regime \citep{CCTSHHLB15,LS16}. \citet{SBN94} described the repeated formation of necks during the pinching of a viscous drop falling from a faucet. They explained this phenomenon in terms of the experimental noise. Long microthreads were observed by \citet{K96} during the breakup of jets around 50 cSt in viscosity. Interestingly, those microthreads stretched until their diameters fell down below approximately 1 $\mu$m. Then, the thinning process stopped, and the microthread broke up due to the growth of capillary waves, which gave rise to micrometer subsatellite droplets.

% Gaseous pinch-off
The pinch-off of an inviscid bubble is qualitatively different from its droplet counterpart. The minimum radius follows a non-universal power law with irrational exponents in the interval 0.53-0.57 \citep{BMSSPL06,KMZN06,TET07}, which are not strictly constant but exhibit logarithmic corrections \citep{OP93,BWT05,GSRM05,EFLS07}. The shape of the singular region evolves from a conical shape to a slender cylinder over many decades in time of the pinch-off process. This behavior bifurcates from that of the droplet for a density ratio around 0.25 \citep{BT08}. If there is a sufficiently intense gas flow across the pinching region so that the convective term in the momentum equation becomes of the order of that of the liquid, then the pinching region may become asymmetric and the scaling exponent becomes 1/3 \citep{GSRM05,DHRMV08}.

% Viscous bath
When the external liquid viscosity, however small, is taken into account, then capillary stresses are asymptotically balanced by viscous stresses in the two liquids, while inertia becomes negligible \citep{LS98,CBEN99}. The linear law (\ref{e0}) holds but with a prefactor which depends on the viscosity ratio. The presence of a viscous outer medium produces an exceptional form of singularity in which a long liquid thread forms before pinch-off, which violates universality and retains an imprint of the initial and boundary conditions \citep{DCZSHBN03,BWT05,TET07}. This behavior was also found in bubbles quasi-statically injected in a moderately viscous bath \citep{BWT05,TET07}, whose pinch-off is characterized by a scaling exponent significantly larger than 1/2. \citet{PSMJ19} have recently shown that the pinch-off dynamics of a bubble confined in a cylindrical capillary go through an early-time self-similar regime which erases the system's memory of the initial conditions and restores universality to bubble pinch-off.

\begin{figure}[h]
\vcenteredhbox{\resizebox{0.48\textwidth}{!}{\includegraphics{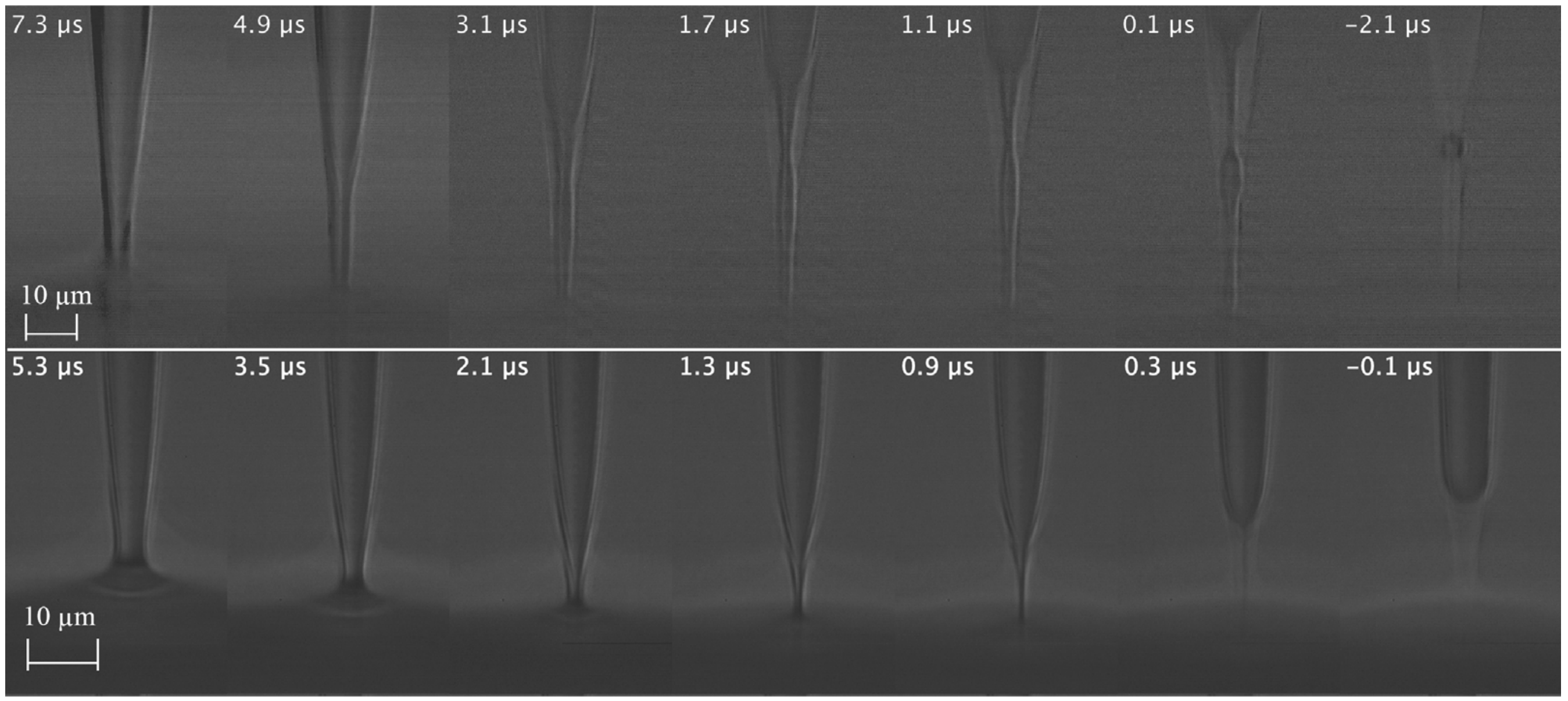}}}
\caption{Pinch-off of a drop of 5-cSt silicone oil (upper images) and glycerol/water 48/52\% (v/v) (lower images) \citep{RPVHM19}. The Ohnesorge number takes practically the same value in the two cases. The labels indicate the time to the pinching with an error of $\pm$100 ns.}
\label{images}
\end{figure}

% Recent results
Recent experiments \citep{RPVHM19} have shown that fluid dynamics in the vicinity of the pinching are much more intricate than what one may expect for apparently Newtonian liquids. Subsatellite droplets arise upon closer inspection during the breakup of pendant droplets of silicone oils. It has been speculated that this phenomenon is produced by bulk viscoelasticity effects caused by the extraordinarily small length and time scales reached as the system approaches the free surface breakup (Fig.\ \ref{images}). For very small surface tensions, the pinch-off of the interface between two liquids can be dominated by diffusion in the bulk \citep{LLMXCLSX19} before thermal fluctuations come into play \citep{HAWWELB06,PRKD12}. \citet{RMSC18} have recently modeled the pinch-off dynamics in the presence of particulate suspensions. \citet{RMPD19} have shown that turbulent flow field freezes in the final stage of the bubble pinching of a bubble, which leads to a self-similar collapse close to that of the unperturbed configuration.

\subsection{Viscoelasticity, electric fields and surfactants}

% 3D Results

% Viscoelasticity. Elasto-capillary regime
The breakup of jets becomes a rich and intricate problem when complex liquids \citep{A16} are considered. The presence of tiny amounts of polymeric molecules drastically alters the breakup dynamics. In this case, two consecutive drops are connected by a thin thread right before the breakup of the viscoelastic jet. These threads are subject to uniform axial stress caused by the elasticity of the dissolved polymeric molecules \citep{CBEM09}. In the elasto-capillary regime, the thinning of the liquid thread is driven by surface tension and resisted by that axial stress \citep{EV08}. Experiments show that the time evolution of the thread minimum radius obeys the exponential function \cite{BER90}
\begin{equation}
\label{eq1}
R_{\textin{min}}(t)= \hat{A}\ \exp{[-t/(3\lambda_{\textin{ext}})]}.
\end{equation}
If one assumes that this evolution can be described by the Olroyd-B model (\ref{h10}), the extensional relaxation time $\lambda_{\textin{ext}}$ coincides with the stress relaxation time $\lambda_s$ (see Sec.\ \ref{sec2}). \citet{EHS19} have recently shown that the thread profile calculated from the Olroyd-B model and rescaled by the thread thickness converges to a similarity solution.

\begin{figure}
\vcenteredhbox{\resizebox{0.48\textwidth}{!}{\includegraphics{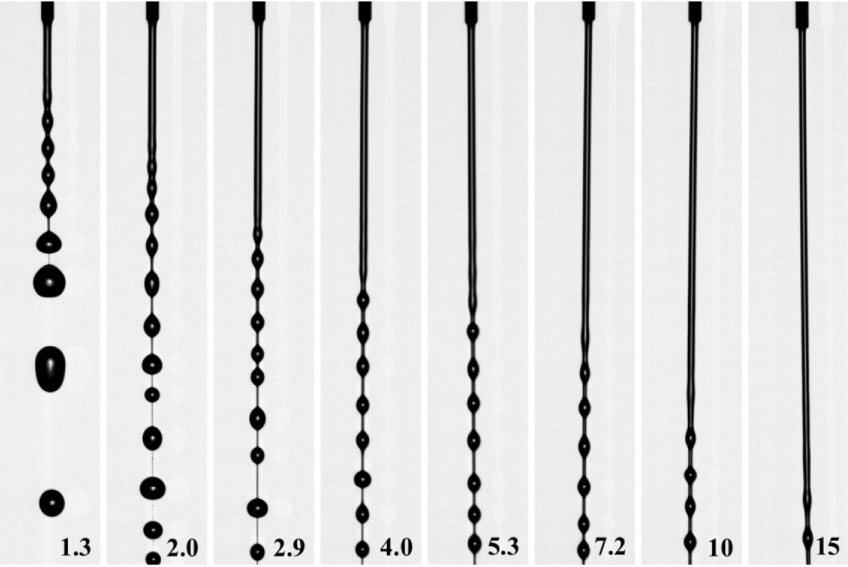}}}
\caption{Breakup of a jet of 0.1 wt.\% PEO solution emitted from a
nozzle 0.15 mm in diameter for different Weber numbers as indicated in the bottom corner \citep{MFMHC18}.}
\end{figure}

% Beads on a string and blistering
The beads-on-a-string structure is a drop-like pattern created on a viscoelastic thread by the capillary instability \citep{CEFLM06,BAHPMB10}. During the later stage of the viscoelastic thread thinning, polymers come close to their full extension, the extensional viscosity reaches an almost constant value, the thread behaves as a viscous Newtonian filament subject to the capillary instability, and tiny beads start to appear on the filament, giving rise to the so-called {\em blistering} instability \citep{OM05}.  However, \citet{SWE08} have observed experimentally growth rates that were greater by several orders of magnitude than those expected from the capillary instability based on a Newtonian extensional viscosity. \citet{E14b} has proposed an alternative explanation of the blistering phenomenon in terms of a demixing instability, which locally relaxes polymeric stress.

% Electric fields
The integration of the full Navier-Stokes equations over the breakup of a perfectly conductor cylindrical jet immersed in a radial electric field shows that nonlinear terms generally delay the jet breakup \citep{CHB07}. The diameter of the primary (satellite) droplet decreases (increases) as the electric strength increases, especially for large values of the Ohnesorge number. \citet{CHB07} described how electrostatic stresses over the interface produce similar effects to those of inertia in producing satellite droplets, which leads to the formation of such droplets even in the Stokes limit. Finite conductivity effects significantly affect the breakup process \citep{W12b}. In this case, electrokinetic effects may considerably alter the distribution of electric charges between primary and satellite droplets \citep{LGPH15}.

The electric field has no significant influence on the local pinch-off dynamics of a conductor liquid thread for low values of the Ohnesorge number \citep{CHB07}. Similar conclusions have been obtained in the Stokes limit \citep{WP11,CMCP11a}. However, the asymptotic behavior might be affected by the electric stresses in this case, because the relative magnitude of the electrostatic stress versus the capillary one increases in the pinch point as the interface breakup approaches \citep{CHB07}. In some configurations, the pinch-off solution cannot be obtained if only the leading order term is used in the electrostatic problem \citep{WP11}.

\begin{figure}[h]
\vcenteredhbox{\resizebox{0.48\textwidth}{!}{\includegraphics{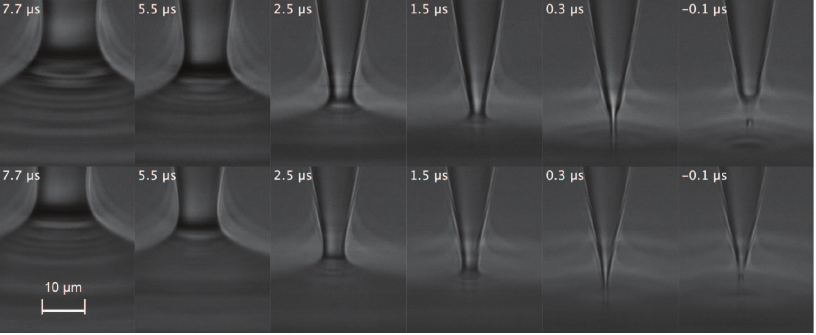}}}
\caption{Pinch-off of a drop of water+SDS 0.8cmc (upper images) and water+SDS 2cmc (lower images)  \citep{PRHME19}. The labels indicate the time to the pinching with an error of $\pm$100 ns. The breakup of that microthread produces a subsatellite droplet 1-2 $\mu$m in diameter.}
\label{images2}
\end{figure}

% Surfactants
Microthread cascades may arise during the breakup of jets loaded with surfactants \citep{MB06} due to the action of Marangoni stresses \citep{KWTB18}. These microthreads give rise to tiny subsatellite droplets following the free surface breakup (Fig.\ \ref{images2}). It is a question of controversy whether surfactants are swept away from the pinching region, and, therefore, the system follows the self-similar dynamics of clean viscous jets at times close to the breakup \citep{CMP02,LSFB04,CMP09,RABK09,CMCP11a,PMHVV17}. \citet{MRSS19} have recently shown the existence of a discontinuous transition at a critical elasticity below which satellite droplets are not formed. Experimental studies have also quantified the effect of surfactants on the size of satellite droplets \citep{KNS16,PMHVV17,KJMS18}.

% Surface viscosities
The breakup of jets of complex fluids is still an open problem. There are factors, such as surface viscous stresses associated with surfactant monolayers \citep{PMHVV17}, whose relevance is yet to be determined. High-resolution experiments and numerical simulations show the importance of surface viscosity in the final stage of the breakup even for quasi-inviscid surfactants \citep{PRHME19}. When Marangoni and surface viscous stresses are taken into account, the surfactant is not swept away from the thread neck. Surface viscous stresses eventually balance the driving capillary pressure in in the pinching region. 

\citet{WWKB20} have recently extended the viscous law (\ref{e1}) to account for the effect of a viscous surfactant monolayer in the limit of infinite surface Peclet number. Bulk viscosity, surface viscosity and capillary pressure compete with each other, which results in the asymptotic law
\begin{equation}
\label{e1m}
R_{\textin{min}}(\tau)=\frac{0.0709}{1+5B_{s0}/3R^*_{\textin{min}0}} \frac{\sigma}{\mu_j} \tau,
\end{equation}
where $B_{s0}=\mu^S/\mu R_0$ is the Boussinesq number, $R_0$ is the initial radius of the liquid thread, and $R^*_{\textin{min}0}$ the initial value of $R_{\textin{min}}$ in terms of $R_0$. When the free surface is flooded by surfactant \citep{MS19}, or in the limit of zero Peclet number \citep{WWKB20}, surface viscous stresses are predicted to cause the exponential thinning of the liquid thread right before its breakup. Surface viscoelasticity effects on the pinching of interfaces covered by surface-active bio-polymers have recently been considered too \citep{GSH20}.

\subsection{The 1D approximation}

% Practical results
The results described above possess great interest at the fundamental level, but they have relatively little consequences in terms of microfluidic applications. At this level, attention is normally paid to the formation and size of satellite droplets. For this purpose, many theoretical studies make use of the 1D (slenderness) approximation (see Sec.\ \ref{sec2.4}) to simulate the breakup of a fluid thread whose length equals $\lambda^{\textin{max}}/2$ ($\lambda^{\textin{max}}$ is the wavelength of the most unstable perturbation). Significant features of that process have been accurately described based on this approximation, even though the jet adopts non-slender shapes before its breakup.

% Viscoelasticity
\citet{BAHPMB10} plotted a phase diagram in the space defined by Deborah $\text{De}=\lambda_s/\hat{t}_{ic}$ and Ohnesorge numbers depicting the regions where the different beads-on-a-string morphologies can be found during the breakage of an Olrodyd-B filament.

\begin{figure}[h]
\vcenteredhbox{\resizebox{0.48\textwidth}{!}{\includegraphics{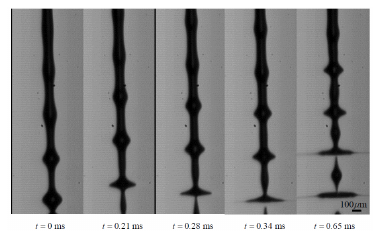}}}
\caption{Formation of spikes on a jet of a mixture of ethanol and glycerol ejected at a flow rate of 3000 $\mu$l/min and subjected to an electric voltage of 6.4 kV \citep{LKYY19}.}
\label{spikes}
\end{figure}

% Electrified jets.
The 1D approximation has been used to examine the nonlinear breakup of electrified jets. \citet{LG04} and \citet{CHB07} calculated the diameter and charge of both the primary and satellite droplets formed after the breakup of a conducting jet subject to a radial electric field. The extended version of this model indicates that ionic surfactants and electrokinetic effects increase the size of satellite droplets but have little influence on the breakup time \citep{CMCP11a}. The breakup of highly-electrified jets can give rise to fascinating structures, including the formation of spikes which emit tens of ultra-fine jets from their periphery (Fig.\ \ref{spikes}) \citep{LKYY19}. If a sufficiently intense radial electric field is applied to the jet, non-axisymmetric perturbations can grow and lead to the formation of Taylor cones evenly distributed at the equator of a droplet \citep{LKXYY20}.

% Surfactants
Numerical solutions of the 1D model have shown that large satellites may form during the breakup of jets laden with strong surfactants at concentrations above the CMC. This phenomenon is driven by Marangoni stresses instead of liquid inertia \citep{CMP09}. Experiments with liquid bridges indicate that the satellite droplet diameter can either increase or decrease due to the presence of surfactant depending on the Ohnesorge number \citep{KJMS18}.

\section{Tip streaming in open systems}
\label{sec8}

% Tip streaming
Viscosity and surface tension are the two forces which oppose the formation of droplets, bubbles and jets in microfluidics. In standard dripping and jetting, the energy necessary to overcome those forces is injected into the dispersed phase somewhere upstream in the feeding system. In tip streaming, that source of energy is replaced by an external agent which gently pushes the fluid towards the tip of a parent drop or stretched meniscus attached to a feeding capillary \citep{Z17,WT25,RM61,T64,S84,ML91,D93,G98a,GG01,G04b,ETS01,DAMHL03a,BS05b,SB06,CJHB08}. In this way, the energy transmitted to the dispersed phase is literally focused on the critical region where the droplet/jet forms. The fluid accelerates in the drop/meniscus tip to the extent of overcoming the intense resistance offered by the viscous stress and capillary pressure, which allows the formation of tiny droplets/bubbles or a very thin jet. The tapering drop/meniscus becomes a complex fluidic structure, the product of a delicate balance between inertia, surface tension and viscosity, depending on the specific situation considered. The existence of a stagnation point right in front of the emission point is a common feature of tip streaming, which may have applications in analytical chemistry and life sciences  \citep{BQ17}.

% Universal features
\citet{ED09} studied numerically the appearance of tip streaming when drops, bubbles and films are deformed by strong viscous flows. They found that the interface near the tip exhibited universal features, independent of the outer flow and the system geometry considered. \citet{TP15} have argued that the tip streaming arising in different configurations is reducible to a common instability that can take place owing to a local convergence of streamlines in the vicinity of a zero-vorticity point or line on the interface.  

% The recirculation pattern
Tip streaming exhibits a rich and interesting phenomenology. For large enough Reynolds numbers, the external driving force typically induces a recirculation pattern with a stagnation point next to the region where the droplets/bubbles or the jet are emitted \citep{HGOBR08,HLGVMP12}. For sufficiently small Reynolds numbers, viscous stresses generally direct the fluid towards the source tip, precluding the growth of such a pattern \citep{MRHG11,PRHGM18}. Interestingly, recirculation can also be found in the Stokes flow appearing in some tip streaming configurations \citep{Z04,SB06}.

% This section
In this section, we present some results about tip streaming realizations in {\em open systems}. In these systems, the external fluid medium is not bounded or it does not affect the tip streaming (electrohydrodynamic tip streaming), and, more importantly, there is no control on the dispersed phase response (e.g., by injecting it at a prescribed flow rate). At least one of these conditions does not hold in the microfluidic configurations described in Sec.\ \ref{sec9}.

\subsubsection{Surfactant-driven tip streaming}

% Bruijn
In his pioneering work, \citet{D93} described the tip streaming occurring in a surfactant-laden droplet submerged in a simple shear flow. He concluded that surfactants trigger tip streaming, which then disappears as surfactants are convected away from the tip. 

\begin{figure}
\centering{\resizebox{0.45\textwidth}{!}{\includegraphics{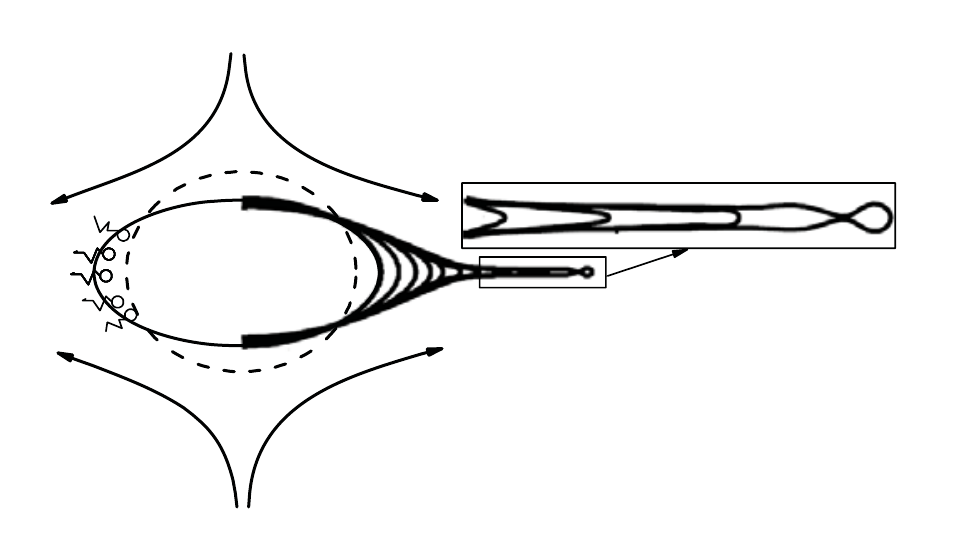}}}
\caption{Breakup of a droplet loaded with an insoluble surfactant in the presence of a linear extensional flow \citep{ETS01}. The contours on the right side were calculated from the numerical simulation of the Stokes flow for $\text{Ca}_G=0.065$, $\Gamma_{\textin{eq}}/\Gamma_{\infty}=0.1$ and $\mu_i/\mu_o=0.05$.}
\label{eggleton}
\end{figure}

% Eggleton
Figure \ref{eggleton} shows the numerical simulation of the breakup of a droplet loaded with an insoluble surfactant in the presence of a linear extensional flow \citep{ETS01}. Surfactant molecules are dragged by the flow towards the poles of the droplet, which reduces the interfacial tension there. When the Capillary number
\begin{equation}
\text{Ca}_G=\frac{\mu_o G a}{\sigma_{\textin{eq}}} \end{equation}
exceeds a critical value, the droplet ejects a thin liquid thread from its poles. Here, $\mu_o$ is the outer viscosity, $G$ is the strain rate of the far-field flow, $a$ is the droplet radius, and $\sigma_{\textin{eq}}$ is the equilibrium interfacial tension. For a given surfactant elasticity, the critical Capillary number in the Stokes regime depends on the inner-to-outer viscosity ratio $\mu_i/\mu_o$ and the (dimensionless) surface surfactant concentration $\Gamma_{\textin{eq}}/\Gamma_{\infty}$. \citet{BS05b} calculated the critical Capillary number when the dispersed phase is a gas.

% Soluble surfactants
\citet{WSR14} studied the effect of the surfactant solubility on the surfactant-driven tip streaming. The ejected filament becomes thinner as the Biot number (the ratio of the desorption time to the characteristic hydrodynamic time) increases. The parameter conditions for the appearance of tip streaming were determined.

% Summary
To the best of our knowledge, tip streaming in the open systems described above has been observed neither experimentally nor numerically without adding surfactants to the interface. Therefore, surfactants are probably necessary in these systems, while they simply facilitate tip streaming in confined hydrodynamic configurations such as coflowing and flow focusing \citep{AM06b,A16,MSPVL16}. This facilitator is important if one wants to produce steady jetting tip streaming with a gaseous dispersed phase \citep{CHLG11}.

\subsubsection{Surfactant-free tip streaming}

% First mechanism
As described above, when a drop/bubble is loaded with an insoluble surfactant, the viscous stresses exerted by an outer shear or extensional stream push the surfactant molecules towards the poles of the drop. This reduces the surface tension in that region, which may result in the ejection of a fluid thread much smaller than the droplet/bubble size. This was one of the first mechanisms used to produce tip streaming in both droplets \citep{D93,ETS01} and bubbles \citep{BS05b}.

\begin{figure}
%\begin{center}
\centering{\resizebox{0.35\textwidth}{!}{\includegraphics{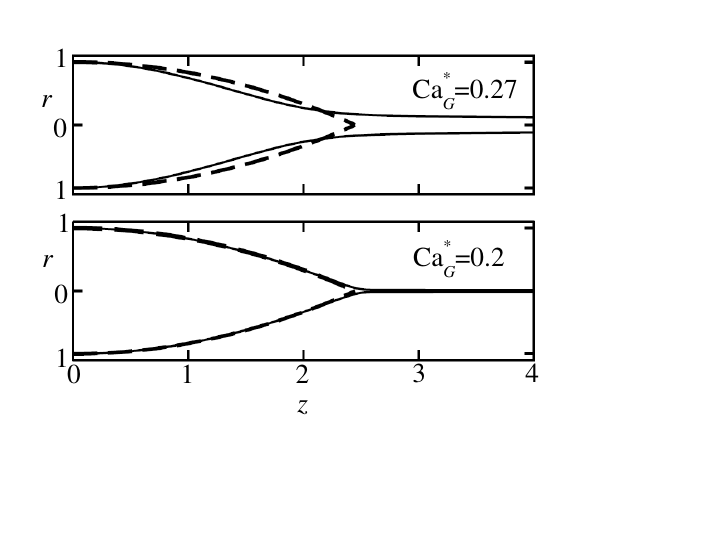}}}
%\end{center}
\caption{Numerical simulation of tip streaming in a droplet submerged in an extensional flow \citep{Z04}. The solid lines represent the solutions when the Capillary number $\text{Ca}_G^*$ is progressively reduced. The critical Capillary number is 0.2. The dashed line outlines the shape for $\text{Ca}_G^*=0.2$ and zero ejected flow rate.}
\label{Zhang}
\end{figure}

% Wendy Zhang
In the absence of surfactants, \citet{Z04} showed that the steady recirculating stream arising in a droplet attached to a capillary of radius $R_i$ and submerged in an extensional (straining) flow evolves towards tip streaming when the Capillary number
\begin{equation}
\text{Ca}_G^*=\frac{2\mu_o G R_i}{\sigma}
\end{equation}
exceeds a critical value (Fig.\ \ref{Zhang}). This transition to tip streaming refers to the sharp reduction of the diameter of the ejected fluid thread when the Capillary number reaches its critical value. The above result indicates that steady tip streaming can be obtained by purely hydrodynamic means if the ejected volume is replaced through the feeding capillary at the appropriate rate (see Sec.\ \ref{sec11}). Other microfluidic techniques, such as electrospray and flow focusing, can be categorized as this type of flow too (see Secs. \ref{sec12} and \ref{sec13}). \citet{Z04} pointed out the importance of reaching tip streaming at the critical condition by reducing progressively the Capillary number, a requisite similar to that experimentally found for the injected flow rate in electrospray and flow focusing. It should be noted that these numerical results were based on the slender body theory, which is an uncontrolled approximation for a conical interface.

\subsubsection{Electrohydrodynamic tip streaming}

\begin{figure}[h]
\vcenteredhbox{\resizebox{0.48\textwidth}{!}{\includegraphics{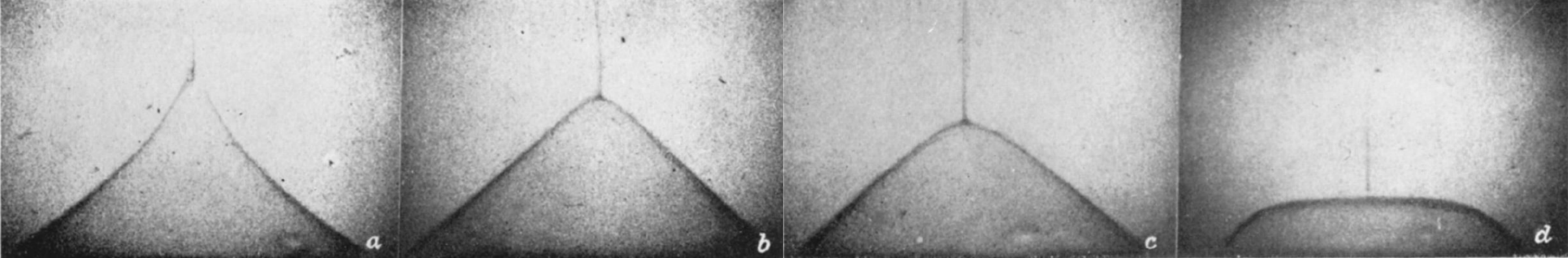}}}
\vcenteredhbox{\resizebox{0.48\textwidth}{!}{\includegraphics{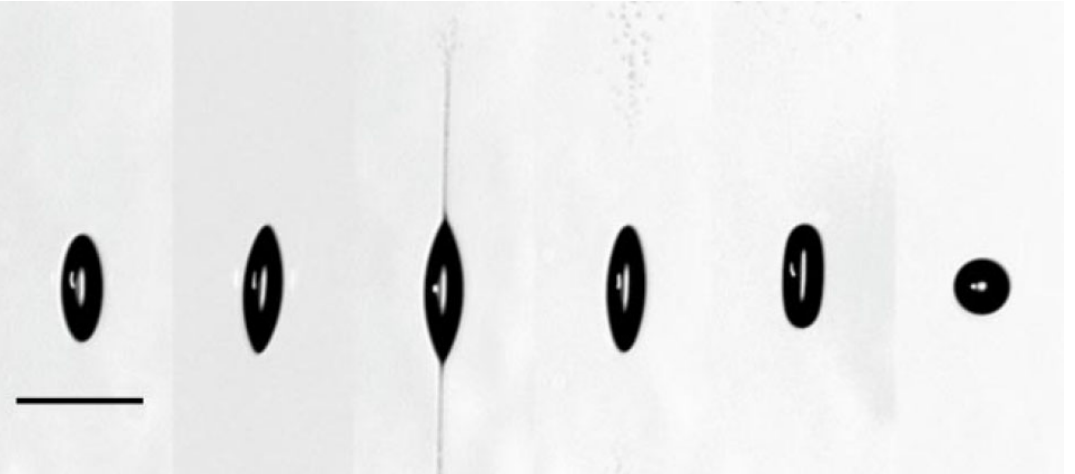}}}
\caption{(Upper images) Jet formation and subsequent collapse of a soap film \citep{T64}. (Lower images) Disintegration of a levitated droplet of ethylene
glycol charged to the Rayleigh limit. Scale bar, 100 $\mu$m \citep{DAMHL03a}.}
\label{duft}
\end{figure}

% Electrohydrodynamic tip streaming
In his pioneering work, \citet{R81} predicted that an incompressible charged liquid droplet becomes unstable due to the growth of the quadrupole oscillation mode when the disruptive Coulomb force equals the attractive surface tension force (the so-called Rayleigh limit). He also claimed that higher multipole oscillations can cause the ejection of very fine jets. \citet{T64} described experimentally the ejection of those jets from the conical points of electrified films. \citet{DAMHL03a} observed the emission of Rayleigh jets from the tip streaming taking place in the poles of levitated droplets (Fig.\ \ref{duft}) \citep{GK64,AMDL05,GGRMHDMLG08}. The problem has been numerically solved considering both the leaky-dielectric model \citep{CSHB13,GMT20} and electrokinetic effects \citep{GLRM16,PBHD16,MMK19}.

% The transient regime
The periodic or steady ejection produced by tip streaming in the microfluidic configurations described in Sec.\ \ref{sec9} is preceded by an intrinsically unsteady process, where the drop/bubble/meniscus stretches until emitting the first droplet/bubble. The characteristics of the latter (size, velocity, electric charge, \ldots) significantly differ from those of the droplets/bubbles ejected when tip streaming is properly established (steady ejection). The size and velocity of the first-emitted droplet shown in, for instance, Fig.\ \ref{eggleton} is expected to be significantly different from those of the droplets emitted under steady conditions. Steady tip streaming would be reached if the ejected liquid were replaced by injecting liquid into the droplet with a feeding capillary. In this case, the injected flow rate becomes a control parameter that allows tuning the size of the emitted jet.
 
\begin{figure*}[hbt]
\centering{\resizebox{0.85\textwidth}{!}{\includegraphics{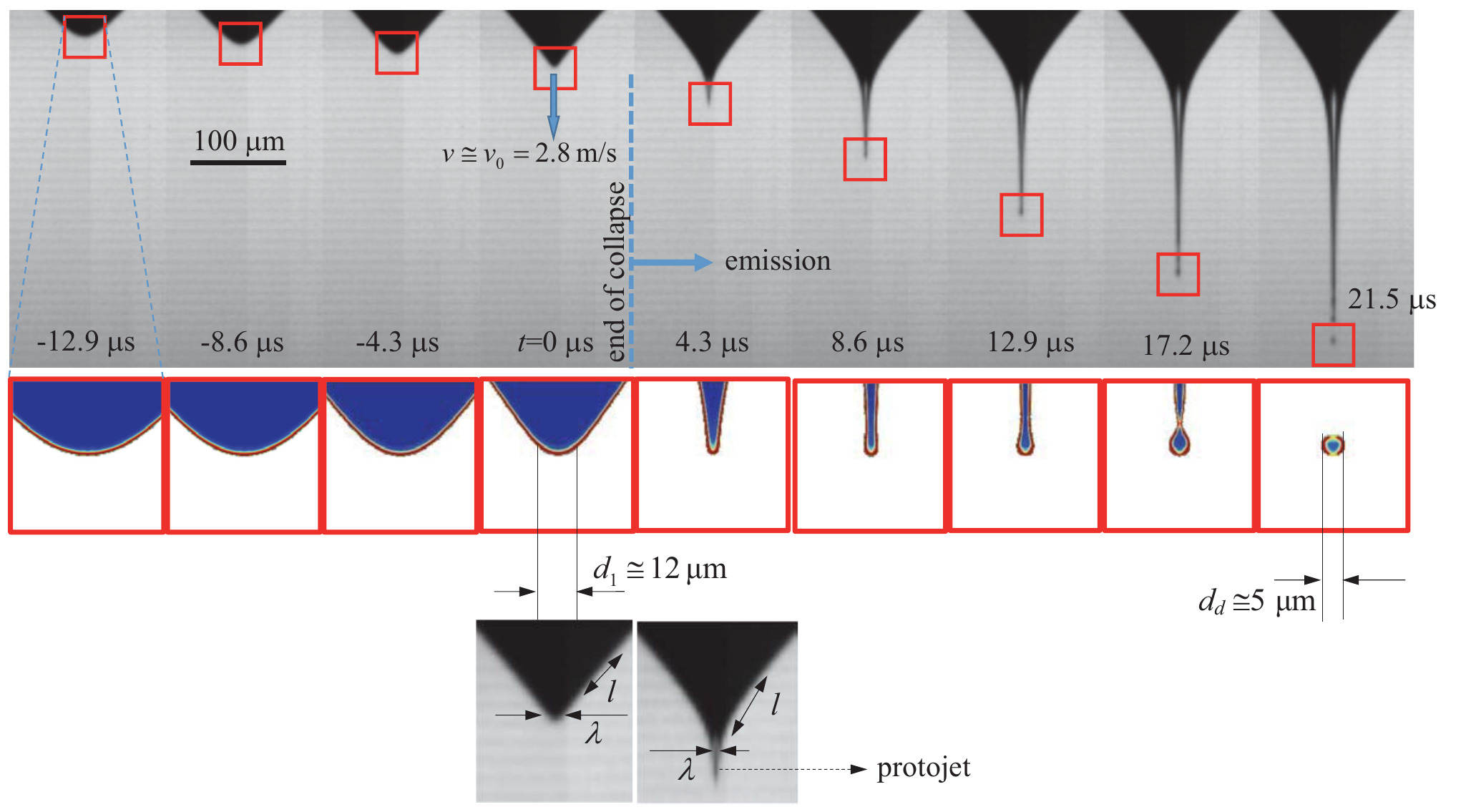}}}
\caption{Global view of the first-droplet ejection process in electrohydrodynamic tip streaming \citep{GLRM16}. The upper images correspond to an experiment with 1-octanol. The central graphs show the evolution of the front from the corresponding numerical simulations. The colors indicate the dimensionless volume charge density from 0 (blue) to 1 (red). The two lower images show the magnification of the meniscus tip for $t=0$ and $t=4.3$ $\mu$s. The process can be divided into two stages: the droplet tip collapse for $t<0$, and the jet emission for $t>0$.}
\label{alfonso-scale0}
\end{figure*}

% The transient regime in electrohydrodynamic tip streaming
The distinction between the first ejection and that taking place in the steady regime of tip streaming has been made clear for electrohydrodynamic tip streaming. Under certain conditions, a low-conductivity pendant droplet subject to a strong electric field ejects a thin liquid thread, which issues a stream of tiny drops (Fig.\ \ref{alfonso-scale0}). The masses and charges of those droplets are essentially determined by the liquid properties and reflect the electrohydrodynamic history leading to them. The first droplet produced by this unsteady process is particularly important due to its very small diameter and large electric charge (per unit volume). After this first phase of the ejection process, the system reaches spontaneously a quasi-steady regime characterized by a natural (intrinsic) flow rate $Q^*$, electric current, and droplet diameter, which essentially depend upon the liquid properties too \citep{GRM13}. If a flow rate $Q\geq Q^*$ is prescribed by, for instance, injecting liquid across a feeding capillary, the above quasi-steady process gives rise to the steady cone-jet mode of electrospray.

\begin{figure*}[ht]
\centering{\resizebox{0.85\textwidth}{!}{\includegraphics{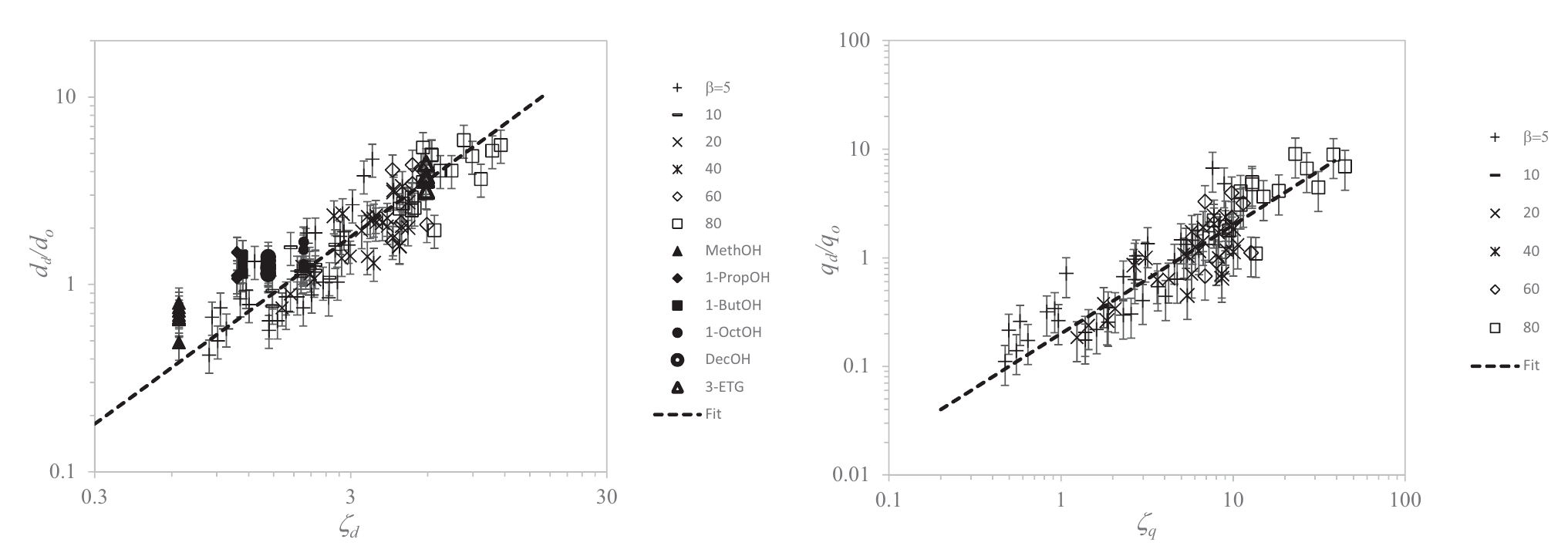}}}
\caption{Dimensionless diameter $d_d/d_o$ (left) and electric charge $q_d/q_o$ (right) of the first droplet ejected in electrohydrodynamic tip streaming as a function of $\zeta_d\equiv \delta_{\mu}^{-1/3}\epsilon^{5/12}$ and $\zeta_q\equiv \delta_{\mu}^{-2/3}\varepsilon^{7/12}$ \citep{GLRM16}. The lines are the scaling laws (\ref{dq}). The open and solid symbols are the results obtained from numerical simulations and experiments, respectively. The labels indicate the value of $\varepsilon$ in the numerical simulations, as well as the liquid used in the experiments.}
\label{alfonso-scale}
\end{figure*}

% Scaling laws
\citet{GLRM16} have studied the onset of electrohydrodynamic tip streaming by assuming that the hydrodynamic and electric relaxation times are commensurate with each other. This assumption, the self-similar collapse of the drop's tip, and the balance between inertia, axial viscous stresses, surface tension, and electrostatic suction during the liquid thread evolution, enable the calculation of the scales for the emitted droplet diameter $d_d$ and electric charge $q_d$:
\begin{equation}
d_d=\delta_\mu^{-1/3}\varepsilon^{5/12} d_o\, ,\quad q_d=\delta_\mu^{-2/3} \varepsilon^{7/12} q_o\, ,
\label{dq}
\end{equation}
where $\delta_{\mu}=\rho d_o v_o/\mu$ is the electrohydrodynamic Reynolds number, $\varepsilon$ is the liquid permittivity in terms of that of vacuum $\varepsilon_o$, and $d_0$, $v_o$, $q_o$ and $E_o$ are the characteristic quantities of the process given by the expressions $d_o=(\sigma \varepsilon_o^2/\rho K^2)^{1/3}$, $v_o=(\sigma K/\rho \varepsilon_o)^{1/3}$, $q_o=\varepsilon_o E_o d_o^2$ and $E_o=(\sigma/d_o\varepsilon_o)^{1/2}$ ($\rho$, $\mu$, and $K$ are the liquid density, viscosity, and conductivity, respectively). Figure \ref{alfonso-scale} shows the experimental and numerical validation of scaling laws (\ref{dq}).

% Other works
The first ejection taking place in electrohydrodynamic tip streaming has also been considered in other works. \citet{FKV08} studied theoretically the unsteady elementary disintegration of charged and neutral conducting drops under externally applied electric fields. The onset of tip streaming in low-conductivity drops after a step-change in the electric field magnitude has been examined both experimentally \citep{ON00,SDYM07,P09b,FLHMA13} and theoretically \citep{CJHB08,CSHB13,FLHMA13,PBHD16,GLRM16,DJE18}. In the latter case, the analysis has been conducted both assuming perfect volumetric charge relaxation (the leaky-dielectric model) \citep{CJHB08,CSHB13,FLHMA13} and considering certain charge relaxation phenomena along the process \citep{PBHD16,GLRM16,DJE18}.

\begin{figure}
\centering{\resizebox{0.45\textwidth}{!}{\includegraphics{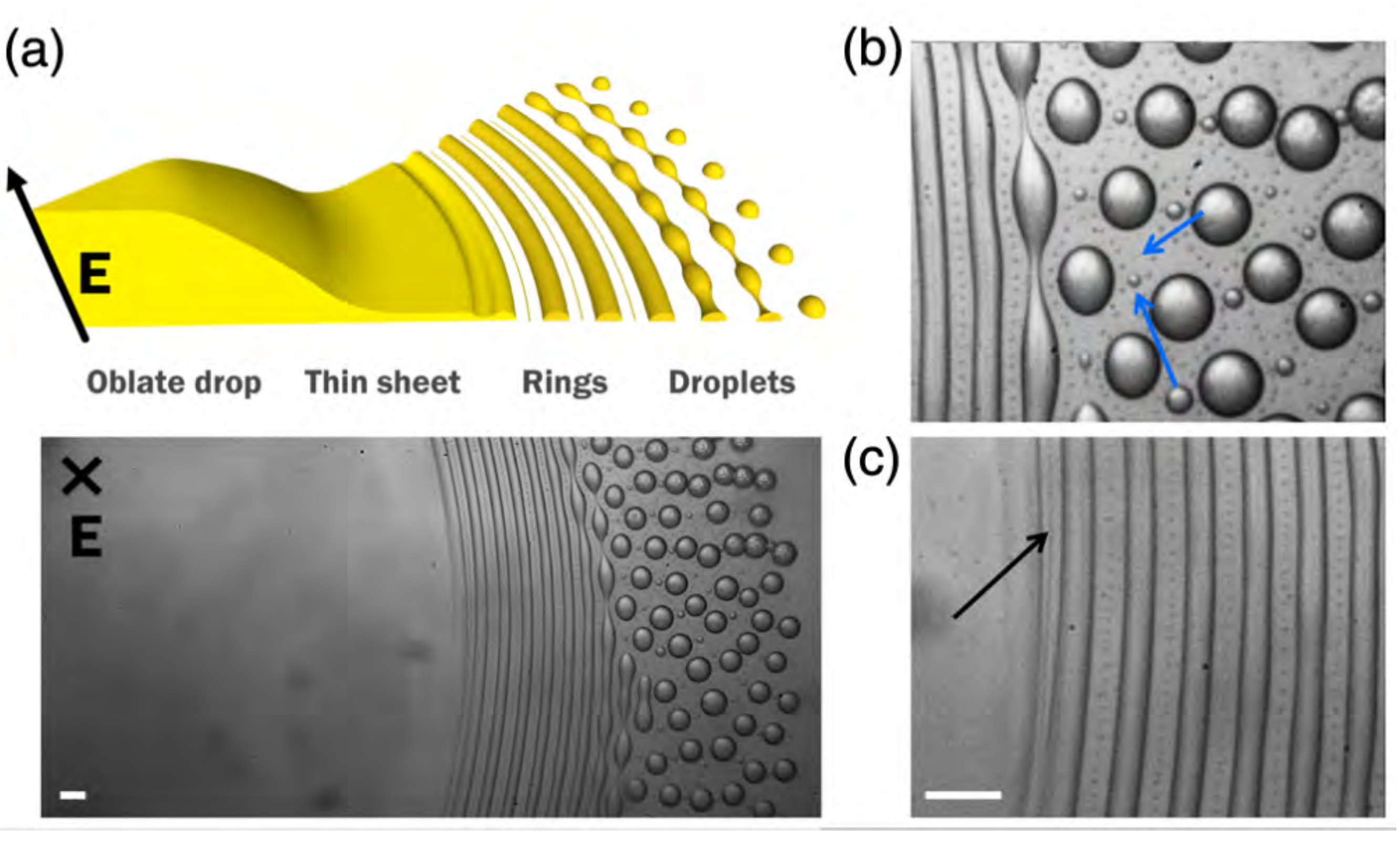}}}
\caption{Electrohydrodynamic equatorial tip streaming \citep{BV17}. (a) Three-dimensional rendering of the phenomenon. The thin sheet is the slightly brighter region right in front of the edge where the ring is forming. (b) Two generations of satellite droplets produced by the ring breakup (blue arrows). (c) Satellite ring that breaks up in droplets (black arrow). The scale bars are 100 $\mu$m.}
\label{equatorial}
\end{figure}

% Vlahovska 2017
Electrohydrodynamic tip streaming from an initially spherical droplet can be triggered by an electric field producing either oblate or prolate deformations on the droplet in the direction of the applied field. These deformations are due to the electrohydrodynamic flow pattern induced by the electric field \citep{T66}. In the Stokes regime, for a liquid droplet suspended in an immiscible liquid environment, the induced surface velocity depends on the ratios of inner to outer viscosities, electrical conductivities and permittivities. \citet{BV17} demonstrated that an oblate deformation can eventually lead to tip streaming from the equatorial rim generated by the sustained non-linear deformation of the drop (Fig.\ \ref{equatorial}). This phenomenon leads to the formation of a thin sheet that gives rise to thin toroidal rings by 2D capillary breakup, which eventually break up in the azimuthal direction \citep{BV17}. This process generates beautiful arrangements of monodisperse droplets in the equatorial plane of the parent drop. Interestingly, the two-dimensional breakup can produce main and satellite rings, leading to a finite variety of droplet sizes.

% Critical conditions. Beroz PRL
\citet{BHB19} have recently calculated the critical electric field at which a conducting droplet or bubble sitting on a conductor plate becomes unstable and emits a tiny jet from its apex. The result is
\begin{equation}
\label{Beroz}
\chi=\frac{2}{\pi}\, \frac{R_i^3}{{\cal V}},
\end{equation}
where $\chi=\varepsilon_o E^2 R_i/\sigma$ is the electric Bond number, $E$ is the applied electrical field, $R_i$ is the triple contact line radius, and ${\cal V}$ the droplet volume. The prefactor $2/\pi$ was determined experimentally.

\begin{figure}
\centering{\resizebox{0.35\textwidth}{!}{\includegraphics{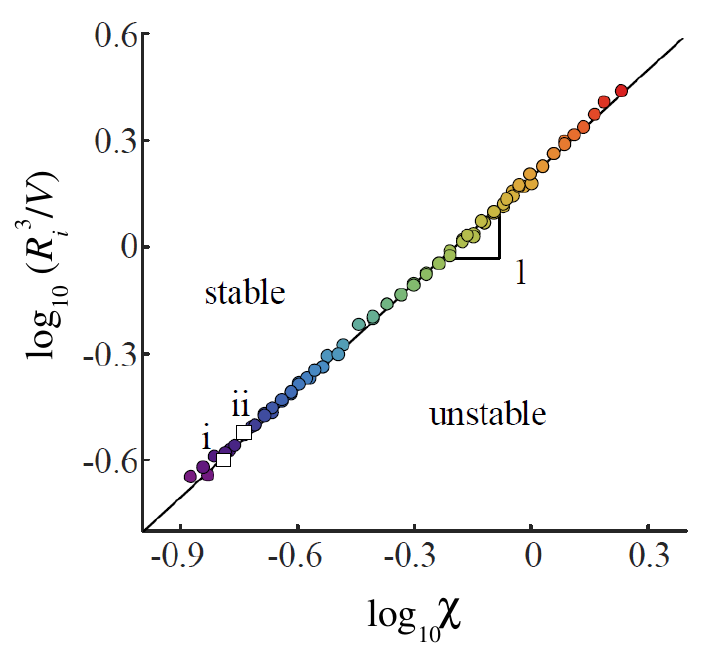}}}
\caption{Critical value of the electrical Bond number $\chi$ as a function of the dimensionless drop volume $V/R_i^3$ \citep{BHB19}. The symbols and the line correspond to the experimental values and the prediction (\ref{Beroz}), respectively. The squares are the results of the experiments conducted by \citet{WT25} (i) and \citet{BS90} (ii).}
\label{BerozPRL}
\end{figure}

% Similarities
The ejection of the first droplet in electrohydrodynamic tip streaming resembles other similar phenomena, like the formation of jets by the collapse of a free surface \citep{ZKFL00,Y06}. Similar arguments to those presented here have been used to derive the universal scaling laws for the size and speed of the drop emanating from the breakup of a liquid jet generated by the collapse of a bubble \citep{G17}.

% Connection
As mentioned above, tip streaming can be sustained over time by replacing the fluid ejected at the appropriate rate using a certain microfluidic configuration. In this way, periodic or steady ejection is achieved. This phenomenon occurs within relatively narrow regions of the parameter space. Understanding the physical mechanisms that bound those regions is still an open problem, and constitutes an important challenge at a fundamental level. However, these aspects of the problem lag behind most applications, where the major interest is precisely the ejecta. In Secs.\ \ref{sec10} and \ref{sec11}, we will focus on this facet of the problem for the microfluidic configurations considered in this review.

\section{Microfluidic configurations and governing parameters}
\label{sec9}

\subsection{Axisymmetric injection}
\label{sec9.1}

\begin{figure}
%\begin{center}
\centering{\resizebox{0.45\textwidth}{!}{\includegraphics{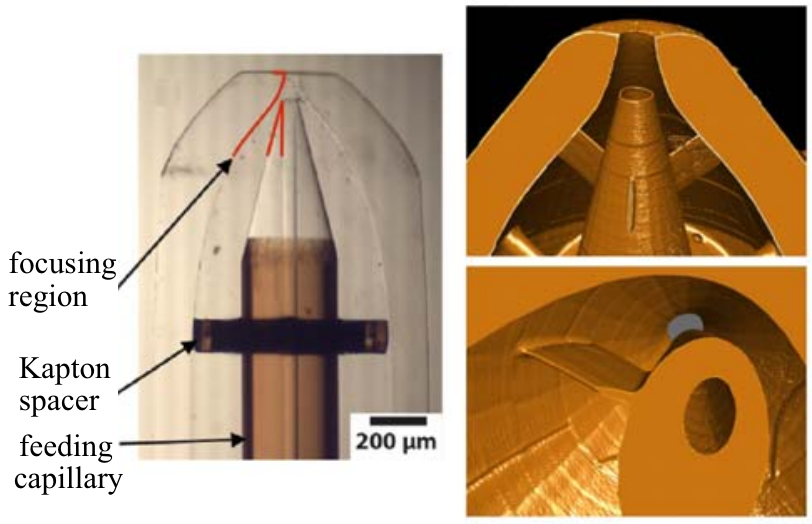}}}
%\end{center}
\caption{Gaseous flow focusing device fabricated from a flame polished glass capillary (left) and the equivalent nozzle printed with two-photon polymerization (Nanoscribe GmbH, Karlsruhe, Germany) (right) \citep{N16}.}
\label{fabrication}
\end{figure}

% Manufacturing axisymmetric devices
Axisymmetric devices have higher throughput and are more robust and resistant to aggressive chemical conditions than their 2D counterparts. They are typically made of hard materials such as metal \citep{G98a,MFRRCCG05,GGRHF07} or borosilicate glass \citep{ULLKSW05,DWSWSSD08}, which do not swell and can easily be functionalized to control surface properties. Glass nozzles with the desired converging or diverging rate can be produced in a controlled manner with the fire shaping procedure \citep{MC18,MGC19}. Needles manufactured by a laser beam welding process have been assembled to fabricate triaxial flow focusing devices \citep{SFLX15}. Alternative fabrication processes of axisymmetric devices include soft-lithography \citep{TKDRCF14}, ceramic \citep{BAHKKWCB15} or polymer (Cellena$\circledR$, OneNeb$\circledR$, Ingeniatrics Tec.\ S.L.) microinjection molding, and, more recently, 2-photon polymerization \citep{N16}, which is a form of high-resolution 3D printing to manufacture devices with submicron resolution (Fig.\ \ref{fabrication}).

\begin{figure}
%\begin{center}
\centering{\resizebox{0.28\textwidth}{!}{\includegraphics{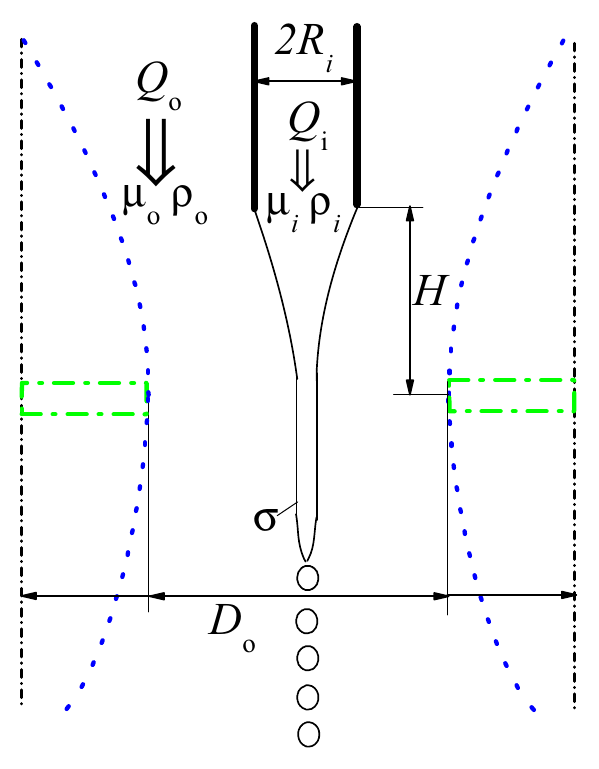}}}
%\end{center}
\caption{Sketch of an axisymmetric microfluidic configuration driven by hydrodynamic forces. The short-dashed (black), dotted (blue) and dashed (green) lines represent the shape of the outer channel in the coflowing, nozzle flow focusing and plate-orfice flow focusing configurations, respectively.}
\label{sketch}
\end{figure}

In an axisymmetric microfluidic device (Fig.\ \ref{sketch}), a fluid stream of density $\rho_i$ and viscosity $\mu_i$ is injected across a feeding capillary (nozzle) of inner radius $R_i$ at a constant flow rate $Q_i$. The surrounding medium is another fluid phase of density $\rho_o$ and viscosity $\mu_o$, immiscible with the former, and separated from it by an interface with a surface tension $\sigma$. To simplify the analysis, we will assume that the end of the feeding capillary is sharpened, and the triple contact line anchors at its edge. It is worth mentioning that care must be taken when injecting gases in a microfluidic application, because pressure fluctuations taking place both upstream and downstream the injection circuit may lead to significant variations of the instantaneous flow rate.

% Dimensionless numbers of the axisymmetric injection
The dimensionless numbers characterizing the fluid injection described above are the Weber and Reynolds numbers defined as
\begin{equation}
\text{We}_i=\frac{\rho_i V_i^2 R_i}{\sigma}, \quad \text{Re}_i=\frac{\rho_i V_i R_i}{\mu_i},
\end{equation}
where $V_i=Q_i/(\pi R_i^2)$ is the mean velocity in the capillary. The problem is also described in terms of the density and viscosity ratios
\begin{equation}
\rho=\frac{\rho_o}{\rho_i}, \quad \mu=\frac{\mu_o}{\mu_i}.
\end{equation}
For Re$_i\ll 1$, the inertia of the inner fluid becomes negligible, and the fluid injection can be described just in terms of the Capillary number
\begin{equation}
\label{ci}
\text{Ca}_i=\frac{\text{We}_i}{\text{Re}_i}=\frac{\mu_i V_i}{\sigma}
\end{equation}
and the viscosity ratio $\mu$.

% External forces
For sufficiently large values of the Weber and Reynolds numbers, inertia overcomes the resistance offered by both surface tension and viscosity during the fluid ejection, and the feeding capillary drips or emits a jet. When this condition does not hold, the collaboration of some kind of mass or superficial driving force is required to produce that effect. Figure \ref{sketch} sketches two of the most commonly used methods to generate that force.

\subsection{Gravitational forces}
\label{sec9.2}

% Gravitational
In the gravitational ejection, the feeding capillary is placed vertically in a still ambient, and the fluid is expelled under the action of gravity $g$. The additional dimensionless number characterizing this process is the gravitational Bond number
\begin{equation}
B=\frac{|\rho_i-\rho_o| g R_i^2}{\sigma},
\end{equation}
which compares the capillary pressure $\sigma/R_i$ with the variation of hydrostatic pressure across the interface, $|\rho_i-\rho_o| g R_i$, due to gravity.

\subsection{Electrical force}
\label{sec9.2b}

At the submillimeter scale and/or for density-matched liquids, interfacial stresses dominate over gravity (the Bond number takes very small values), and the fluid ejection cannot rely on that force. The driving mass force can be augmented by applying an external electric field parallel to the feeding capillary.

\subsubsection{Electrospray and electrospinning applications}

Electrospray is an atomization technique characterized by the interaction between electrostatic, capillary and viscous stresses on a liquid meniscus, which gives rise to the formation of charged jets and droplets. Electrospray can operate in dripping, jetting and tip streaming \citep{RGL18}. The steady cone-jet mode of electrospray and electrospinning is a classical example of tip streaming induced by electric fields.

% Electrospray. Applications
The cone-jet mode of electrospray has been applied to produce charged droplets in mass spectrometry \citep{FMMWW89}, to generate electric propulsion for spacecraft \citep{GH01}, to form capillary shapes and structures for drug encapsulation \citep{MC07,CLAL09} and tissue engineering \citep{SR08}, for high-resolution electrohydrodynamic jet printing \citep{P07a}, to enhance the quality and resolution in Electrostatic Inkjet Printing \citep{OSFAR15} with nanoparticle inks \citep{IYXS18}, to form emulsions and microparticles in food industry, and to fabricate compound and hollow fibers \citep{BL07}, among many other applications. New applications are continuously emerging. For instance, \citet{B19} have recently shown that, under certain conditions, electrospray can constitute an advantageous alternative to flow focusing for sample injection for single-particle imaging with X-ray lasers, because it reduces both the size and polydispersity of the aerosol droplets and, consequently, the nonvolatile contaminants.

\begin{figure}
%\begin{center}
\centering{\resizebox{0.5\textwidth}{!}{\includegraphics{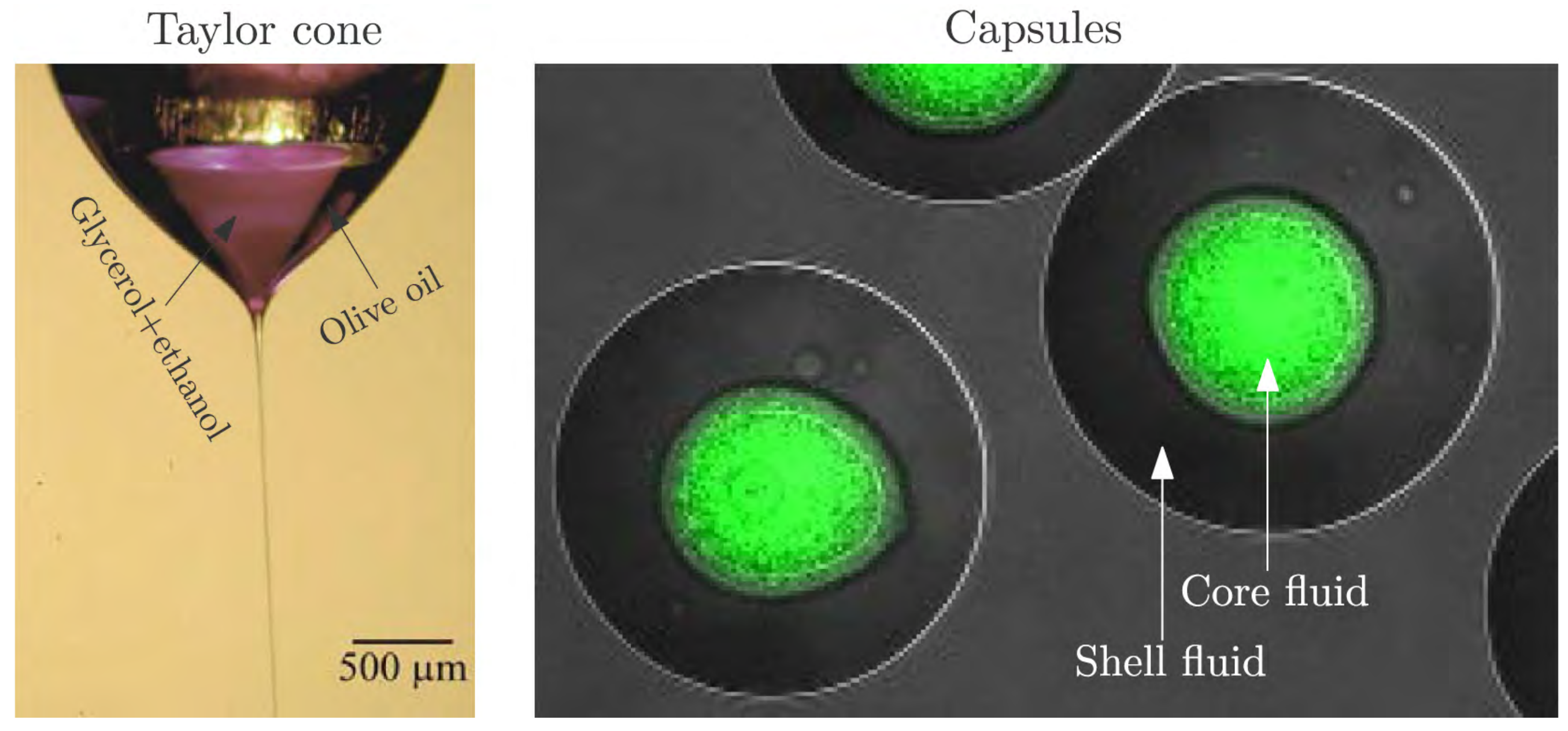}}}
%\end{center}
\caption{Microencapsulation via AC coaxial electrohydrodynamic atomization \citep{GP20}.}
\label{AC}
\end{figure}

% Coaxial electrospray. Applications
Coaxial electrohydrodynamic atomization \citep{LBGCMG02,MLMB07,BL07,Y11} has been used for the formation of polymer coated starch-protein microspheres \citep{PE06}, preparation of suspensions containing microbubbles \citep{FZESS07}, and production of organic and inorganic macro and nano-sized emulsions \citep{BL07}, among other applications. Recent advances in coaxial electrohydrodynamic atomization have facilitated the use of this technique for fabrication of micro- and nano-sized drugs loaded biodegradable polymeric particles for controlled drug release and biochemical applications \citep{CLAL09,EASE10}. In this case, AC actuation can be used to reduce the detrimental effect of charge content on some drug delivery applications, and enhance the stability of the ejection process \citep{GP20} (Fig.\ \ref{AC}).

% Electrospinning. Applications
Electrospinning of polymers has become a highly recognized method for the preparation of polymer fibers with diameters ranging from tens of microns down to a few nanometers \citep{RYFK00,YKR01,YKR01b,TZY04,YKR05,HRY07,TCYR07,HYR08,RY08,MSY09,Y11,YPR14}. A wide range of complex architectures and morphologies of nanofibers and nonwovens can be produced with electrospinning. The applications of this method include medical areas such as tissue engineering and drug delivery \citep{KB10}, as well as technical fields demanding nanofibers with specific electronic, photonic, photocatalytic and magnetic properties. Nanofiber-based architectures have a positive influence on the development of fuel cells, lithium ion batteries, solar cells, electronic sensors, energy storage systems. Other areas like textile and filter applications have benefited from electrospinning as well \citep{AGW13}.

\begin{figure}
%\begin{center}
\centering{\resizebox{0.4\textwidth}{!}{\includegraphics{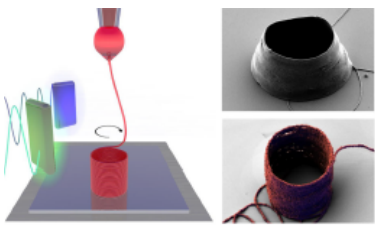}}}
%\end{center}
\caption{Ultrafast 3D printing of PEO cylindrical microstructures achieved by controlling the voltage applied to electrodes located around the jet emitted in electrospinning \citep{LRC20}.}
\label{joan}
\end{figure}

% Near-field eledctrospinning. Applications
In traditional electrospinning, fibers are produced chaotically. This feature limits the applications of electrospinning in devices that demand arranged or patterned micro/nanoscale fibrous structures. Near-field electrospinning has been developed to deposit ultrafine fibers with unique physical and chemical properties in a direct, continuous, and controllable manner \citep{SCLL06}. Fibers fabricated with near-field electrospinning can be used in  electronic components, flexible sensors, energy harvesting, and tissue engineering \citep{HZYYYNL17}. Sophisticated versions of this technique are continuously emerging. For instance, \citet{LRC20} have recently shown that ultrafast 3D printing with submicrometer resolution can be achieved by electrostatically deflecting the emitted jet with electrodes located around it (Fig.\ \ref{joan}).

% The rol of polymer concentration
Two key parameters controlling the quality of the electrospinning outcome are the polymer molecular weight and concentration, which somehow play similar roles. It has been observed that lower molecular weights and concentrations lead to thinner fibers, which is desirable because it enhances their functionality. However, decreasing the values of these parameters reduces both the solution viscosity and elasticity (stress relaxation time), two inhibitors of the capillary instability (see Sec.\ \ref{sec5}) responsible for the appearance of the undesired droplets or beads. Therefore, uniform (bead-free) fibers are produced at the expense of increasing their diameters, and {\em viceversa}.

\subsubsection{Electric dimensionless numbers}

The effect of the electric field can be quantified through the electric Bond (Taylor) number (already introduced in Sec.\ \ref{sec8})
\begin{equation}
\label{eb}
\chi=\frac{\varepsilon_o E^2 R_i}{\sigma},
\end{equation}
where $E$ is a characteristic electric field (typically, the ratio of the applied voltage to the characteristic length $R_i$) and $\varepsilon_o$ the electrical permittivity of the outer medium (insulator). The electric Bond number is essentially the electric strength (\ref{strength}) used in the linear stability of jets. It measures the electrostatic pressure $\varepsilon_o E^2$ in terms of the capillary pressure $\sigma/R_i$.

% Electric properties of the dispersed phase
The electrical properties of the dispersed phase are taken into account through the relative permittivity and dimensionless electrical conductivity:
\begin{equation}
\label{kk}
\varepsilon=\frac{\varepsilon_i}{\varepsilon_o}, \quad \widehat{K}_i=K_i\left(\frac{\rho_i R_i^3}{\sigma \varepsilon_o^2}\right)^{1/2}=\frac{\varepsilon t_{ic}}{t_e},
\end{equation}
where $t_{ic}=(\rho_i R_i^3)/\sigma$ is the characteristic inertio-capillary time, and $t_e=\varepsilon_i/K_i$ is the electric relaxation time. The dimensionless electrical conductivity $\widehat{K}_i$ is essentially the inverse of the dimensionless electric relaxation time $\tau_{ej}$ [Eq.\ (\ref{tk})] used in the linear stability of jets.

% Electrified. Conductors, leaky-dielectric, dielectric
For perfect conductor fluids, $t_e\ll t_{ic}$ ($\widehat{K}_i\gg 1$), the net free electrical charge accumulates ``instantaneously"\ in the interface to cancel the voltage variations throughout the dispersed phase, and the problem becomes essentially independent from $\widehat{K}_i$. In a leaky-dielectric liquid \citep{S97}, $\widehat{K}_i$ is sufficiently small to affect the rate at which the net free charge is transferred to the interface. In this case, $\widehat{K}_i$ is one of the governing parameters of the problem.

\subsection{Coflowing}
\label{sec9.3}

% Hydrodynamic forces
Hydrodynamic forces can be employed to control the pinching of the fluid-liquid interface, and to modify the droplet/bubble size, production frequency, etc. Microfluidic devices designed for this purpose can be grouped into three main classes: cross-flowing systems, co-flowing streams, and flow focusing \citep{CA07}. Only the last two classes can be realized in an axisymmetric configuration.

% Coflowing liquid-liquid
In a coflowing device, an outer liquid stream is coaxially injected across a tube of radius $R_o$ to produce drops/bubbles. Tangential viscous stresses exerted by the outer stream constitute the major source of momentum in the liquid-liquid configuration. For this reason, the injection is typically described in terms of the inner Capillary number $\text{Ca}_i$ [Eq.\ (\ref{ci})], while the intensity of the driving force is quantified through the outer Capillary number
\begin{equation}
\label{co}
\text{Ca}_o=\frac{\mu_o V_o}{\sigma},
\end{equation}
where $V_o$ is a characteristic velocity of the outer stream. The response of the dispersed phase also depends on the viscosity ratio $\mu$.

The ratio of radii
\begin{equation}
\label{rat}
R=R_i/R_o
\end{equation}
is the only parameter that characterizes this simple geometry. The parameters of the problem can be combined to replace (\ref{co}) with the ratio $Q_r$ of the outer flow rate $Q_o$ to the inner one $Q_i$ \citep{SB06}:
\begin{equation}
\label{qr}
Q_r=\frac{Q_o}{Q_i}=\frac{R^2-1}{\mu} \frac{\text{Ca}_o}{\text{Ca}_i},
\end{equation}
where the thickness of the feeding capillary wall has been neglected. One can distinguish between coflowing configurations with $R$ of order unity \citep{SB06,GCUA07,GCA08}, and those characterized by $R\ll 1$ \citep{CFW04,UFGW08}. Their behavior is substantially different due to the stabilizing role of confinement and the effect of this factor on the nonlinear phase of the jet breakup.

\begin{figure}
%\begin{center}
\centering{\resizebox{0.35\textwidth}{!}{\includegraphics{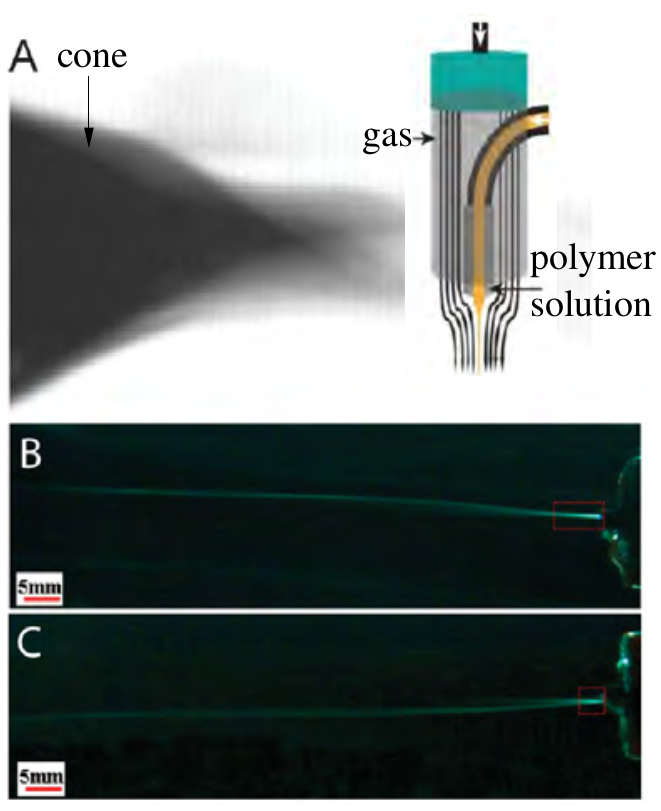}}}
%\end{center}
\caption{(A) Image of a polymer solution cone at the tip of a SBS device \citep{OMCAMOM11}. Image of the polymer jet produced by SBS at gas pressures of 1.121 bar (B) and 1.363 bar (C) \citep{LHW14}. The sketch in the upper-right corner shows the SBS configuration \citep{DBSK16}.}
\label{SBS}
\end{figure}

% Coflowing liquid-gas
The production of liquid jets assisted by an outer gaseous stream in a coflowing geometry is much less popular than its liquid-liquid counterpart due to the limited force exerted by the gas in this geometry. Submicrometer fibers can be spun from polymeric solutions using the Solution Blow Spinning (SBS) \citep{OMCAMOM11,DBSK16} or airbrushing technique \citep{OMCAMOM11}, which constitutes an interesting example of a gaseous coflowing geometry. 
SBS has recently become a popular technique. The physics governing the flow in this configuration has been recently reviewed \citep{LSZPY20}. In SBS, a polymer solution is extruded across an inner nozzle at a constant flow rate. A pressure drop accelerates the gas current released from an outer nozzle. This current expands around the inner nozzle and creates a liquid meniscus similar to Taylor's cone in electrospinning (Fig.\ \ref{SBS}). For high enough applied pressure drops, a liquid jet tapers from the meniscus apex and flies towards the collection target. Then, the solvent evaporates leaving behind polymer fibers. The Gas Jet Process \citep{BJR12} relies on a similar mechanism, although in this case the polymer solution drop is exposed to a high-speed gas stream ejected independently. Nanofibers can also be produced from gas-assisted polymer melt electrospinning \citep{LYYWLGLW08,ZCJ10}, where the drag force exerted by the coflowing gas can significantly reduce the fiber diameter.

% Coflowing, gas-liquid
A coflowing liquid stream can also be used to control the generation of bubbles both in the bubbling and jetting regimes. This configuration produces bubbles considerably smaller than those formed without coflow \citep{SGM05a}.

\subsection{Flow focusing}
\label{sec9.4}

\subsubsection{Axisymmetric geometry and injection method}

% Flow focusing general
In flow focusing, the coflowing stream is forced to cross an orifice of diameter $D_o$ located in front of the feeding capillary at a distance $H$ from its end (Fig.\ \ref{sketch}). This originates favorable pressure gradients \citep{G98a} and both shear \citep{GM09} and extensional viscous stresses which stretch the fluid meniscus attached to the feeding capillary. The meniscus tip emits either drops/bubbles or a thin jet that coflows with the outer stream.

% Alternative axisymmetric geometries.
Different axisymmetric geometries have been employed to take advantage of the flow focusing principle. The original plate-orifice configuration \citep{G98a} has been implemented in planar silicon microfluidic chips with multiple orifices keeping the axisymmetric geometry per orifice \citep{LPEMGQ07}. Microfluidic devices with the axisymmetric flow focusing geometry have also been manufactured in PDMS \citep{TGWW05} using stereolithography \citep{MTT09}. Experimenters have utilized glass \citep{DWSWSSD08,AFMG12,VHDSW12} or ceramic \citep{BAHKKWCB15} nozzles, and have made use of two-photon polymerization to manufacture the microfluidic device \citep{N16}. \citet{YMOR06} have used an orifice with a cusp-like edge to maximize the stress exerted by the outer stream, which ensures the controlled breakup of droplets for a wide range of flow rates. The focusing effect can also be produced by locating the feeding capillary in front of another, both placed inside an external tube through which the focusing current is injected \citep{ZKLTKW16}. In the so-called non-embedded coflow-focusing configuration \citep{DRVSS20}, a converging nozzle is placed in front of a collector tube to produce microemulsions in the dripping and jetting modes.

% Injection methods
Different injection methods have been proposed to enhance the tip streaming stability in flow focusing. One of the essential ideas is to eliminate the recirculation pattern arising in the meniscus of the original configuration, which seems to stabilize the flow \citep{AMFHG12,ARMGV13,MSPVL16}. Stable jets with diameters down to 2.9 $\mu$m were emitted at speeds up to 80 m s$^{-1}$ using conical tips and focusing the liquid with Helium \citep{W18}.

\subsubsection{Flow focusing applications}

\begin{figure}
%\begin{center}
\centering{\resizebox{0.45\textwidth}{!}{\includegraphics{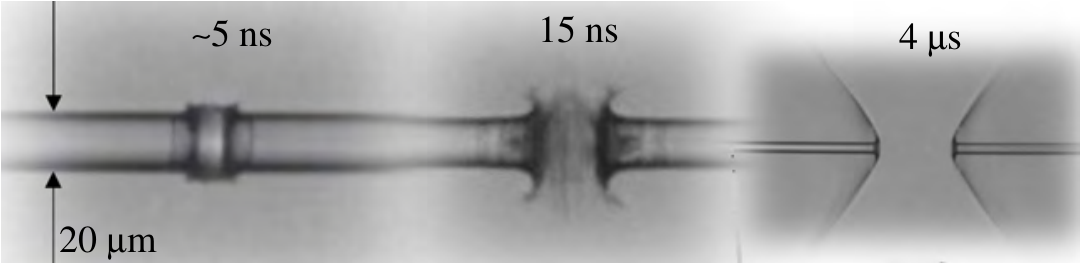}}}
%\end{center}
\caption{Evolution of the blast induced by a strong X-ray laser pulse (photon energy 8.2 KeV, 0.75 mJ, duration 30 fs) on a liquid microjet of 20 $\mu$m in diameter and discharged in a vacuum \citep{S16,G19}.}
\label{Xray}
\end{figure}

% Flow focusing liquid-gas. Applications
The configurations mentioned above include the focusing of liquid streams with either a high-speed gaseous current or another liquid stream with a similar velocity. Gaseous flow focusing has several important applications. Among them, it is of special relevance its use as a sample delivery system in the Serial Femtosecond Crystallography \citep{Cetal11,O17}, which has revolutionized the molecular determination of complex biochemical species by avoiding radiation damage through femtosecond-duration X-ray pulse (Fig.\ \ref{Xray}). Sub-micron jets with speeds exceeding 160 m/s have recently produced with gaseous flow focusing \citep{K20}, which enables the use of megahertz Serial Femtosecond Crystallography repetition rates.

\begin{figure}
%\begin{center}
\centering{\resizebox{0.3\textwidth}{!}{\includegraphics{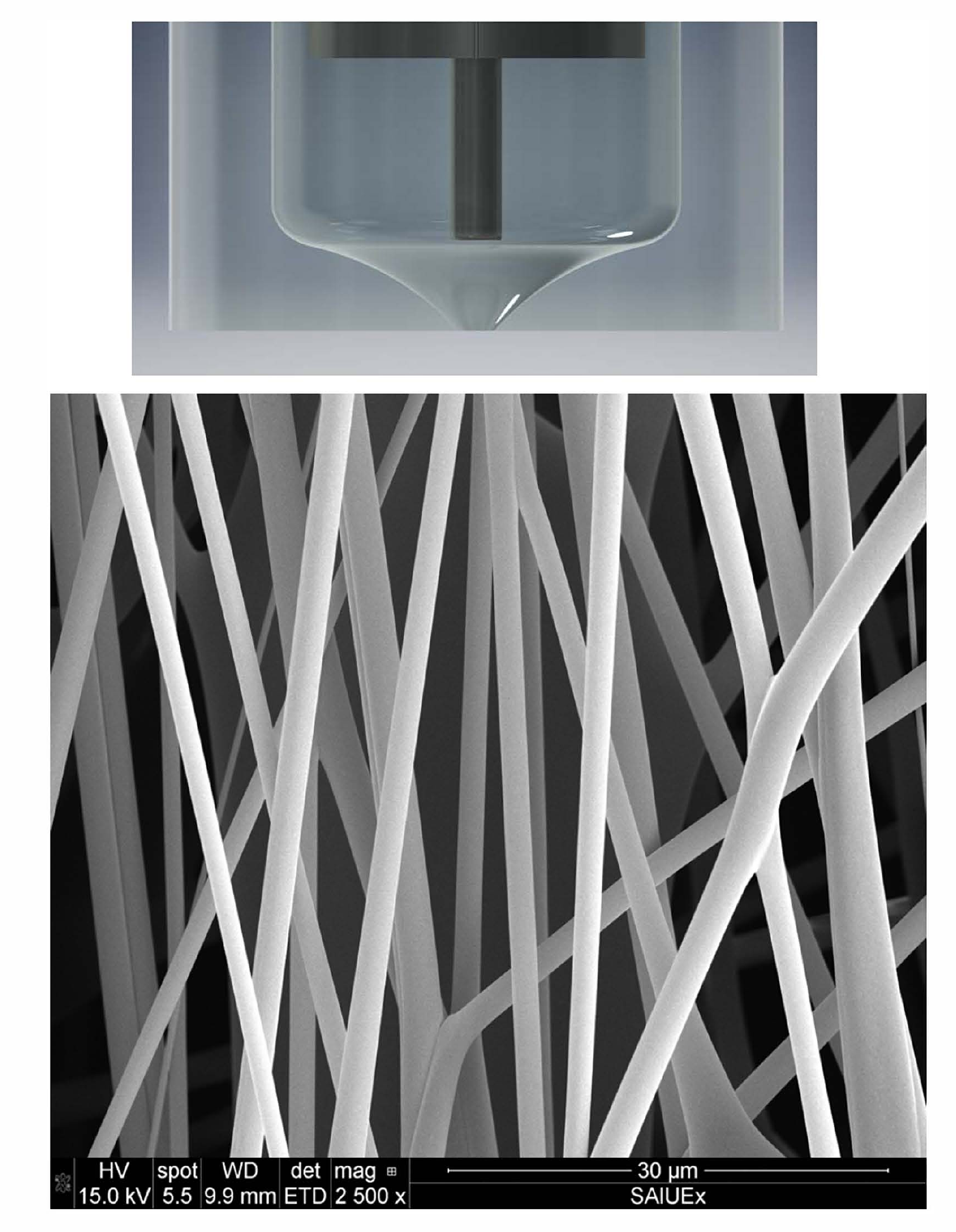}}}
%\end{center}
\caption{SEM images of fibers produced with gaseous flow focusing for a flow rate $Q=5$ ml/h and an applied pressure drop $\Delta P=115$ mbar \citep{PORRVM19}. The ejector is shown in the upper image. The liquid was a solution of 10 wt \% of PVP in ethanol. The average diameter is 1.87 $\mu$m and the standard deviation is 0.7 $\mu$m.}
\label{GVFF}
\end{figure}

Gaseous flow focusing has also been used to produce microparticles of complex structures \citep{GGRHF07,GMMF13,GCFM14} by injecting coaxially two immiscible liquid streams \citep{GGRHF07}. More recently, the same principle has been applied to form viscoelastic threads \citep{PMVG16,HKHSTF18} for smooth printing and bioplotting \citep{PVCM17}, and to fabricate fibers with diameters ranging from a few microns down to hundreds of nanometers \citep{PORRVM19,VKVKKMT19} (Fig.\ \ref{GVFF}). \citet{SLWZLX16} have developed a gaseous co-flow focusing process to produce stimuli-responsive microbubbles that comprise perfluorocarbon suspension of silver nanoparticles in a lipid. Gaseous flow focusing has been implemented not only in the original axisymmetric configuration but also in 2D and 3D geometries \citep{TTH18}.

% Kinetic energy
Gaseous flow focusing offers several attractive advantages over other atomization methods. One of them is that the collimated droplet streams acquire the velocity of the outer gas current in times of the order of $t_s=\rho_1 d_d^2/(18\mu_g)$, where $d_d$ is the droplet diameter and $\mu_g$ the gas viscosity. This means that a micrometer water droplet in a co-flowing subsonic air stream moving at the speed, say, $U_g=$200 m/s can reach kinetic energies (per unit volume) as high as 20 MPa within distances from the source as small as $t_s U_g\sim 1$ mm .

% Electro-flow focusing
Electro-flow focusing \citep{GLR06} is a combination of electrospray and flow focusing where a relatively small voltage is applied to the focused liquid to charge the droplets produced via flow focusing. These droplets, with the proper combination of size, speed, and electrical charge, are perfect candidates for applications such as surface sample desorption \citep{FS14a,FS14b}.

% Flow focusing liquid-liquid
The liquid-liquid flow focusing configuration has received more attention than the gaseous one because of its very diverse applications. Here, we list some of them. \citet{LHKHHL01} built a flow cytometer by focusing a liquid stream injected together with an outer one, both flowing across coaxial converging nozzles. Axisymmetric liquid-liquid flow focusing was proposed for the encapsulation and release of actives in the pioneering work of \citet{ULLKSW05} and subsequent experimental studies \citep{TGWW05,SSHW10}. \citet{TMT10} described a method based on axisymmetric flow focusing to produce monodisperse cell-encapsulating microgel beads composed of a self-assembling peptide gel for 3D cell culture. Biodegradable Poly(Lactic Acid) particles have been fabricated using this technique \citep{VHDSW12}. Double flow focusing has been successfully used to produce monodisperse multiple-emulsions \citep{CLSSW11,CWZ15}, which are ideal microreactors or fine templates for synthesizing advanced particles \citep{WWH11}. \citet{GDM11} have reviewed the use of liquid-liquid flow focusing to produce emulsions with high throughput for encapsulation, chemical synthesis and biochemical assays. \citet{WYLXZSX17} have proposed multiplex coaxial flow focusing for single-step fabrication of multicompartment Janus microcapsules.

The applications of the liquid-liquid flow focusing are not limited to the fields mentioned above and can be extended to many others. For instance, the multiple-emulsions produced with a variant of this technique also allow the formation of PDMS microcapsules with tunable elastic properties \citep{NAMMCD17}. \citet{ZWHXZYDSX18} have manufactured multi-compartment polymeric microcapsules in an axisymmetric flow focusing device for magnetic separation and synergistic delivery. Micrometer fibers can be formed when a viscoelastic liquid is focused with a Newtonian one \citep{ESMRD06,MGEPG17} or a viscoelastic stream \citep{JKKLMB04}.

% Counter-flow focusing
Single-step generation of multi-core double emulsion droplets can be achieved by a specific class of axisymmetric flow focusing \citep{WYLXZSX17}. \citet{ULLKSW05} first assembled a counter-current double emulsion flow focusing device by aligning glass capillaries. This configuration has been recently examined experimentally by \citet{NVM17}. Monodispersed emulsions down to 2 $\mu$m in size were produced at a frequency of 20 kHz by using a similar configuration, the so-called ``impinging flow-focusing"\ \citep{WLDCH17}. The counter-current flow focusing can exhibit a second-order transition which leads to extremely thin jets \citep{DMFESR18}, which gives rise to silicone oil emulsions with a submicron particle radius.

\subsubsection{Flow focusing dimensionless numbers}

% Geometrical dimensionless numbers
Two geometrical parameters enter the problem owing to the existence of a discharge orifice in flow focusing: the orifice diameter $D_o$ and the orifice-to-capillary distance $H$. The diameter $D_o$ is frequently chosen as the characteristic length because it determines the size of the region where the focusing effect takes place. Among the dimensionless parameters characterizing the flow focusing device, the ratio $\widehat{H}=H/D_o$ is the most important because it considerably affects the meniscus slenderness and, therefore, its stability \citep{VMHG10}. The ratio $2R_i/D_0$ generally takes values around unity and is not typically taken into account. The ratio (\ref{rat}) between the radii of the feeding and outer capillaries has little influence on the flow focusing effect, and is not considered in the analysis either.

% Flow focusing liquid-liquid
The non-geometrical dimensionless numbers characterizing the Newtonian liquid-liquid flow focusing configuration are typically the same as those of a coflowing device, i.e., the outer Capillary number $Ca_o$ [Eq.\ (\ref{co})], or the flow rate ratio $Q_r$ [Eq.\ (\ref{qr})], and the viscosity ratio $\mu$. Inertial effects can be important in flow focusing because larger speeds can be reached. For this reason, the Reynolds number or the Weber number may enter the problem too. The density ratio always takes values of the order of unity, and, therefore, it plays a secondary role in the analysis.

% Flow focusing liquid-gas.
The focusing of a liquid stream by an outer high-speed gas current \citep{G98a} constitutes a phenomenon considerably different from that in the liquid-liquid case. The pressure drop $\Delta P$ caused by the gas acceleration becomes the force driving the liquid ejection in front of the discharge orifice. In the so-called monosized dripping mode \citep{CMG16}, the magnitude of this force is quantified through the ratio of the pressure drop $\Delta P$ to the capillary stress $\sigma/R_i$, i.e.,
\begin{equation}
{\cal P}=\frac{R_i \Delta P}{\sigma}=\frac{R_i}{d_{\sigma}},
\end{equation}
where $d_{\sigma}=\sigma/\Delta P$. The steady jetting regime is typically analyzed considering the parameter plane defined by the Reynolds and Weber numbers
\begin{equation}
\label{re}
\text{Re}_{\textin{FF}}=\frac{\rho_i V_{\textin{FF}} R_{\textin{FF}}}{\mu_i},\quad \text{We}_{\textin{FF}}=\frac{\rho_i V_{\textin{FF}}^2 R_{\textin{FF}}}{\sigma},
\end{equation}
where $V_{\textin{FF}}=Q_i/(\pi R_{\textin{FF}}^2)$ and
\begin{equation}
\label{SJ}
R_{\textin{FF}}=\left(\frac{\rho Q_i^2}{2\pi^2\Delta P}\right)^{1/4}
\end{equation}
are the jet's velocity and radius calculated in terms of the injected flow rate $Q_i$ and applied pressure drop $\Delta P$ assuming mass and energy conservation \citep{G98a,GM09}.

% Flow focusing liquid-gas. Jetting mode. Viscoelasticity
The formula (\ref{SJ}) does not apply to viscoelastic liquids. In this case, $\widehat{H}\gg 1$, which means that the focusing effect is confined within a region much smaller than the liquid thread formed between the feeding capillary and the discharge orifice. The axial stress generated by the air stream next to the orifice is transmitted upstream by the stretched polymers, providing the fiber with the necessary consistency to avoid its breakup. That axial stress survives downstream far away from the orifice so that the liquid thread flies freely in front of the discharge orifice along distances hundreds of times its diameter \citep{PMVG16}.

% Flow focusing gas-liquid
Focusing a gaseous stream with an outer liquid current leads to a monodispersed bubbling regime \citep{GG01,G04b,JSB06,VAMHG14}. The strong oscillatory character of this phenomenon suggests the use of the flow rate ratio $Q_r$ as the major governing parameter. The flow is essentially inviscid, and the density ratio takes very small values. Therefore, neither the Reynolds number nor the density ratio comes into play.

\subsection{Selective withdrawal and electrified films}
\label{sec9.5}

\begin{figure}
%\begin{center}
\centering{\resizebox{0.33\textwidth}{!}{\includegraphics{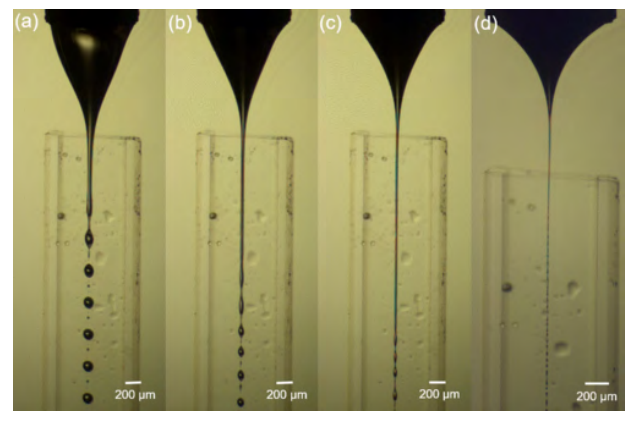}}}
%\end{center}
\caption{Shells produced with the confined selective withdrawal geometry for different speeds in the extraction tube. The shell diameter decreases as the speed decreases \citep{HCGLG19}.}
\label{confined}
\end{figure}

% Selective withdrawal
The production of drops/bubbles does not necessarily require a feeding capillary through which the dispersed phase is injected. In the selective withdrawal technique, a lower liquid layer is withdrawn by an upper one, which is suctioned across a cylindrical tube placed in front of the interface between them \citep{L89,CLHMN01,ED09,CN07,BZ09}. The interface forms a hump in front of the tube due to the viscous stresses exerted by the upper-layer liquid. When the upper-layer flow rate exceeds a certain threshold, those stresses overcome the resistance offered by surface tension, and the hump emits from its tip a jet much thinner than the collector tube. This visco-capillary phenomenon is governed by the Capillary number \citep{BZ09}
\begin{equation}
\label{co2}
\text{Ca}_o^{\textin{sw}}=\frac{\mu_o Q_o}{4\pi \hat{H} \sigma},
\end{equation}
where $\mu_o$ and $Q_o$ are the upper-layer viscosity and flow rate, respectively, and $\hat{H}$ is the distance between the capillary and the undisturbed interface. The selective withdrawal technique can be applied to produce monodisperse emulsions of micrometer droplets resulting from the capillary breakup of the jet. It can also be used to coat microparticles present in the withdrawn liquid \citep{CLHMN01}. A confined selective withdrawal geometry has been used to produce bubbles \citep{ECG15}, emulsions \citep{ECG16}, double emulsions and nematic shells \citep{HCGLG19} (Fig.\ \ref{confined}), and micro-sized PDMS particles \citep{MSPVL16} in the tip streaming regime. In this geometry, the dispersed phase is injected at a constant flow rate through a capillary/needle located in front of the collecting tube.  

% Electrified films
An electrohydrodynamic process analogous to the selective withdrawal takes place when an electrified collector tube is placed in front of a liquid film of finite conductivity \citep{ON00,CJHB08}. For small applied voltages, the free surface forms a hump in front of the tube owing to the action of the electric stresses. Above a certain critical voltage, the hump ejects a very thin liquid thread that moves into the tube. In this technique, the role of the Capillary number Ca$_o^{\textin{sw}}$ [Eq.\ (\ref{co2})] is played by the electric Bond number $\chi$ [Eq.\ (\ref{eb})].

\section{Dripping and bubbling}
\label{sec10}

% Next sections
We devote this and the next section to present some of the major results obtained using the droplet continuous production methods described above. We consider the dripping/bubbling and jetting modes separately. In each of those modes, we will distinguish the direct ejection method from that reached via tip streaming. Droplets/bubbles and jets are formed right after the feeding capillary in the former case. In tip streaming, the fluid is directed by some external actuation towards the tip of a deformed film, drop or stretched meniscus attached to a feeding capillary. Then, that tip emits small drops/bubbles either directly (Fig.\ \ref{tip2}) or through the breakage of a very thin jet (Fig.\ \ref{tip}). In many cases, there is a gradual transition between dripping/bubbling and jetting on one side, and between direct ejection and that taking place via tip streaming on the other side. Then, the distinction between the different ejection modes is not always evident.

\subsection{Direct dripping and bubbling}
\label{sec10.1}

% Gravitational dripping/bubbling. Quasi-static process
As mentioned in Sec.\ \ref{sec9.2}, gravitational dripping/bubbling is obtained for sufficiently large Bond numbers and small enough Weber numbers. At the incipience of a drop formation, a nearly spherical volume of fluid hangs on a thin filament attached to the feeding capillary. This filament eventually becomes unstable and breaks up driven by the surface tension force. For sufficiently small Weber numbers, the quasi-static dripping and bubbling processes are identical. In this regime, monodisperse collections of droplets/bubbles are produced with sizes of the order of or much larger than that of the feeding capillary.

\begin{figure}
%\begin{center}
\centering{\resizebox{0.35\textwidth}{!}{\includegraphics{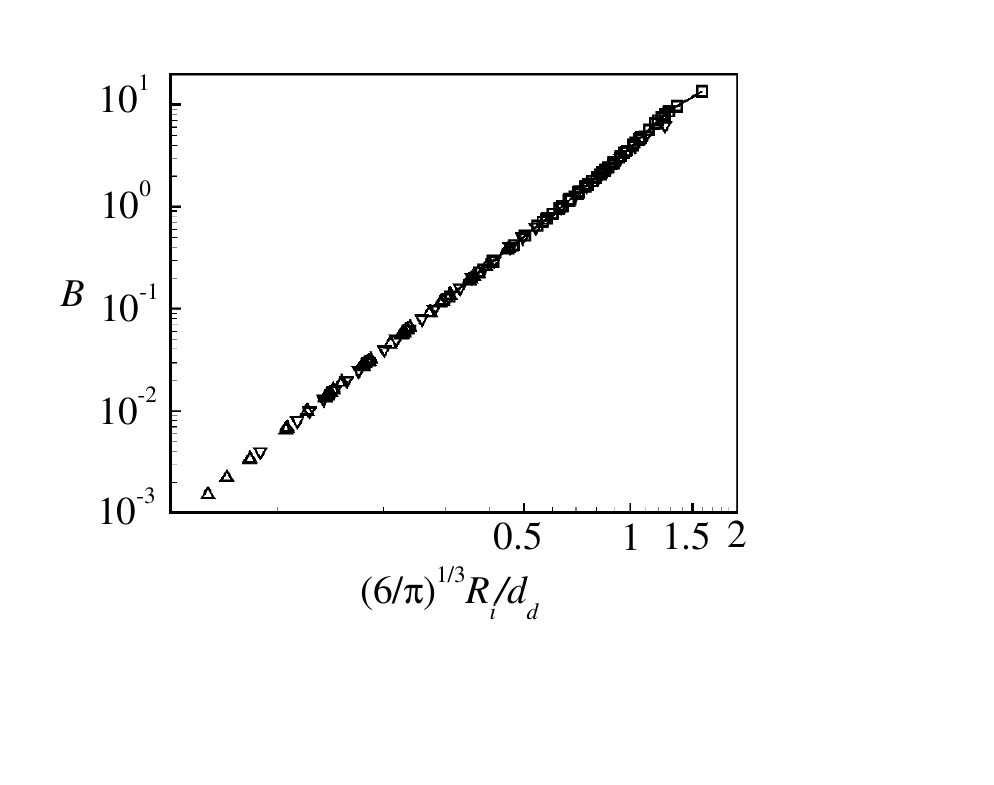}}}
%\end{center}
\caption{Diameter $d_d$ of the primary droplet falling down from a vertical capillary as a function of the gravitational Bond number $B$ \citep{YXB05}. The squares, up-triangles and down-triangles correspond to the experimental data obtained by \citet{HB19} and \citet{W72}, and the 1D numerical simulations conducted by \citet{YXB05}, respectively. The fitting to the data yields (\ref{dw}).}
\label{DWfig}
\end{figure}

The diameter $d_d$ of the primary droplet/bubble that quasi-statically falls down from the capillary is given by the approximate formula
\begin{equation}
\label{dw}
\frac{d_d}{R_i}= \alpha B^{-\beta},
\end{equation}
where $\alpha\simeq 4.46$ and $\beta\simeq 0.356$ \citep{YXB05} (Fig.\ \ref{DWfig}). In the drop weight method for measuring the interfacial tension \citep{HB19,W72}, Eq.\ (\ref{dw}) provides the value of that quantity as a function of the droplet diameter measured in the experiment. There is a clear similarity between Eqs.\ (\ref{Beroz}) and (\ref{dw}), which characterize the instability conditions for droplets under the action of an electrical field and the gravitational force, respectively.

% Gavitational dripping/bubbling. Dynamical effects
As the Weber number increases, dripping and bubbling become significantly different. Dripping evolves from a period-1 response to jetting, either directly or through several period doubling (halving) bifurcations \citep{APB00}. On the contrary, bubbling persists producing bubbles with diameters $d_b$ which depend on the injected gas flow rate $Q_i$ as \citep{G04b}
\begin{equation}
d_b\propto Q_i^{2/5}.
\end{equation}

% Electrified dripping. Quasi-static process
For sufficiently large values of the electric Bond number $\chi$, a quasi-static dripping regime similar to the gravitational one can be produced in the absence of gravity. Consider a conductor droplet anchored to a feeding capillary placed on a face of a circular parallel-plate capacitor. If the intensity of the electric field is progressively increased, the droplet will detach from the capillary for $\chi\gtrsim 0.3$ \citep{BS90}.

% Electrified dripping. Dynamical effects. AC field
A complex scenario appears when dynamical effects arise in electrified dripping. If the electric potential is increased while keeping the flow rate constant, the frequency of droplet formation increases and the drop volume decreases. If the electric Bond number is further increased, an electrified meniscus forms and emits varied liquid shapes operating in several periodic and aperiodic pulsating modes \citep{MNV07}. The periodic modes are the so-called electro-dripping, spindle, and intermittent cone-jet modes. A detailed description of them can be found in the recent review by \citet{RGL18}. Electrified dripping can also be produced by applying AC electric fields with characteristic times much larger than those of the system \citep{CGMS09,RYTZ04,RYZB06,TBNK09,TBYO11}.

% Coflowing. Dripping
Viscous stresses exerted by the outer liquid stream in the coflowing configuration detach droplets from the feeding capillary in the dripping regime. For small values of the outer-to-inner flow rate ratio, $Q_r$, drop formation occurs in a slug flow regime in which drops are elongated axially and occupy almost the entire cross-section of the outer tube before their ejection. For larger flow rate ratios, the system adopts a proper monodisperse dripping regime \citep{UPW00}. The volume of the emitted droplets decreases as $Q_r$ increases \citep{UPW00,SB06}.

% Coflowing. Bubbling
In a typical gas-liquid coflowing configuration, the gas is injected with a velocity larger than that of the outer liquid stream, which leads to the periodic formation of bubbles near the nozzle exit \citep{GSM07}. The bubble diameter is typically larger than that of the nozzle. For a given velocity ratio, the bubbles produced when the injection pressure is kept constant are larger than those formed under the constant gas flow rate condition \citep{RSMG15}. For outer stream velocities similar to that of the gas injection, the jetting regime can be produced \citep{SGM05a}. Gaseous filaments can be elongated and pinched in converging microchannels \citep{AP07}. The existence of a coflowing stream hinders the coalescence of bubbles, which constitutes an important technological advantage.

\subsection{Dripping and bubbling from tip streaming}
\label{sec10.2}

% Gravitational
Gravity alone does not seem to be capable of producing dripping or bubbling from tip streaming, probably because of its inability to focus the flow towards the tip. \citet{DE06} conducted a nice experiment where microbubbles were formed when the interface between a viscous liquid and air was deformed by a sink flow to form a sharp tip. Experiments did not elucidate whether bubbles were produced in the bubbling or jetting mode. This configuration can be categorized as a variant of selective withdrawal driven by gravity.

\begin{figure}
%\begin{center}
\centering{\resizebox{0.325\textwidth}{!}{\includegraphics{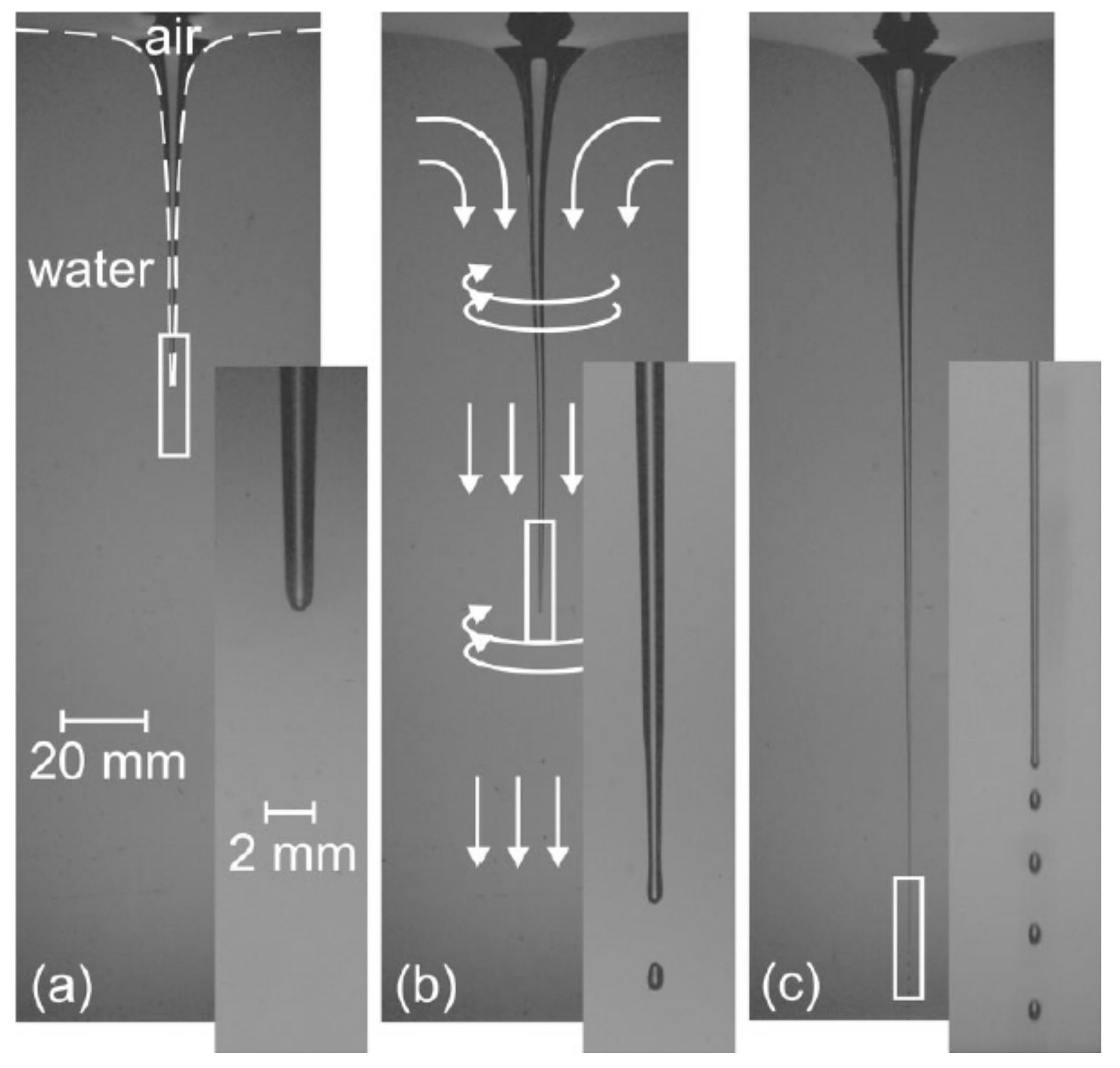}}}
%\end{center}
\caption{Bathtub vortex formed when the container rotates at 13.5 rpm (a), 24 rpm (b) and 30.5 rpm (c) \citep{BAMB09}.}
\label{bathtub}
\end{figure}

% Gravity plus centrifugal
Gravity combined with the centrifugal force arising in a rotating cylindrical container produces a ``bathtub vortex"\ as the liquid drains out through a small hole \citep{ABSRL03,BAMB09}. Bubbling from tip streaming takes place at the tip of the needle-like surface depression for sufficiently high rotation speeds (Fig.\ \ref{bathtub}). The frequency of bubble
formation increases and the bubble size decreases as the rotation speed increases.

% Electrospray
Under certain operational conditions, the tip of an electrified meniscus can drip \citep{JK99,NB99}, giving rise to a monodisperse collection of microdrops with diameters much smaller than that of the feeding capillary. \citet{HLIH15} produced drops using this procedure with diameters down to $0.1R_i$. The diameter and generation frequency of the microdrops can be controlled by appropriately selecting the flow rate and applied voltage \citep{WLTXZWT19}.

% Flow focusing liquid-liquid
The planar liquid-liquid flow focusing configuration has been massively used since \citet{ABS03} implemented this technique using microchannels. In most cases, this configuration works in the dripping/bubble mode to produce emulsions \citep{ABS03}, microparticles \citep{XHDHKWLA09} and bubbles \citep{GGDWKS04} with a very high degree of monodispersity. The use of the corresponding axisymmetric configuration has been much less frequent \citep{VKN12}.

\begin{figure}
%\begin{center}
\centering{\resizebox{0.375\textwidth}{!}{\includegraphics{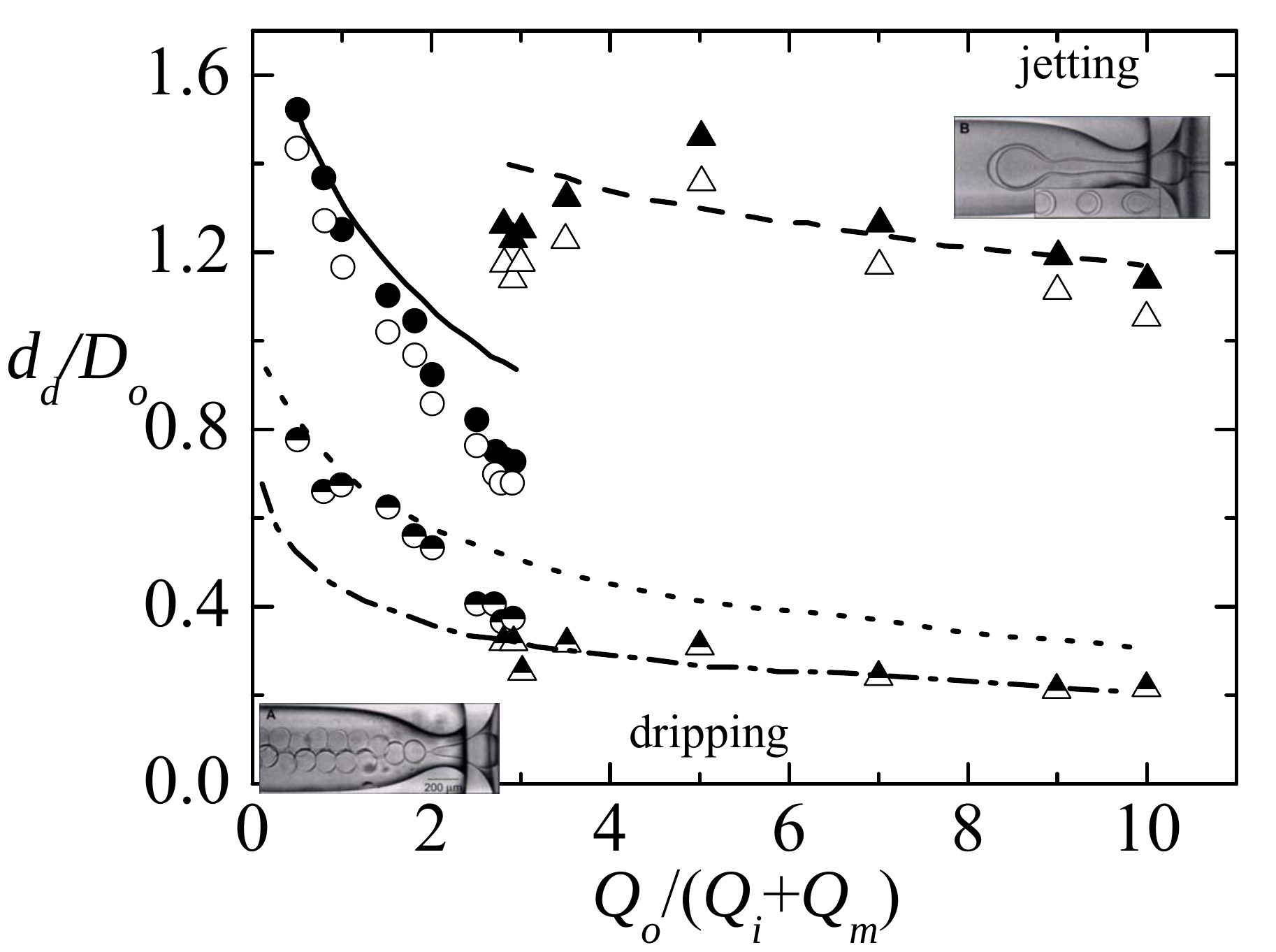}}}
%\end{center}
\caption{Diameter $d_d$ of the droplets produced in a flow focusing device in terms of the orifice diameter $D_o$ as a function of the outer liquid flow rate $Q_o$ divided by the sum of the inner and middle liquid flow rates, $Q_i+Q_m$ \citep{ULLKSW05}. The half-filled, white and black circles correspond to the precursor liquid thread diameter and the inner and outer droplet diameters in the dripping regime, respectively. The dotted and solid lines are the theoretical predictions (\ref{utada1}) and (\ref{utada3}) for the liquid thread and outer droplet diameter in this regime, respectively. The half-filled, white and black triangles correspond to the precursor jet diameter and the inner and outer droplet diameters in the jetting regime, respectively. The dashed-dotted and dash lines are the predictions for the liquid thread and outer droplet diameter [Eq.\ (\ref{utada5})], respectively.}
\label{utadaf}
\end{figure}

% Utada
\citet{ULLKSW05} applied the flow focusing principle to produce double emulsions in both the dripping and jetting regimes in glass micronozzles. The transition between the two regimes took place for Capillary numbers around unity. The diameter $d_d$ of the droplets produced in the dripping mode was estimated as
\begin{equation}
\label{utada3}
d_d\simeq 1.9 d_j,
\end{equation}
where $d_j$ is the diameter of the short liquid thread formed at the discharge orifice. This quantity can be easily estimated from the continuity equation assuming a flat velocity profile,
\begin{equation}
\label{utada1}
Q_r^{-1}+1=\frac{d_j^2}{D_{\textin{o}}^2-d_j^2}.
\end{equation}
Interestingly, Rayleigh's prediction (\ref{utada3}), commonly used to predict the droplet diameter in the jetting mode, satisfactorily fitted experimental data of the dripping mode too \citep{ULLKSW05}, which indicates the sometimes subtle difference between the two regimes. The droplet inner diameter was slightly smaller than the outer one. Figure \ref{utadaf} shows the diameters of both the precursor liquid thread and the resulting droplets as a function of the outer liquid flow rate $Q_o$ divided by the sum of the inner and middle liquid flow rates, $Q_i+Q_m$. The transition from the dripping to the jetting regime takes place for $Q_o/(Q_i+Q_m)\simeq 0.3$.

% Flow focusing liquid-gas
The axisymmetric flow focusing configuration was originally conceived to produce droplets from the breakage of a liquid jet expelled due to the action of a high-speed gas stream \citep{G98a}. However, \citet{CMG16} have been recently shown that monodisperse dripping can also be obtained within relatively narrow intervals of viscosities, injected flow rates and applied pressure drops. In this regime, the applied pressure drop is balanced by the local acceleration of the fluid particle, which yields the scaling law for the droplet diameter
\begin{equation}
\frac{d_d}{R_i}=2.4 {\cal P}^{-1/2}.
\label{fit}
\end{equation}
This monosized dripping mode of axisymmetric flow focusing has been found for viscosities $1\lesssim \mu\lesssim 25$ mPa, and leads to diameters smaller than that of the feeding capillary in most cases \citep{CMG16}. The ejection process differs substantially from that taking place in the micro-dripping mode of electrospray, where the electrified meniscus stretches and shrinks more notably.

\begin{figure}[hbt]
%\begin{center}
\centering{\resizebox{0.325\textwidth}{!}{\includegraphics{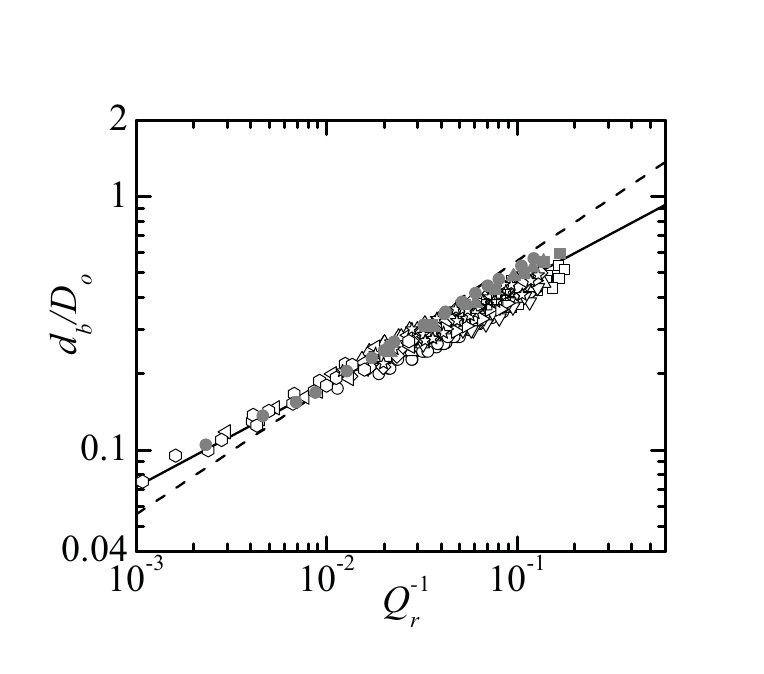}}}
%\end{center}
\caption{Diameter $d_b/D_o$ of the bubbles produced with flow focusing as a function of the flow rate ratio $Q_r$ \citep{VAMHG14}. The white and grey symbols correspond to experiments conducted with the plate-orifice \citep{G04b} and nozzle \citep{VAMHG14} configuration, respectively. The solid and dashed lines correspond to Eq.\ (\ref{bubbling}) and $d_b/D=1.77 Q_r^{-0.5}$, respectively. The latter can
be regarded as the version for this configuration of the classical flow-focusing formula \citep{G98a}.}
\label{diameter}
\end{figure}

% Flow focusing gas-liquid
As explained in Sec.\ \ref{sec4}, gaseous jets injected into a quiescent pool of liquid are prone to absolute instability owing to the high values of the capillary wave velocity. For this reason, most flow focusing experimental realizations produce bubbles in the bubbling mode \citep{GG01}. The bubble diameter can be obtained by assuming that the unsteady and convective terms of the momentum equation for the gas phase are commensurate with each other, which yields \citep{G04b}
\begin{equation}
\label{bubbling}
\frac{d_b}{D_o}=1.1 Q_r^{-0.4}.
\end{equation}
This scaling law remarkably agrees with experimental data obtained using both the plate-orifice \citep{G04b} and nozzle \citep{VAMHG14} configurations (Fig.\ \ref{diameter}). \citet{JSB06} analyzed numerically the bubbling in an axisymmetric flow focusing device. The results for the bubble volume were consistent with those obtained experimentally by \citet{GGDWKS04} for the 2D topology.

\section{Jetting}
\label{sec11}

\subsection{Direct jetting}
\label{sec11.1}

\subsubsection{Gravitational direct jetting}

% Gravitational. Transition
Capillary liquid jets can be formed by simply ejecting liquid across a vertical feeding capillary at high enough Weber numbers. \citet{CL99} extended Taylor's model \citep{T59} for the recession speed of a free edge to obtain the critical injection velocity leading to the dripping-to-jetting transition in an inviscid gravity-driven flow. The Weber number above which dripping gives rise to jetting is given by the expression
\begin{eqnarray}
\text{We}_{i}&=&2\left(\frac{B_o}{B}\right)^{1/2}\left\{1+K (B_oB)^{1/2}\right.\nonumber\\
&&\left.-\left[(1+K (B_oB)^{1/2})^2-1\right]^{1/2}\right\}^2,
\end{eqnarray}
where $B_o$ is the Bond number based on the outer radius of the feeding capillary, and $K$ is a dimensionless constant which depends on the fluids involved ($K=0.72$ for water injected in the air). The viscosity of both the inner and outer fluids can significantly modify this result \citep{BBB17}. The gravitational dripping-to-jetting transition has also been analyzed in terms of the global stability of the base flow calculated from the 1D approximation (see Sec.\ \ref{sec2.4}) \citep{SB05a,RSG13}.

% Jetting. Clarke
For flow rates sufficiently larger than the minimum leading to jetting, the liquid jet adpots a slender shape which can be accurately described by the 1D model \citep{EV08}. If one neglects the surface tension force, the exact solution of this model is \citep{C66,MHFVG11}
\begin{equation}
v=\frac{2^{-1/3} \text{Ai}^2(r)}{\text{Ai}'^2(r)-r\text{Ai}^2(r)}\; , \quad r=2^{-1/3} (x-k)\; ,
\label{usol}
\end{equation}
where $v=w/v_0$ and $x\equiv z/L_0$ are the velocity and distance to the feeding capillary scaled with $v_0\equiv (L_0 g)^{1/2}$ and $L_0\equiv 3^{2/3}(\mu_i^2/\rho_i^2 g)^{1/3}$, respectively \citep{C66}. In addition, $\text{Ai}$ and $\text{Ai}'$ are the Airy function and its derivative, respectively, and $k= 2.9455\ldots$ is the first zero of $\text{Ai}(-k/2^{1/3})$. Intriguingly, the power of this solution has been hardly exploited to date in microfluidics.

% Jetting. Breakup length
\citet{JEBHR13} examined the breakup of a viscous gravitational jet using a WKB-type approach, in which disturbances locally have the form of plane waves. They showed that viscosity plays completely independent roles in the axial momentum balance of the steady base state and in the growth of perturbations about that state. Specifically, viscosity does affect the perturbation growth rates, although it does not influence the base flow in most of the jets formed in many experiments. The same idea underlies, for instance, the stability analysis of a viscous electrified jet emitted in the cone-jet mode of electrospray \citep{IYXS18}.

\begin{figure}[hbt]
%\begin{center}
\centering{\resizebox{0.3\textwidth}{!}{\includegraphics{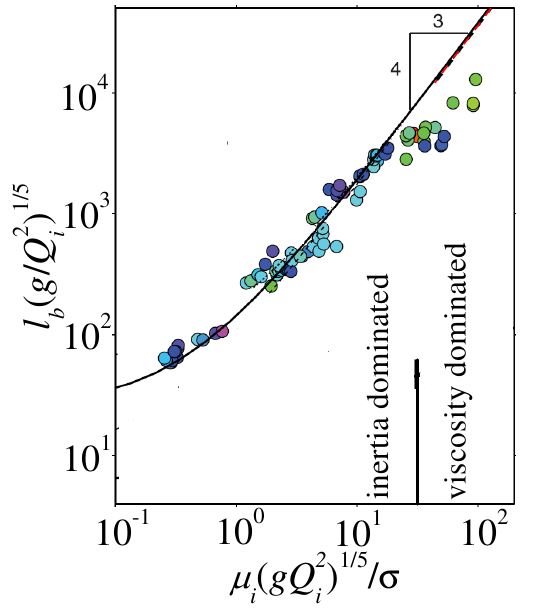}}}
%\end{center}
\caption{Breakup length $l_b$ measured experimentally (symbols) and calculated theoretically (solid line) \citep{JEBHR13}. The dashed line is the prediction (\ref{grav}) for the high-viscosity limit.}
\label{jadavi}
\end{figure}

The breakup length of a falling viscous jet can be estimated from the natural extension of Eqs.\ (\ref{esti}) to an accelerated jet. In the high-viscosity limit, the growth rate $\omega_i$ of an axisymmetric perturbation scales as the inverse of the viscous-capillary time (\ref{tvc}). Because most of the jet accelerates in the inertial regime, $R_j\sim (Q_i^2/gz)^{1/4}$ and $t_f\sim (z/g)^{1/2}$, where $z$ is the distance to the nozzle exit and $t_f$ the time for a fluid element to reach that distance. Considering the scalings for $R_j$ and $t_f$, and the visco-capillary time (\ref{tvc}), one derives a simple scaling law for the jet breakup length:
\begin{equation}
\label{grav}
l_b\sim \left(\frac{gQ_i^2\mu_i^4}{\sigma^4}\right)^{1/3}.
\end{equation}
The comparison between the prediction (\ref{grav}) and experimental data shows remarkable agreement when the base flow is dominated by inertia (Fig.\ \ref{jadavi}).

\subsubsection{Electrified direct jetting}

% Electrospray
When a liquid is emitted through a feeding capillary at Weber numbers in the range 1-10, the classical jetting regime is obtained. This regime can be manipulated by applying an electrical potential to the feeding capillary, which leads to the so-called simple-jet mode of electrospray \citep{ATYBFM12,AYFM12,RGL18}. For relatively small values of the electric Bond number $\chi$ [Eq.\ (\ref{eb})], this mode does not substantially differ from the non-electrified case in terms of the jet diameter and velocity, breakup distance, and size of the emitted droplets. The major effect is that Ohmic conduction charges the jet's free surface. That charge accumulates in the droplets resulting from the varicose jet breakup. As the electric Bond number increases, the electrostatic repulsion experienced by the droplets forms a plume \citep{ATYBFM12}. Above a critical electric Bond number, which grows with the Weber number, the transition from axisymmetric breakup to whipping takes place.

\subsubsection{Coflowing direct jetting}

% Coflowing, liquid-liquid
Liquid-liquid coflowing devices normally operate either in the direct dripping mode of in the direct jetting regime. \citet{CFW04} showed experimentally that the limiting length of drops produced in these devices increases as the flow rate of the outer fluid increases. When this parameter exceeds a certain threshold (while keeping the rest constant), the system experiences a transition from dripping to jetting. The size of the droplets produced by a coflowing device in the jetting mode can simply be derived by neglecting inertia. In this case, the velocity profiles of both the continuous and dispersed phases are parabolic, and the dominant perturbation wavelength can be estimated from Tomotika's dispersion relationship \citep{T35}. Using these two ingredients, one obtains, for instance, $d_d/D_o=1.44 Q_r^{-1/2}$ for $\mu=1$ \citep{VDGVVL11}.

% Low Reynolds numbers. Guillot
In the jetting regime of the coflowing configuration, the jet tapers directly from the feeding capillary, and, therefore, the system stability reduces to that of the jet itself. \citet{GCUA07} identified three dimensionless parameters governing the system behavior at low Reynolds numbers. The critical (re-defined) Capillary number for the convective-to-absolute instability transition (Sec.\ \ref{sec3}) marks the border between the dripping and jetting modes \citep{GCUA07}. The global stability analysis leads to almost the same results for sufficiently high external flow rates \citep{AFG18}. 

\begin{figure}[hbt]
%\begin{center}
\centering{\resizebox{0.4\textwidth}{!}{\includegraphics{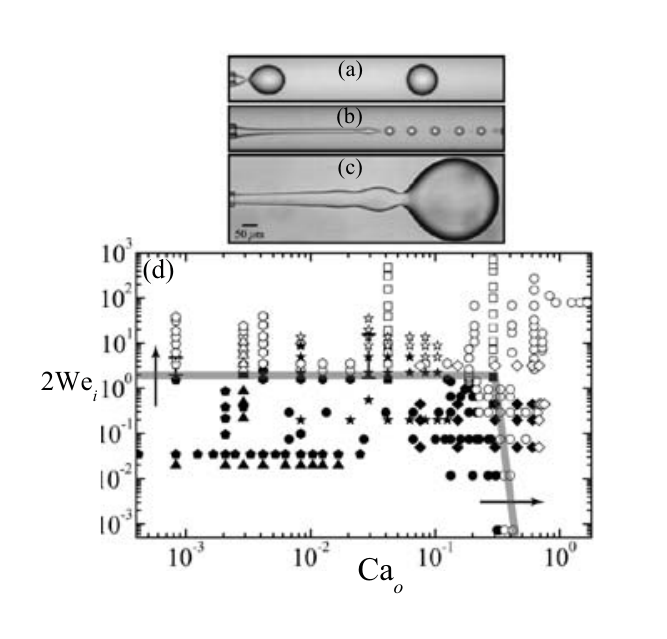}}}
%\end{center}
\caption{Dripping (a), narrowing (b) and widening (c) regimes. State diagram showing the dripping (solid symbols) and narrowing/widening (open symbols) experimental realizations for varied liquid-liquid configurations (d) \citep{UFSW07}.}
\label{coflowing}
\end{figure}

% Moderate Reynolds numbers. Utada. The widening regime
\citet{UFSW07} experimentally found that the dripping-to-jetting transition in coflowing liquid streams with moderate Reynolds numbers can be characterized by a state diagram that depends on both the Weber number We$_i$ of the inner fluid and the Capillary number $\text{Ca}_o$ of the outer one (Fig.\ \ref{coflowing}). Dripping takes place when both parameters take small enough values, while jetting is obtained if the sum $\text{Ca}_o+\text{We}_i$ is at least of order unity, i.e. when inner inertia plus outer viscosity overcomes surface tension. The produced jets can either narrow or widen downstream depending on the values of $\text{Ca}_o$ and $\text{We}_i$ (Fig.\ \ref{coflowing}). For Weber numbers at least of order unity, jets in the widening mode break up due to an absolute instability occurring for sufficiently small values of $\text{Ca}_o$ \citep{UFGW08}. This instability produces droplets with diameters much larger than that of the precursor jet. \citet{CGFG09} found that the droplet diameter in both the narrowing and widening regimes can be approximately obtained from (\ref{esti})-right, the only difference resides in the dependence of $\lambda^{\textin{max}}$ and $d_j$ on the governing parameters.

% Coflowing, gas-liquid
The diameter of the bubbles produced in the jetting mode of a gas-liquid coflowing device can be estimated by assuming negligible inertia \citep{VDGVVL11}. In this case, the bubble volume is calculated from the conservation of mass during the gaseous jet breakup under constant flow rate conditions, which gives
\begin{equation}
d_b\sim Q_r^{\beta},
\end{equation}
where the exponent $\beta$ ranges from $-0.5$ to $-5/12$. These values correspond to the limits $Q_r^{-1}\to 0$ and $\mu^{-1}\to 0$, respectively. The exponent $-5/12$ has been observed experimentally by \citet{CHLG11}.

\subsubsection{Flow focusing direct jetting}

% Flow focusing. Liquid-liquid
\citet{ULLKSW05} studied the jetting regime arising in the axisymmetric liquid-liquid flow focusing configuration for Capillary numbers Ca$_o$ [Eq.\ (\ref{co})] larger than unity (Fig.\ \ref{utadaf}). The formation of droplets in this regime was similar to that caused by the end-pinching mechanism rather than to the Rayleigh capillary instability (see Sec.\ \ref{sec5}). The droplet diameter $d_d$ was calculated by assuming that the droplet is inflated by the precursor jet during a time which scales with the viscous-capillary time (\ref{tvc}). This leads to the expression
\begin{equation}
\label{utada5}
d_d=\left[\frac{60 (Q_i+Q_o)}{\pi}\frac{d_j \mu_o}{\sigma}\right]^{1/3},
\end{equation}
where $d_j$ can be obtained from mass conservation by assuming parabolic velocity profiles \citep{ULLKSW05}. The slowing down of the outer stream caused by the diverging character of the nozzle made the jet break up at a fixed location \citep{T36}, which resulted in higher monodispersity degrees \citep{AFMG12}.

% Figure Utada
Figure \ref{utadaf} shows both the diameters of the precursor liquid jet and the inner and outer diameters of the resulting droplets formed in the double-emulsion flow focusing device developed by \citet{ULLKSW05}. The dash-dotted and dashed lines are the predictions obtained from mass conservation and Eq.\ (\ref{utada5}), respectively. The experimental results show that the liquid thread diameter in the dripping regime smoothly evolves into that of the jetting mode. The small difference between the trends in the two cases may be attributed to the development of the velocity profile rather than a true bifurcation. The dripping-to-jetting transition at $Q_O/(Q_i+Q_m)\simeq 0.3$ becomes apparent due to the sharp increase of the ratio of the droplet diameter to that of the precursor thread.

% Flow focusing. Liquid-liquid. Viscoelastic
When a viscoelastic stream is focused by a Newtonian outer current, the tensile stresses generated by the stretched polymers preclude the formation of a proper tip streaming regime. On the contrary, the viscoelastic liquid forms a slender jet which accelerates progressively until crossing the discharge orifice \citep{ESMRD06,PMVG16,MGEPG17}. The outer stream keeps the dissolved polymers stretched, increasing the jet's extensional viscosity. This effect favors the damping of the varicose free surface perturbations.

\subsection{Jetting from tip streaming}
\label{sec11.2}

\subsubsection{Gravity-driven steady tip streaming}

% Gravitational
A natural question is whether steady tip streaming can be produced by simply ejecting a liquid thread in a passive ambient under the action of the gravitational force. This force does not seem to be able to produce accelerations sufficiently localized to give rise to such a phenomenon. In fact, the recirculation patterns characteristic of tip streaming \citep{TP15} cannot be produced by the gravity mass force. However, liquid shapes fairly similar to some of those regarded as tip streaming can be found very close to the system's global stability limit \citep{RSG13}.

% The variant of selective withdrawal
As described in Sec.\ \ref{sec10}, gaseous tip streaming can also be produced by a variant of selective withdrawal driven by the gravitational force \citep{DE06}. In this case, microbubbles only form when the tip is allowed to enter the tube. We will come back to this point in Sec.\ \ref{sw}.

\subsubsection{Coflowing steady tip streaming}

\begin{figure}
%\begin{center}
\centering{\resizebox{0.35\textwidth}{!}{\includegraphics{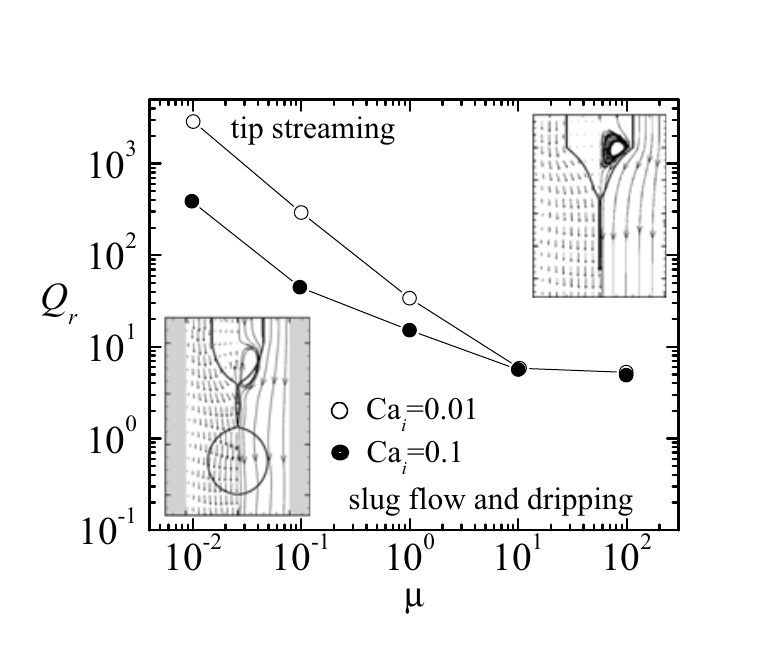}}}
%\end{center}
\caption{Critical value of the flow rate ratio $Q_r$ as a function of the viscosity ratio $\mu$ calculated numerically \citep{SB06}.}
\label{suryo}
\end{figure}

% Coflowing
\citet{SB06} analyzed the transition from dripping to tip streaming in a coflowing configuration. Viscous stresses exerted by the outer liquid increase as the ratio $Q_r$ of the outer flow rate to the inner one increases. For sufficiently large values of this quantity, those stresses make the growing drop adopt a conical shape within which a recirculation pattern appears. For a relatively narrow region of the parameter space, a jet with a radius a few orders of magnitude smaller than that of the feeding capillary emanates steadily from the drop tip \citep{CGFG09,GSC14}. Simple scaling analyses show that this transition takes place for \citep{SB06}
\begin{equation}
Q_r\gtrsim (\mu\ \text{Ca}_i)^{-1}.
\end{equation}
Numerical simulations agree with this prediction for small $\mu$ (Fig.\ \ref{suryo}). For moderate and large values of this parameter, the transitional flow rate ratio becomes independent of the viscosity ratio.

\citet{GSC14} concluded that the parameter conditions to obtain monodisperse micro-emulsions via tip streaming in coflows are those leading to the asymptotic global stability of this flow. It worths mentioning that it is not simple to distinguish tip streaming in a coflowing configuration from the narrowing jetting regime described in Sec.\ \ref{sec11.1}. In some experimental configurations, there seems to be a gradual transition from one mode to another, rather than a sharp steady flow bifurcation. This kind of bifurcation does take place in other tip streaming configurations. In fact, \citet{DMFESR18} has recently introduced the concept of {\it second-order transition} to describe this phenomenon in a variant of flow focusing, which provides an intriguing insight into the physics of the problem.

\subsubsection{Selective withdrawal steady tip streaming}
\label{sw}

% Summary of results
Selective withdrawal has been studied on fewer occasions than other techniques such as electrospray or flow focusing. \citet{CLHMN01} proposed selective withdrawal to coat small particles with polymer films. The method proved to be flexible in terms of the chemical composition and thickness of the conformal coatings.

\begin{figure}
%\begin{center}
\centering{\resizebox{0.45\textwidth}{!}{\includegraphics{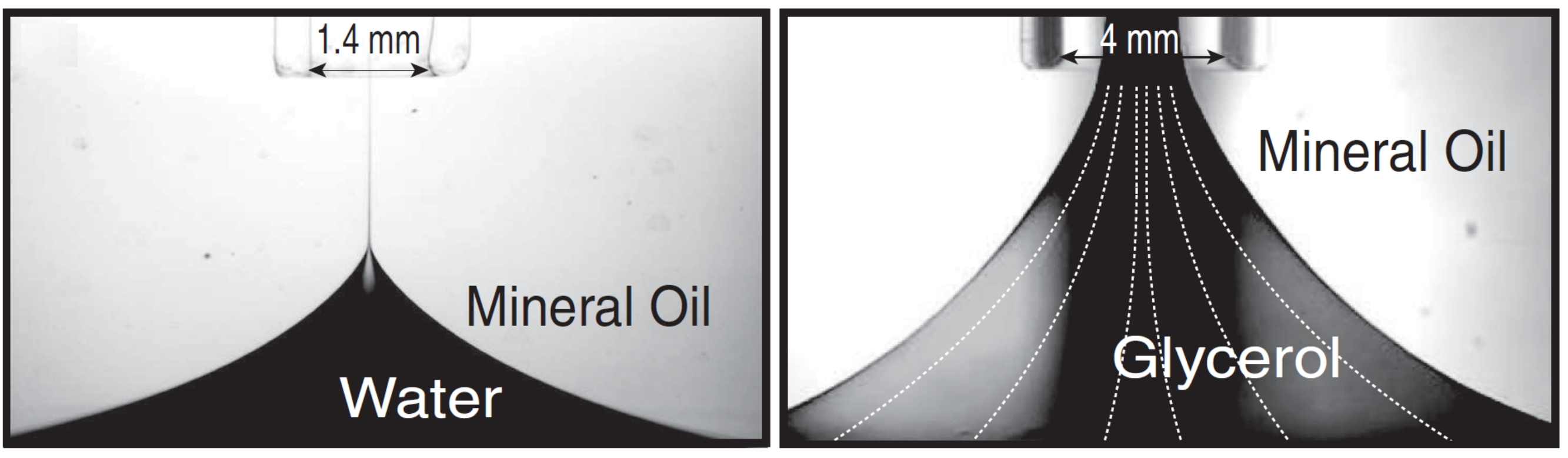}}}
%\end{center}
\caption{Selective withdrawal for $\mu_i/\mu_o=0.005$ and 20 \citep{CN07}.}
\label{selective}
\end{figure}

The force exerted by the withdrawal stream produces a hump of the interface for flow rates smaller than that leading to tip streaming. \citet{L89} modeled the flow in this configuration through a point sink in a two-layer unbounded system in which viscous forces dominate. In this way, he determined the conditions for the fluids to be withdrawn. \citet{CN02} showed that the steady-state profiles for humps at different flow rates and tube heights can be scaled onto a single similarity profile. This profile does not depend on the lower fluid viscosity \citep{C04}. \citet{CN07} discussed the influence of the viscosity ratio on the diameter of the spout at the transition. True tip streaming can be produced only for sufficiently small values of the lower-to-upper viscosity ratio (Fig.\ \ref{selective}). Due to its local character, the flow in selective withdrawal exhibits universal features next to the interface tip \citep{ED09}. 

% Gaseous dispersed phase
In the classical selective withdrawal configuration, the tip formed by the dispersed phase does not enter the tube. \citet{DE06} showed that in this case a gaseous tip never becomes unstable. Their numerical simulations \citep{ED09} for this configuration agreed remarkably well with the experiments \citep{DE06}. \citet{ZF10a} examined the role played by the polymer stress in the viscoelastic case. They found that there is a critical transition where the surface forms a cusp from which an air jet emanates toward the suction tube.

% Controversy
There has been certain controversy about the mechanism causing tip streaming in selective withdrawal. According to \citet{BZ09}, the transition is not determined by the local flow in the interface tip, but by a global balance between the upward force exerted by the withdrawal flow and the downwards force owing to interfacial tension. A similar idea was proposed by \citet{MRHG11} to explain the critical conditions for tip streaming in gaseous flow focusing operating with viscous liquids. The theory of \citet{BZ09} assumes that the failure of the interface is insensitive to the nature of the entrained phase. According to \citet{EC10}, this assumption is not valid because there is no transition if the entrained phase is air, something corroborated by the experiments of \citet{ZF10a}.

% Confined selective withdrawal
The strong stretching experienced by the simple or compound liquid jets in confined selective withdrawal (Fig.\ \ref{confined}) allows reducing the critical Capillary number above which steady tip streaming is obtained \citep{ECG16}. The diameter of the bubbles generated with this configuration have been calculated in terms of the Weber and Reynolds numbers, and the gas to liquid flow rate ratio \citep{ECG15}. 

% Connection
Electrospray and flow focusing are probably the two techniques most frequently used to produce jets via steady tip streaming. We devote the next two sections to these configurations.

\section{Electrospray steady tip streaming}
\label{sec12}

\subsection{The DC cone-jet mode of electrospray}

Electrostatic fields have been used to spray liquids since the XVIIth century. In the most typical configuration of electrospray, the liquid is injected into a passive dielectric medium across a metallic capillary hundreds of microns in diameter at flow rates in the range $10^{-3}$-10 ml. One imposes in the feeding capillary an electric potential of the order of kilovolts with respect to a grounded electrode located some millimeters downstream from the capillary. When both the liquid properties and control parameters are selected adequately, the cone-jet mode of electrospraying is obtained. This flow is a way to escape the singularity arising in the tip of the so-called Taylor cone, which results from the exact static balance between surface tension and electric forces. Among the varied electrospray regimes \citep{JK99}, the cone-jet mode has attracted attention because of its applicability in very diverse fields. The steady cone-jet is observed over a huge range of length scales. The increase of the conductivity allows reducing the jet diameter from tens of microns to a few nanometers \citep{BSJ85,DDNP95}.

In the cone-jet mode, the meniscus adopts a stable quasi-conical shape whose apex steadily ejects a thin jet, which eventually breaks up into droplets due to capillary forces \citep{R79a}. In general, the superficial charge is not relaxed to its electrostatic value within the cone-jet transition region \citep{FL94,ML94a,PRHGM18,GM18}, so that inner electric fields arise in that region. However, those electric fields are not large enough to invalidate the scaling laws presented below.

\begin{figure}
%\begin{center}
\centering{\resizebox{0.45\textwidth}{!}{\includegraphics{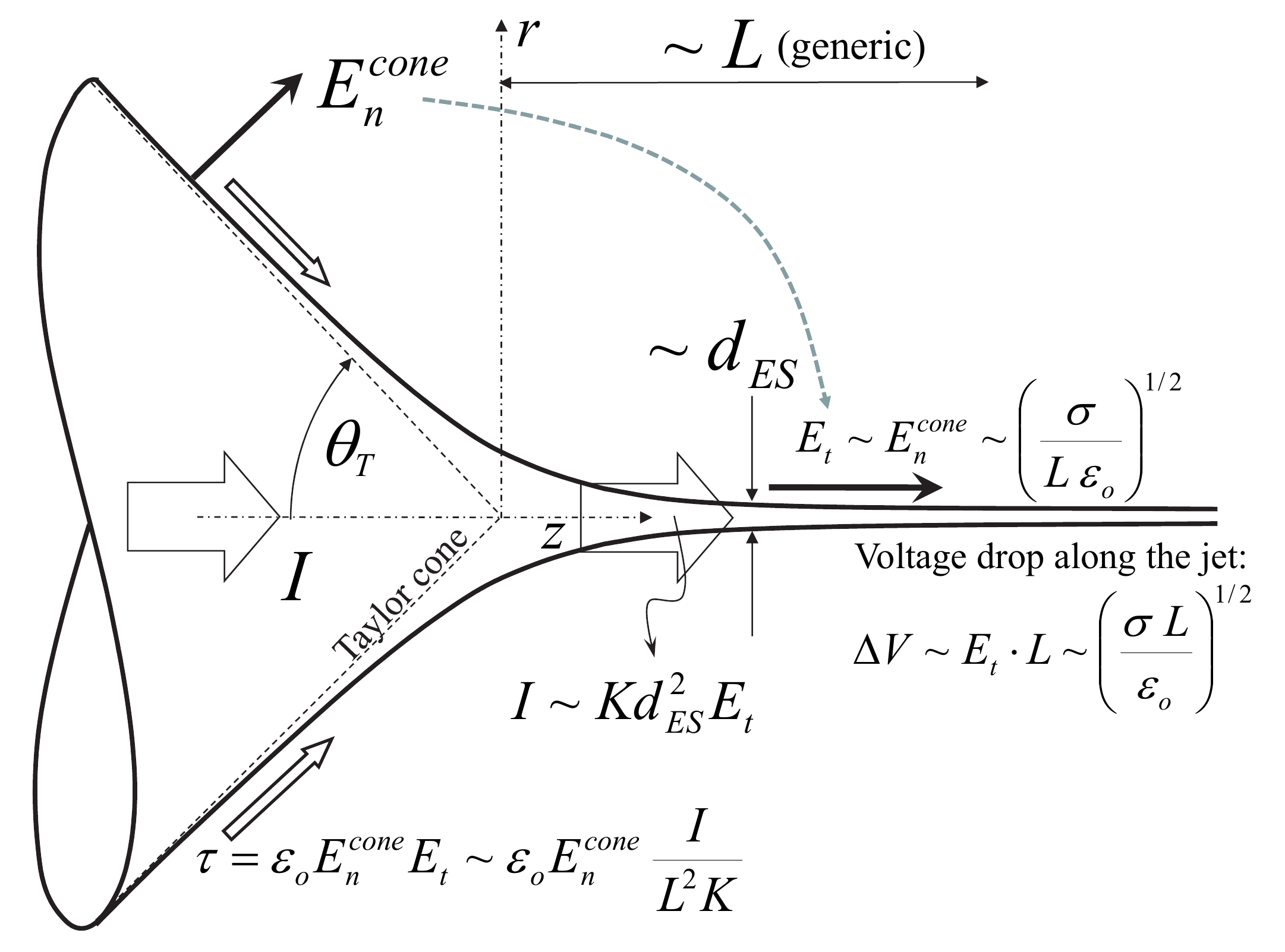}}}
%\end{center}
\caption{Sketch of the generic electrospray cone-jet configuration in which electrohydrodynamic steady tip streaming takes place. Its axial length scale $L$ is small compared to the feeding capillary diameter.}
\label{sketchES}
\end{figure}

% Characteristic quantities
\citet{GLHRM18} have recently reviewed both the conditions for the cone-jet mode to appear and the scaling laws for the droplet diameter $d_d$ and emitted electric current $I$. The locality of the ejection process allows to describe it in terms of the liquid properties exclusively (independently of geometrical and electrical parameters). On this condition, the characteristic quantities of the problem are $d_{\textin{ES}}=[\sigma \varepsilon_o^2/(\rho_i K_i^2)]^{1/3}$, $v_{\textin{ES}}=[\sigma K_i/(\rho_i \varepsilon_o)]^{1/3}$ and $I_{\textin{ES}}=(\sigma K_i v_{\textin{ES}} d_{\textin{ES}}^{2})^{1/2}$, and the dimensionless governing parameters are the relative permittivity $\varepsilon$, the electrohydrodynamic Reynolds number $\delta_\mu=\rho_i v_{\textin{ES}} d_{\textin{ES}}/\mu_i$, and the relative flow rate $Q_i/Q_{\textin{ES}}$, where $Q_{\textin{ES}}=v_{\textin{ES}} d_{\textin{ES}}^{2}$ \citep{G99b,G04a}. Naturally, the characteristic diameter and velocity $d_{\textin{ES}}$ and $v_{\textin{ES}}$ coincide with the corresponding quantities $d_o$ and $v_0$ introduced in Sec.\ \ref{sec8} to describe the onset of electrospray. Figure \ref{sketchES} sketches the cone-jet mode of electrospray and the quantities involved in its analysis.

\subsubsection{Droplet diameter and electric current}

\begin{figure}
%\begin{center}
\centering{\resizebox{0.335\textwidth}{!}{\includegraphics{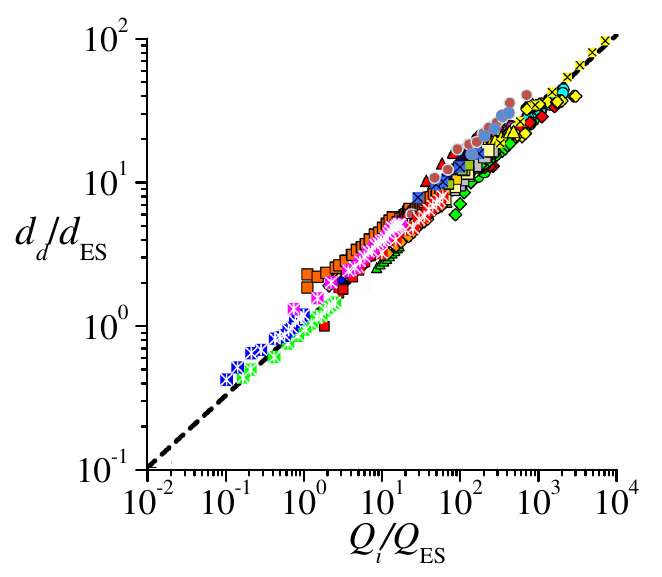}}}
%\end{center}
\caption{Diameter $d_d$ of the droplet emitted in the cone-jet mode of electrospray and measured by different autors \citep{GLHRM18}. The dashed line is the scaling law (\ref{GC}).}
\label{des}
\end{figure}

\begin{figure}
%\begin{center}
\centering{\resizebox{0.335\textwidth}{!}{\includegraphics{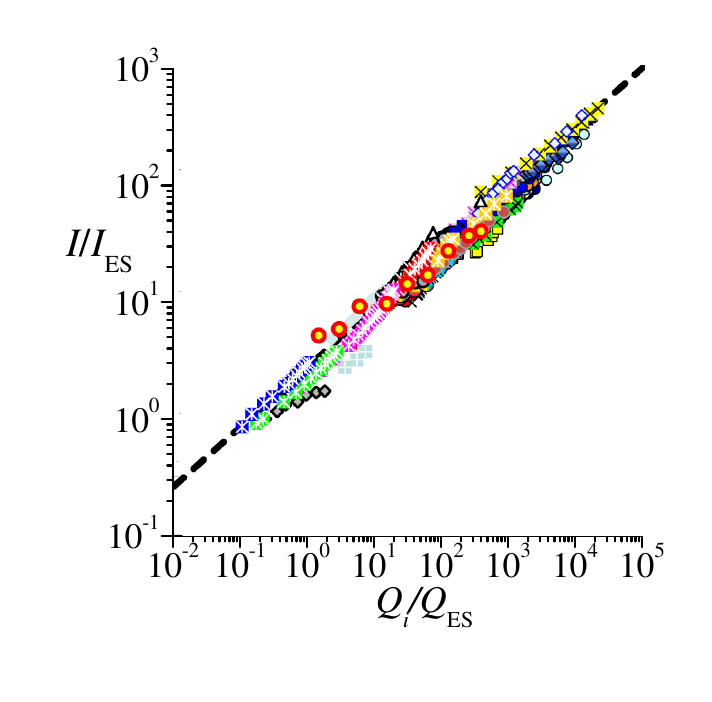}}}
%\end{center}
\caption{Current intensity $I$ emitted in the cone-jet mode of electrospray and measured by different autors \citep{GLHRM18}. The dashed line is the scaling law (\ref{GC}).}
\label{int}
\end{figure}

Most cone-jet mode realizations of electrospray correspond to the ``I-E regime"\ identified by \citet{G04a}, where the liquid electrical permittivity and viscosity have little influence on the outcome. In this regime, the scaling laws for the droplet diameter $d_d$ and emitted electric current $I$ are:
\begin{equation}
\frac{d_d}{d_{\textin{ES}}}\simeq \left(\frac{Q_i}{Q_{\textin{ES}}}\right)^{1/2}, \quad \frac{I}{I_{\textin{ES}}}\simeq 2.5 \left(\frac{Q_i}{Q_{\textin{ES}}}\right)^{1/2};
\label{GC}
\end{equation}
or, in dimensional form:
\begin{equation}
d_d\simeq \left(\frac{\rho_i \varepsilon_o Q_i^3}{\sigma K_i}\right)^{1/6}, \quad I \simeq 2.5 \left(\sigma K_i Q_i \right)^{1/2}.
\label{GCD}
\end{equation}
The scaling for the emitted current was first derived by \citet{GBP93}, while that for the droplet was given by \citet{G97a}. Both scalings were also obtained from dimensional analysis \citep{G99b} and numerical simulation \citep{HBCMS99}. These laws exhibit remarkable agreement with experimental data gathered from many authors (Figs.\ \ref{des} and \ref{int}) \citep{GLHRM18}. Significant deviations from these laws can be expected in the viscous and polar regimes, corresponding to small and large values of $\delta_{\mu}$ and $\beta$, respectively.

% Fernandez de la Mora
The scaling laws for the droplet diameter and current differ from those derived by \citet{FL94}, who assumed that the flow becomes a spherically symmetric sink in a region of the cone tip with size of the order of $d_j$, and that the electric relaxation time is of the same order of magnitude as that of the fluid particle residence time in that region. This led these authors to include the permittivity of the liquid in their general scaling laws for electrospray. However, that is only valid for very polar liquids and close to the minimum flow rates for steady cone-jet mode, as shown by both experiments and dimensional analysis \citep{G04a}.

% Electrospray-flow focusing analogy
\citet{GM09} showed that there is a simple analogy between electrospray in the I-E regime and gaseous flow focusing. The jet in electrospray is powered by an effective pressure drop
\begin{equation}
\Delta P= k_p \left(\frac{\sigma^2 K_i^2 \rho_i}{\varepsilon_0^2}\right)^{1/3},
\end{equation}
where $k_p$ is a constant of order unity. Interestingly, the pressure drop is not the major source of energy in many realizations of electrospray and gaseous flow focusing. However, its capacity to stretch the tapering meniscus tip and extrude the liquid filament from it renders this force an essential element of these two techniques.

\subsubsection{Breakup length and whipping}

\begin{figure}
%\begin{center}
\centering{\resizebox{0.35\textwidth}{!}{\includegraphics{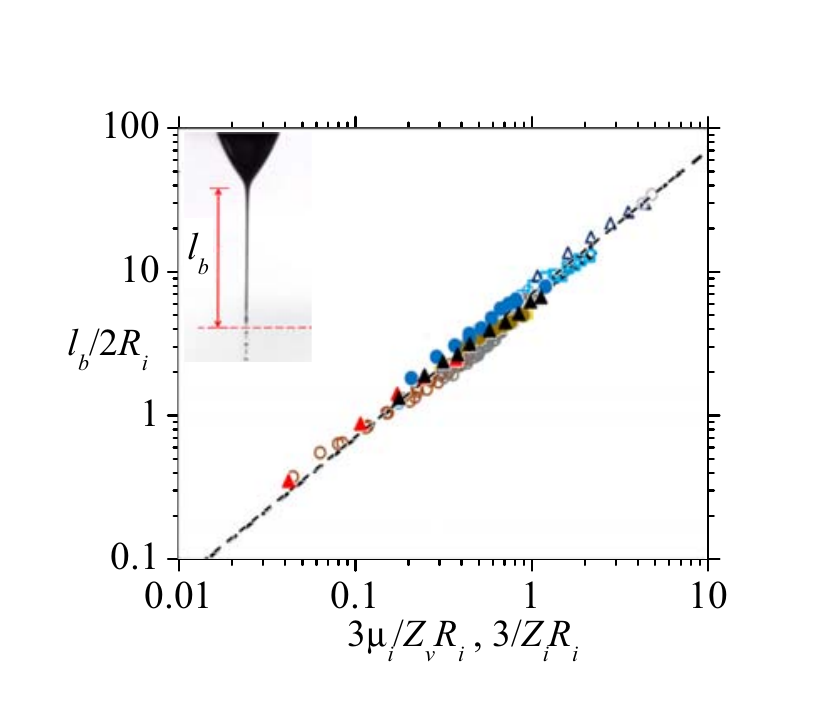}}}
%\end{center}
\caption{Experimental values of the breakup length $l_b$ of a jet emitted in the cone-jet mode of electrospray for low (open symbols) and high (solid symbols) Reynolds numbers \citep{IYXS18}. The dashed line is the law (\ref{lbes}).}
\label{lbelectrospray}
\end{figure}

The breakup length $\ell_b$ of a jet emitted in the cone-jet mode of electrospray has been estimated by using the approximation (\ref{esti}) and the maximum growth rate calculated by \citet{S71b} for a leaky-dielectric jet moving in an axial electric field \citep{IYXS18}. The result for low and high Reynolds numbers is
\begin{equation}
\label{lbes}
l_b\simeq 42\ Z_v^{-1}, \quad l_b\simeq 42\ Z_i^{-1},
\end{equation}
respectively, where
\begin{equation}
Z_v\equiv\mu_i^{-1}\left[\left(\frac{\rho_i\sigma^5 \varepsilon_0}{Q_i^3 K_i}\right)^{1/6}-\frac{1}{4\pi}\left(\frac{\rho_i^2 \sigma K_i}{\varepsilon_0}\right)^{1/3}\right]
\end{equation}
and
\begin{equation}
Z_i\equiv\left[\left(\frac{\sigma^5 \varepsilon_0}{\rho_i^5 Q_i^9 K_i}\right)^{1/6}-\frac{1}{4\pi}\left(\frac{\sigma K_i}{\rho_i Q_i^3\varepsilon_0}\right)^{1/3}\right]^{1/2}.
\end{equation}
The above theoretical predictions agree with experimental data of liquids with low polarity (Fig.\ \ref{lbelectrospray}). Using the scaling laws (\ref{GCD}), \citet{XIYS19} have found experimentally that whipping arises in the cone-jet mode of electrospray when the dimensionless number $G=\Gamma_e^2 \delta_{\mu}^{1/3}$ takes values above the threshold $G\simeq 155$.

\subsubsection{Minimum flow rate}

\begin{figure}
%\begin{center}
\centering{\resizebox{0.4\textwidth}{!}{\includegraphics{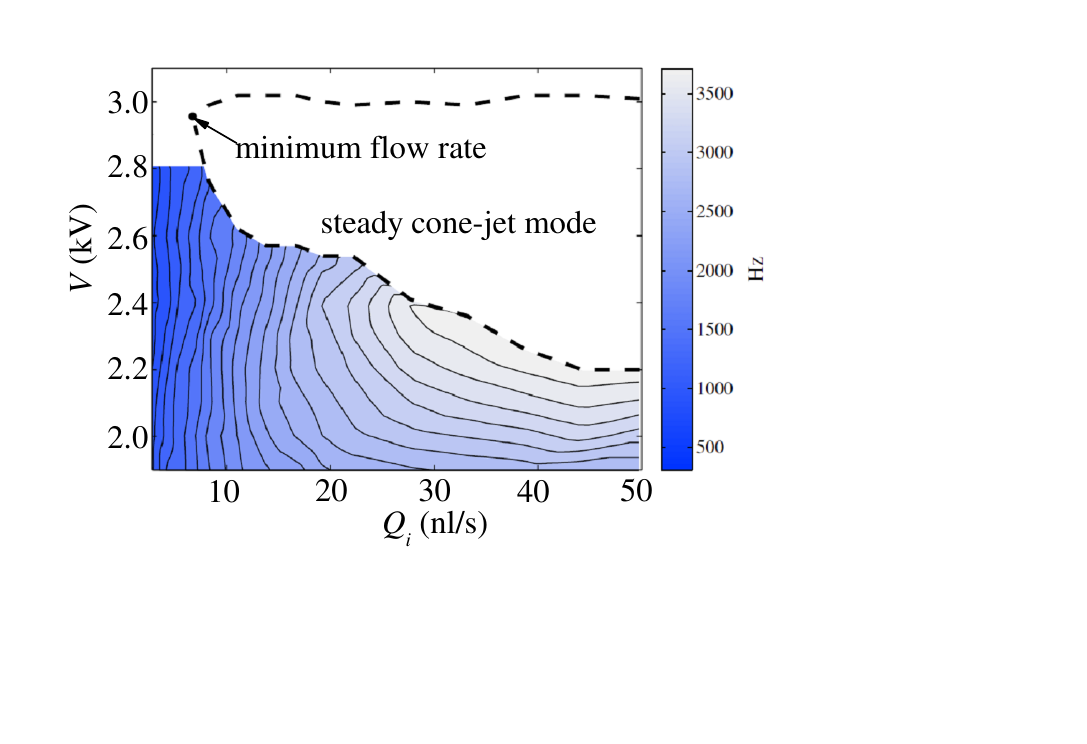}}}
%\end{center}
\caption{Parameter window for the steady cone-jet mode observed by \citet{BC11}. The contours show the frequency of the dripping mode (intermittent cone-jet) for flow rates below the minimum flow rate stability limit. The arrow indicates the absolute minimum flow rate for the configuration analyzed.}
\label{islandES}
\end{figure}

\begin{figure}
%\begin{center}
\centering{\resizebox{0.35\textwidth}{!}{\includegraphics{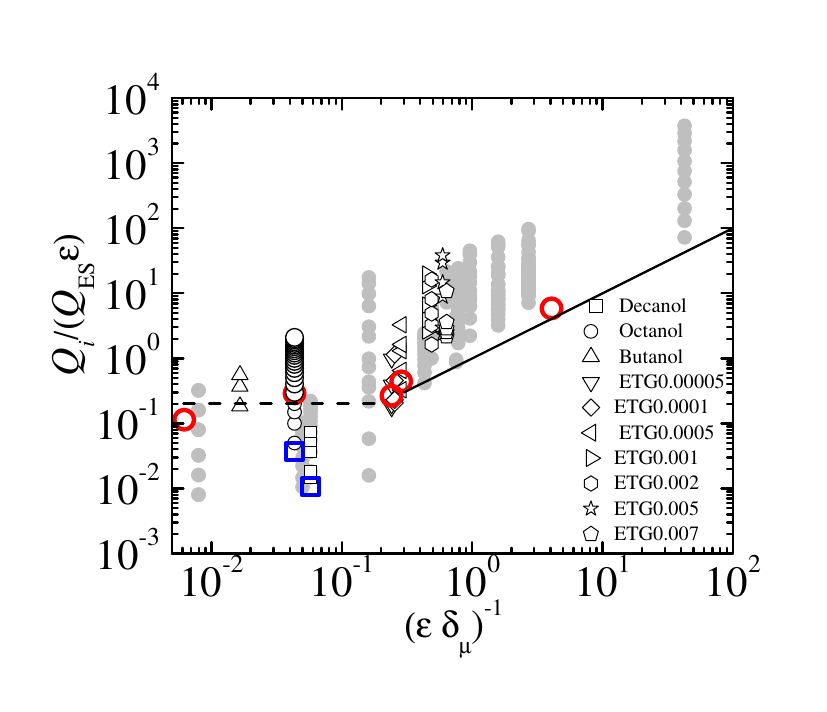}}}
%\end{center}
\caption{(Color online) Steady cone-jet mode realizations \citep{PRHGM18}. The grey symbols are the data gathered from different authors. The white symbols correspond to the experimental results obtained with a capillary $R_i=105$ $\mu$m in radius. The big white circles show the experiments conducted with 1-octanol and $R_i=550$ $\mu$m. The red big circles and blue big squares are marginally stable basic flows corresponding to $R_i=105$ $\mu$m and $R_i=550$ $\mu$m, respectively. The dashed and solid lines are the scaling laws $Q_{\textin{min}}/Q_{\textin{ES}}=0.2\varepsilon$ and $Q_{\textin{min}}/Q_{\textin{ES}}=\delta_{\mu}^{-1}$ for the polar and viscous limits, respectively.}
\label{qmin}
\end{figure}

% Minimum flow rate. Nosotros
The steady cone-jet mode of electrospray is adopted within a narrow parameter window \citep{CP89,BC11} (Fig.\ \ref{islandES}). For a given applied voltage, if the flow rate falls below its minimum value, then the balance between the electric tangential stress driving the liquid ejection and the polarity/viscosity ones opposing it breaks down in the cone-jet transition region. As can be seen in Fig.\ \ref{qmin}, experimental results are consistent with the simple scaling laws
\begin{equation}
Q_{\textin{min}}/Q_{\textin{ES}}\sim \varepsilon, \quad Q_{\textin{min}}/Q_{\textin{ES}}\sim \delta_{\mu}^{-1}
\end{equation}
for the minimum flow rate $Q_{\textin{min}}/Q_{\textin{ES}}$ of the cone-jet mode in the polar ($\varepsilon\gg \delta_{\mu}^{-1}$) and viscous ($\varepsilon\ll \delta_{\mu}^{-1}$) limits, respectively \citep{GRM13}. The figure also shows the stabilizing effect of the feeding capillary ($Q_{\textin{min}}/Q_{\textin{ES}}$ decreases as $d_{\textin{ES}}/R_i$ increases) \citep{SC14}.

% Minimum flow rate. Gamero
\citet{GM19} examined both experimentally and numerically the limit of the minimum flow rate of the cone-jet mode of electrospray. The major finding of their experiments is the fact that the regime dominated by polarity was not reached even for realizations with $\varepsilon\gg \delta_{\mu}^{-1}$ because of the significant influence of viscosity in those cases. The numerical solution of the leaky-dielectric model allowed them to hypothesize that instability occurs because viscous dissipation takes place before shear electric stresses inject energy into the system.

% Global stability analysis
\citet{DC16} conducted the global linear stability analysis of a leaky-dielectric liquid in the framework of the 1D model. The jet thinning produces a stabilizing effect due to both the variation of the surface charge density and the extensional viscous stresses in the base flow. \citet{PRHGM18} have recently shown that the minimum flow rate stability limit can be predicted from the global linear stability analysis of the solution to the full leaky-dielectric model (Fig.\ \ref{qmin}) \citep{S97}.

% Stabilization methods
Different methods have been proposed to stabilize the cone-jet mode in electrospray. These can be categorized into those modifying the applied electric field \citep{PKK04,MPGC12} and the emitter geometry. In this latter case, use has been made of carbon fiber emitters \citep{SDKL06}, ballpoint pen nozzles \citep{LCYMD14} or hemispherical caps \citep{MRRS16}. \citet{CWHBCG19} have recently proposed a way to stabilize steady micro/nanoliquid jets issuing from Taylor cones by applying a coflowing gas stream. That stream is shown to promote the global stability of the Taylor cone-jet, thus reducing very significantly the minimum flow rate for a given set of geometrical parameters.

% Maximum flow rate
The maximum flow rate limit is very relevant in those applications where the droplet production rate must be maximized. Its analysis is complicated because it involves a wide variety of complex issues. The excess of the level of charge carried by the liquid jet limits the electric field on the cone surface \citep{CP89,CP94} and can cause the surrounding gas ionization. There can also be polarity effects related to the different mobility and nature of ions that transport the charge \citep{PKDKHM16}.

\subsection{Coaxial electrospray}

\begin{figure}
%\begin{center}
\centering{\resizebox{0.485\textwidth}{!}{\includegraphics{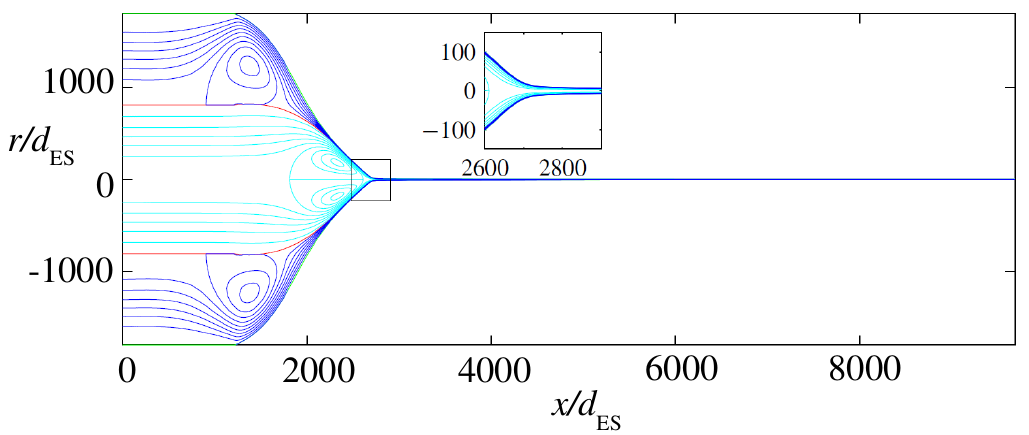}}}
%\end{center}
\caption{Shape and streamlines of a coaxial electrospray using deionized water and tributyl phosphate as the inner and outer fluids, respectively \citep{LHGG20}. The inner and outer flow rates are 234 $\mu$l/h and 60 $\mu$l/h, respectively, while the applied voltage is 4570 V.}
\label{coaxial}
\end{figure}

Coaxial electrified jets \cite{LBGCMG02} have been produced in the cone-jet mode for producing complex fibers \citep{LX2004,GW07}, and, to a lesser extent, complex particles (see, e.g., \citep{MLMB07}). The working modes of this configuration have been described by \citet{CJYCL05} in connection with those appearing in the single-phase electrospray. The ejection is driven by the liquid with the largest electrical conductivity, which typically is the external one. In this case, the differences between the coaxial and single-phase electrospray are minimal. In particular, the diameter of the compound jet is similar to that of the single-phase electrospray for the same flow rate as that of the driver, provided that the viscosities are commensurate with each other. \citet{LHGG20} have recently solved the 2D leaky-dielectric model for the cone-jet mode of coaxial electrospray (Fig.\ \ref{coaxial}). They have shown an unexpected feature exhibited by this complex system: the possibility of the appearance of a segment of opposite charge (negative, if the polarity of the applied potential is positive, and vice versa) on the inner interface at the transition region.

\subsection{AC electrospray}

The cone-jet mode of electrospray can also be produced with AC electric fields at frequencies greater than 50 kHz. In this case, the cone is more slender and is sustained by the entrainment of net space free charge within the cone. This physical mechanism is completely different from that of DC electrospray, and cannot be described in terms of the leaky-dielectric model \citep{CMC08}. AC electrospray can be used as an ionization source for mass spectrometry of biomolecules \citep{CCGC10}.

\subsection{Electrospinning}

% Global stability
\citet{LSZPY20} have recently reviewed the theoretical description and numerical simulation of electrospun jets from polymeric solutions, paying special attention to the critical role played by instability phenomena of electrical and hydrodynamic origin. Interested readers are referred to that excellent work to gain insight into the physical mechanisms governing the flow. Here, we only describe some results obtained from the global stability analysis of electrospinning. 

\begin{figure}
%\begin{center}
\centering{\resizebox{0.27\textwidth}{!}{\includegraphics{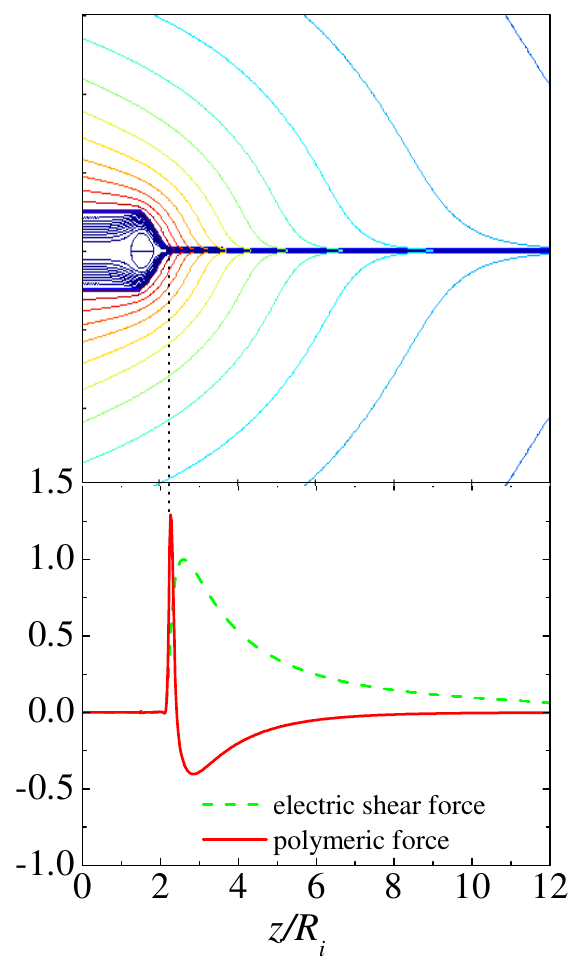}}}
%\end{center}
\caption{(Upper graph) Streamlines (inner domain) and equipotential lines (outer domain) calculated from the 2D leaky-dielectric Oldroyd-B model for $\varepsilon=10$, $\delta_{\mu}=2.29$, $Q_i/Q_{\textin{ES}}=7.72$, $\lambda_s^{(i)}=2.656\, t_e$ and $\lambda_r^{(i)}=0.0926\lambda_s^{(i)}$. (Lower graph) Electric shear force and polymeric force both per unit volume and normalized with the maximum value of the former \citep{BHGM19}.}
\label{electrospinning}
\end{figure}

When the electric field is applied to a viscoelastic liquid, the cone-jet mode of electrospinning can be produced \citep{Y11,TZY04,TCYR07,RY08}. To analyze the linear stability of this mode, \citet{DC17} extended their previous analysis for electrospray \citep{DC16} incorporating rheological effects described by the Olroyd-B and XPP models. As in the Newtonian case, the results showed the stabilizing effect of the jet thinning. \citet{BHGM19} have recently calculated the axisymmetric global modes of the cone-jet mode described by the 2D hydrodynamic equations and the Olroyd-B constitutive model. They have considered the physical properties of 1-octanol and added polymeric stresses characterized by relatively small relaxation times. The simulations of the base flow show that stress relaxation times of the order of some hundreds of microseconds are commensurate with the residence time in the cone-jet transition region. In this case, fluid particles undergo an extensional deformation in front of the meniscus tip intense enough for the dissolved polymers to continuously stretch, which prevents their relaxation to the coiling state. This produces axial polymeric stresses much larger than those caused by the solvent viscosity. These stresses pull from the liquid in front of the meniscus tip (Fig.\ \ref{electrospinning}). As a consequence, the liquid accelerates much faster than in the Newtonian case, which makes the electrified liquid meniscus shrink. The kinetic energy gained by the meniscus is lost in the jet region, where polymers relax to their coiling state. One may say that the liquid borrows energy from the polymers in the most unstable region and returns it when safely moving downstream. This stabilizes the flow and explains the reduction of the minimum flow rate. This theoretical study is just a first attempt to describe the electrospinning of viscoelastic liquids at a fundamental level. Factors like the possible anisotropy of electrical conduction or the finite extensibility of polymers must be taken into account in future studies.

\section{Flow focusing steady tip streaming}
\label{sec13}

\subsection{Liquid-liquid flow focusing}

% Flow focusing liquid-liquid
Liquid-liquid flow-focusing devices induce a strong hydrodynamic focusing effect in front of the discharge orifice owing mainly to the action of high viscous stresses in that region, which favors the transition from dripping to steady tip streaming. In fact, the steady tip streaming regime is found in flow focusing for outer Capillary numbers much smaller than those leading to the dripping-to-jetting transition in an equivalent coflowing configuration \citep{LWZC17,WLZC17}. The viscous stresses in the focusing region scale as $D_o^{-2}$. As a consequence, the droplet formation is considerably affected by the orifice diameter, while it is practically insensitive to the orifice length \citep{WLZC17}.

% Flow focusing liquid-liquid. The second-order transition
According to \citet{DMFESR18}, the use of the ``opposed flow focusing"\ configuration may lead to a continuous reduction of the jet radius down to vanishing scales in what they termed as {\it second-order jetting transition}. This kind of transition was experimentally observed by \citet{GGRHF07} using an axisymmetric liquid-liquid-gas coaxial flow focusing configuration. As explained in Sec.\ \ref{sec4}, the existence of such a vanishing jetting condition was theoretically demonstrated for parallel flows from the analysis of the convective-to-absolute instability transition \citep{GGRHF07,G08a}. \citet{Z04} was the first to demonstrate theoretically the existence of the second order transition in the case of a non-parallel flow (tip streaming), but without interrogating the found solution about its stability. The experimental evidence provided by \citet{DMFESR18} provides definitive support for this topological transition.

% Flow focusing liquid-liquid. Stability. JFM2006
The theoretical analysis of the axisymmetric liquid-liquid flow focusing configuration has received little attention. \citet{GR06} studied both theoretically and experimentally the jetting-to-dripping transition taking place at the minimum flow rate stability limit. They compared the experimental minimum flow rates with those leading to the convective-to-absolute instability transition. As mentioned in Sec.\ \ref{sec3.1}, this comparison is pertinent only if the instability origin is located in the jet. The experimental stability limit in the (Re$_j$,We$_j$) plane exhibited an ``elbow"\ for Weber numbers around unity. This elbow has probably to be explained in terms of the global stability analysis of the problem or the superposition at short times of asymptotically stable global modes. Preliminary results \citep{RRHCM20} indicate the existence of parameter islands in the $(Q_r,\mu)$ plane within which the base flow is globally stable. These islands are delimited by oscillatory/non-oscillatory and convective/absolute instabilities, and their location depends on the geometry and the rest of the governing parameters.

% Flow focusing liquid-liquid. Stability. Others
The stability of the axisymmetric liquid-liquid flow focusing configuration has been recently studied by \citet{MDS18} both experimentally and theoretically. They distinguish between the instability originated in the tapering meniscus and that localized in the emitted jet, as previously proposed for the gaseous configuration \citep{VMHG10,MRHG11}. In this way, they have successfully explained their experimental observations and direct numerical simulations.

% Viscoelasticity and surfactants
Numerical solutions of the FENE-CR model show that viscoelasticity delays the dripping-to-jetting transition \citep{NIM16}, which coincides with the prediction from the local stability analysis prediction for relaxed coflowing Oloroyd-B jets (Sec.\ \ref{sec4}) \citep{MG08b}. \citet{MWA12} have proposed a model to predict the appearance of tip streaming in presence of soluble surfactant. Essentially, tip streaming occurs when the mass of surfactant adsorbed to the interface is that needed to maintain the interfacial conical shape. \citet{WBSW18} have modeled the outer velocity field arising in an axisymmetric flow focusing configuration by combining an imposed uniaxial extension flow at infinity with two transverse, coaxial, annular baffles placed symmetrically to either side of the drop. They have described how a soluble surfactant and the focusing effect produced by the baffles collaborate to produce tip streaming.

\subsection{Gaseous flow focusing}

\subsubsection{Jet shape and size}

\begin{figure}
%\begin{center}
\centering{\resizebox{0.35\textwidth}{!}{\includegraphics{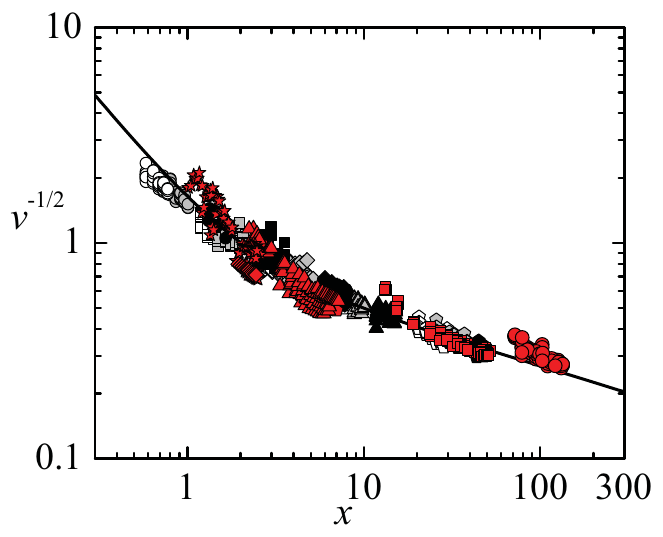}}}
%\end{center}
\caption{Universal jet shape $v^{-1/2}$ as a function of the scaled axial coordinate $x$ \citep{AFMG12}. The symbols correspond to different liquids. The white, grey and black symbols were measured with the plate-orifice configuration \citep{GFM11}, while the red ones were obtained by focusing the liquid stream in a glass nozzle \citep{AFMG12}. The solid and dashed lines correspond to (\ref{usol}) and (\ref{fin}), respectively (they practically overlap).}
\label{jsize}
\end{figure}

% Flow focusing liquid-gas. Jet profile and size
When focusing a liquid meniscus with a high-speed gaseous current, the major source of energy generally comes from the pressure drop $\Delta P$ induced by that current. If one neglects the interfacial and viscous sinks of energy, the jet's radius is given by Eq.\ (\ref{SJ}) \citep{G98a}. The above result can be refined by considering the viscous stresses exerted by the outer stream, and the loss of kinetic energy due to the resistance offered by the liquid viscosity. \citet{GFM11} found that, when the experimental results are expressed in terms of conveniently scaled jet velocity and streamwise direction coordinate $z$, they approximately match the universal solution (\ref{usol}) of the 1D momentum equation for a constant driving force \citep{C66} (Fig.\ \ref{jsize}). This solution can be approximated by the simple expression \citep{GFM11}
\begin{equation}
v=\left[\frac{1}{(2x)^{1/2}}+\frac{2}{x^2}\right]^{-1}\; ,
\label{fin}
\end{equation}
where
\begin{equation}
v=2\left(\frac{6\sqrt{2}R_\mu}{D_o}\right)^{-1/3}\left(\frac{R_{\textin{FF}}}{R_j}\right)^2, \quad x=2 \left(\frac{6\sqrt{2}R_\mu}{D_o}\right)^{-2/3}\frac{z}{D_o},
\end{equation}
are the scaled velocity and section coordinate, respectively, and $R_\mu=[\mu_i^2/(\rho_i \Delta P)]^{1/2}$. For $R_{\mu}\ll 1$ ($x\gg 1)$ and $z=D_o$, Eq.\ (\ref{SJ}) is recovered. For $R_{\mu}\gg 1$ ($x\ll 1)$ and $z\sim D_o\sim H$, the viscous scaling law \citep{GPLG04}
\begin{equation}
R_j\sim \left(\frac{\mu_i Q_i}{\Delta P H}\right)^{1/2}
\end{equation}
is obtained.

\subsubsection{Breakup length and whipping}

\begin{figure}
%\begin{center}
\centering{\resizebox{0.425\textwidth}{!}{\includegraphics{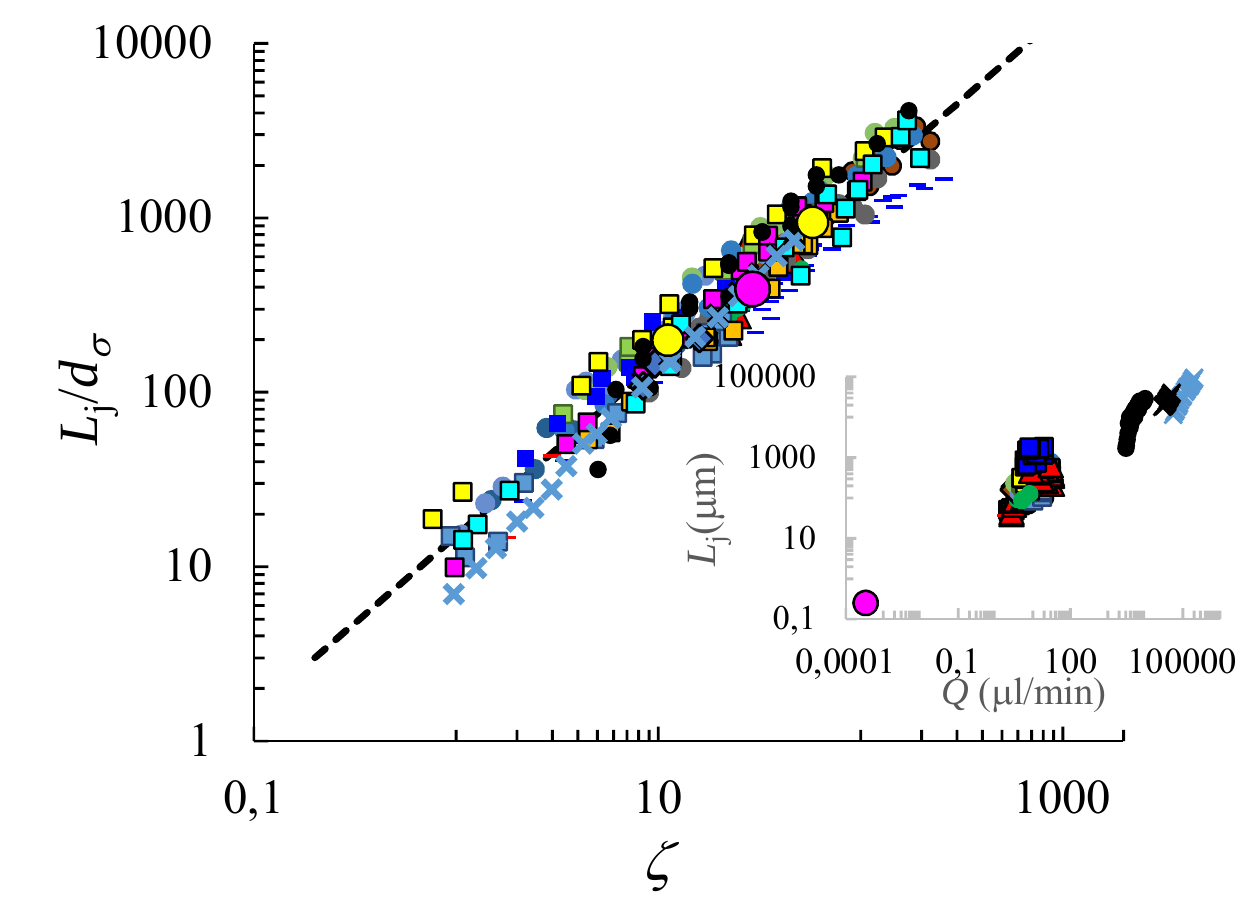}}}
%\end{center}
\caption{Breakup length $l_b$ for different liquids and nozzles in both flow focused and ballistic jets experiments \citep{GCHWKDGHLCBM19}. The figure shows the experimental data of \citet{U16}, additional experiments for water issuing from capillary tubes, and a numerical simulation for this latter configuration.}
\label{length}
\end{figure}

% Breakup length: Scaling law
The jet breakup length $l_b$ is of great importance in applications such as Serial Femtosecond Crystallography \citep{Cetal11}. Very recently \citep{GCHWKDGHLCBM19}, a scaling law for $l_b$ in gaseous flow focusing has been derived from the analysis of the energy balance taking place in the jet breakup region. The essential idea is that the breakup length is determined by the transient growth of perturbations coming from the surface energy excess at breakup \citep{U11,U16,UOSNMKTODSSNO20}. Based on this idea, \citet{GCHWKDGHLCBM19} proposed the scaling law
\begin{equation}
\label{ble}
l_b/d_{\sigma}=\alpha_{\rho}\xi,
\end{equation}
\begin{equation}
\xi=\text{We}_{\textin{FF}}^2\left[\left(\text{We}_{\textin{FF}}+\alpha_{\mu}^2 \text{Ca}_{\textin{FF}}^2\right)^{1/2}-\alpha_{\mu} \text{Ca}_{\textin{FF}}\right]^{-1},
\end{equation}
$\text{We}_{\textin{FF}}=d_G/d_{\sigma}$, $d_G=(8\rho_i Q_i^2/(\pi^2\Delta P)^{1/4}$ and $\text{Ca}_{\textin{FF}}=(\mu_i^2\Delta P/(\sigma^2\rho_i)^{1/2}$. The constants $\alpha_{\rho}= 15.015$ and $\alpha_{\mu}=0.53$ have been determined experimentally. In their study, they gathered a collection of careful measurements from flow-focused jets \citep{G98a} and capillary jets directly ejected from orifices (inertial jets). Figure \ref{length} shows the breakup length measured from the nozzle exit for different liquids and nozzles in both flow focused and ballistic jets experiments. Strikingly, the scaling law (\ref{ble}) accurately predicts the breakup length of a nanojet \citep{ML00} as well.

% Breakup length: DNS
Direct numerical simulations of liquid microjets in free expanding high-speed co-flowing gas streams provide accurate predictions for both the tip streaming stability conditions and the jet diameter. However, they overestimate the breakup jet's length \citep{ZBBS18}. This is probably due to the existence of external perturbations in the experiments.

% Whipping
Experimental studies on the whipping instability in flow-focused jets are still scarce. \citet{AFMG12} have distinguished two types of whipping instabilities: that leading to the bending of the emitted jet while keeping the tapering meniscus stable, and that in which both the jet and the meniscus oscillate laterally (absolute whipping). Preliminary results \citep{BRCHGM20} show the importance of the focusing geometry in the appearance of absolute whipping, and that this kind of instability is linked to the growth of the dominant $m=1$ global mode.    

\subsubsection{Minimum flow rate}

\begin{figure}
%\begin{center}
\centering{\resizebox{0.35\textwidth}{!}{\includegraphics{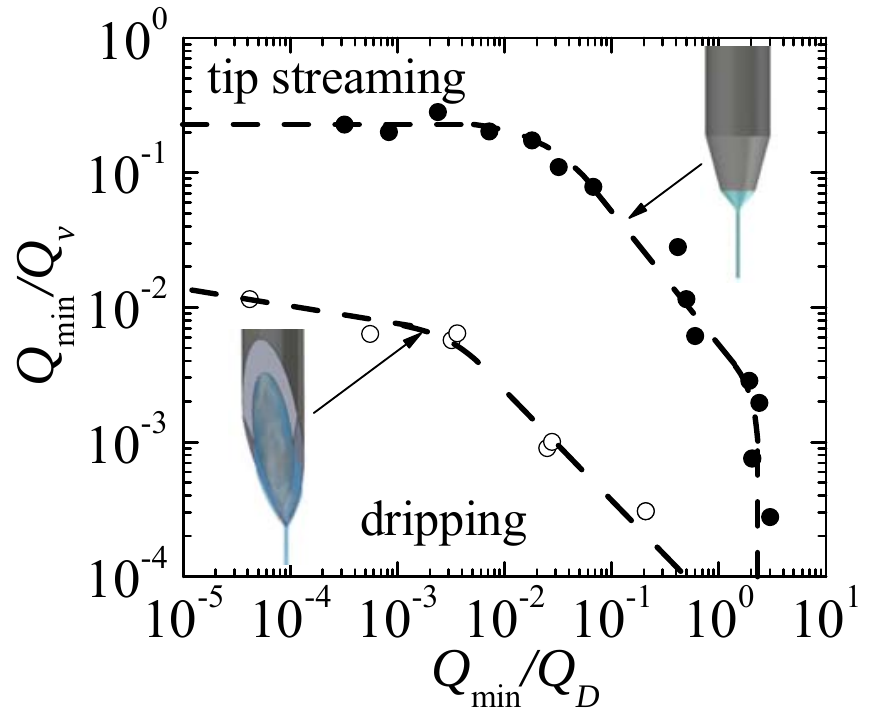}}}
%\end{center}
\caption{Minimum flow rate $Q_{\textin{min}}$ in gaseous flow focusing \citep{ARMGV13}. The solid symbols are the results obtained when the liquid stream tapers from a capillary located in front of a plate orifice \citep{MRHG11} or inside a glass nozzle \citep{AFMG12}. The open symbols are the results obtained when the liquid is injected through a hypodermic needle inside a glass nozzle \citep{ARMGV13}. The dashed lines are to guide the eye.}
\label{ffstab}
\end{figure}

% Flow focusing liquid-gas. Stability
The stability of a liquid stream focused by a high-speed gas current has been studied both theoretically \citep{HGOBR08,SLYY09,SLYY10,MDS17,ZBS18} and experimentally \citep{VMHG10,MRHG11}. For a fixed applied pressure drop $\Delta P$, jetting becomes unstable when the injected flow rate $Q_i$ falls below a critical value $Q_{\textin{min}}$. For small pressure drops, the instability is originated in the jet and has been described in terms of the convective-to-absolute instability transition (Fig.\ \ref{caelec}) \citep{G08a,SLYY09}. If $\Delta P$ takes sufficiently large values, the instability is associated with the flow in the tapering meniscus. In this case, the minimum flow rate $Q_{\textin{min}}$ for sufficiently small and large Reynolds numbers scales as
\begin{equation}
Q_{\textin{min}}\sim Q_v\equiv \frac{R_i^2 \sigma}{\mu_i}\quad \text{and} \quad Q_{\textin{min}}\sim Q_D\equiv \frac{D_o\mu_i}{\rho_i},
\end{equation}
respectively (Fig.\ \ref{ffstab}). In the former case, the stability limit was explained in terms of the axial component of the surface tension force \citep{MRHG11}, while the latter was linked to the growth of the recirculation cell inside the liquid meniscus \citep{HGOBR08}.

% Vega 2010
\citet{VMHG10} measured the minimum flow rates for low-viscosity liquids as a function of the capillary-to-orifice distance $H$ in the original plate-orifice configuration. The minumum attainable flow rate $Q_{\textin{opt}}$ and the corresponding capillary-to-orifice distance $H_{\textin{opt}}$ obeyed the scaling laws:
\begin{equation}
Q_{\textin{opt}}/Q_D\simeq 2.5 (2 R_i/D_o)^{1/3}\; , \quad H_{\textin{opt}}/D_o\simeq 0.6 (2 R_i/D_o)^{1/4}\; .
\end{equation}
These results are consistent with the numerical simulations conducted by \citet{ZBS18}.

% Cruz-Mazo 2017
As mentioned in Sec.\ \ref{sec4}, \citet{CHGM17} found good agreement between the experimental minimum flow rates and those calculated from the asymptotic global stability for large applied pressure drops. As occurs in the liquid-liquid configuration, the projection of the experimental stability limit onto the (Re$_j$,We$_j$) plane shows an ``elbow"\, for Weber numbers around unity. This elbow can be explained neither in terms of the convective-to-absolute instability transition nor by the growth of the dominant global eigenmode. In this region of the parameter plane, instability is caused by the superposition of asymptotically stable global modes at short times \citep{CHGM17} (Fig.\ \ref{ipv}). This superposition makes the jet break up without perturbing the tapering liquid meniscus, which indicates that somehow the instability is originated in the jet. Interestingly, this result coincides with the prediction obtained from the convective-to-absolute instability analysis \citep{SLYY09,VMHG10}, which shows that the jet becomes absolutely unstable for Weber numbers around unity.

% Stabilization
The minimum flow rates can be greatly reduced if the liquid meniscus is replaced with a film sliding over the tip of a hypodermic needle \citep{ARMGV13}. Figure \ref{ffstab} shows the minimum flow rates obtained when the liquid stream tapers from a capillary located either in front of a plate orifice or inside a nozzle, and when the liquid is injected through a hypodermic needle. A similar liquid injection system has been recently used by \citet{MGALHS19} to produce nanometre-sized droplets (comparable to those obtained by electrospray ionization) from a new operational regime for gaseous flow focusing.

\subsection{Gaseous stream focused by a liquid current}

% Flow focusing gas-liquid
Microbubbles were produced in an axisymmetric flow focusing device from the breakage of an air filament driven by a stream of ethanol-water \citep{GHG06} and glycerol-water \citep{FMFL09}. \citet{VAMHG14} examined the production of microbubbles in the tip streaming regime for moderate Reynolds numbers. A gaseous thread ejected from the meniscus tip broke up into a quasi-monodisperse collection fo bubbles with diameters smaller than that of the discharge orifice. Those diameters are given by the formula
\begin{equation}
\label{jb}
\frac{d_b}{D_o}=1.77 Q_r^{-\beta},
\end{equation}
where $\beta$ takes a value between 0.4 and 0.5 depending on the Reynolds number. Equation (\ref{jb}) with $\beta=0.5$ can be regarded as the version for this fluid configuration of the
classical flow-focusing formula \citep{G98a}. It can be readily derived under two conditions: (i) neglecting both the energy dissipation in the gas meniscus and the gas kinetic energy at the feeding capillary end, and (ii) assuming that both the pinching location and the proportion between the precursor jet and resulting bubble diameters do not significantly depend on $Q_r$ \citep{VAMHG14}.

The experimental validation of Eq.\ (\ref{jb}) is shown in Fig.\ \ref{diameter}, where results for bubbling and jetting are mixed. In fact, it is difficult to determine whether a certain realization of this kind of experiment must be regarded as bubbling or jetting due to the short length of the precursor gaseous threads.

\section{Prospectives and futures}
\label{sec14}

The variety and complexity of phenomena associated with the presence of capillary, viscous, inertia and electric forces, bulk and interfacial rheology or certain geometrical conditions offer a multiverse for research, where just a few of its universes have been explored. The insights are impelled by either demanding applications or sheer curiosity and esthetic appeal. However, if one tries to find one overarching reason justifying the strong interest that this field stirs up, the common driver would very probably be our deep need to understand and control a wealth of phenomena normally appearing at the same scales as those of living matter, where not only liquids, soft matter and interfaces prevail, but also where all our experiences and needs originate. We conclude this review by enumerating a few lines where future research would be well justified.

% Transient ejections
\subsection{Transient ejections}

\begin{figure}
%\begin{center}
\centering{\resizebox{0.48\textwidth}{!}{\includegraphics{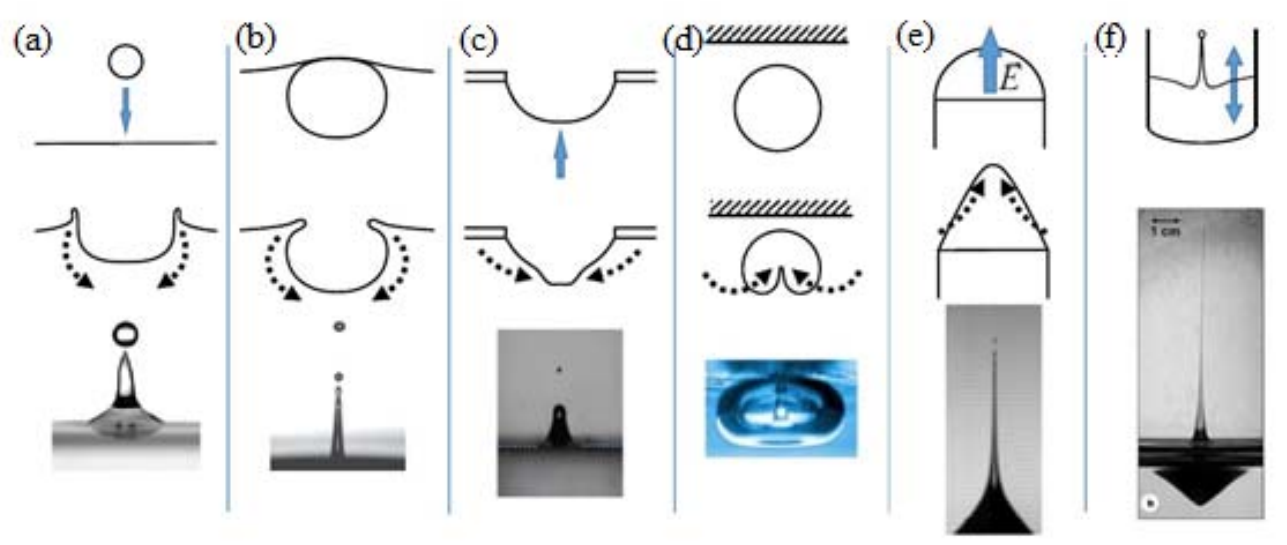}}}
%\end{center}
\caption{Examples of transient and fast ejections following the release of potential energy associated with the presence of an axisymmetric liquid free surface. The images show the collapse of a cavity produced after a droplet impact on a deep liquid pool (a), after bubble bursting on a free surface (b) and the hole on a plate (c), the bubble collapse by surrounding liquid volume oscillations close to a solid surface (d), the conical collapse of a suddenly electrified liquid droplet (e), and the collapse due to over-driven Faraday waves or the sudden vertical acceleration of a semi-enclosed cylindrical liquid volume (f).}
\label{project}
\end{figure}

Owing to their elusive character, one of the areas with higher potential for newer developments is for sure the transient ejections due to the collapse, with stronger or weaker momentum, of axisymmetric converging radial flow patterns in the presence of a free surface (Fig.\ \ref{project}). Paradigmatic examples that can be included in this category are the ejections after (i) surface bubble bursting \citep{BW57,DPJZ02,GLAS16,G17,DGLDZPS18}, (ii) surface acceleration \citep{ZKFL00,B02a,ABLV07}, (iii) cavity collapse \citep{CCM12,IGCHC18}, and (iv) conical collapse of a suddenly electrified liquid droplet \cite{GLRM16}. Although the collapsing angle of the converging flow may differ in these examples, the radial component of the liquid velocity causes a rapid change of the flow around a stagnation point located in the vicinity of the free surface. A fundamental question about this phenomenon is whether a spatiotemporal singularity is hidden somewhere in the parameter space governing the flow. In the case of bubble bursting, the existence of that singularity is relatively well established through the demonstrated existence of self-similar solutions \citep{LED18}, although the value of the governing parameter (the Ohnesorge number in this case) for which the singularity appears is still a matter of debate \citep{WHB15,GLAS16,G17,BBWFYB18,GR19}.

\citet{G17} has postulated that the balance between a precise fraction of the driving energy and viscous dissipation should lead to the above-mentioned singularity. Interestingly, when this balance is not perfect, either by excess or defect of one of the factors, the flow may still show time intervals where self-similar solutions can be experimentally observed \citep{LED18}. One may then postulate that each of the different above-mentioned flow classes, and probably many others leading to rapid transient ejections, would stem from a single class of singularity around which, closer or farther away in the parametrical sense, different flow realizations take place driven by different energy sources, geometries and initial conditions.

Inertia, viscous and surface tension forces compete in the presence of a liquid free surface during transient ejections. In contrast to what most researchers currently believe based on their experience, there might be a perfect balance only between inertia and viscosity asymptotically close to the spatiotemporal singularity (surface tension becomes negligible). This hypothesis is based on the sheer consistency of the time power law that each term exhibits when a self-similar solution is sought for. This was neatly demonstrated by \citet{E93} in his pioneering work on the axisymmetric free surface pinch-off driven by capillarity. This problem differs from free surface collapse in three aspects: (i) the singularity comes up independently of the values of the parameters involved, (ii) the surface tension force is retained in the asymptotic regime because the free surface radius vanishes at the singularity, and (iii) the existence of an outer medium \citep{LS98,CBEN99} and bulk or interfacial rheology \citep{CEFLM06,DBS17,PMHVV17,MFMHC18,RPVHM19} prevents the system from reaching that regime.

Demonstrating the very existence of a time singularity in the parameter space for a given system and flow configuration would be extremely useful in the search for thinner and faster jets emissions for increasingly demanding applications in industry and research \citep{YPR14,BGB13,W18}.

% Natural breakup of steady capillary jets
\subsection{Natural breakup of capillary jets}

Despite being a classic problem, the determination of the average natural breakup length of a capillary jet has resisted the insight of researchers. Experience teaches us that while global stability analysis predicts breakup lengths much above those experimentally observed, the initial value problem (IVP) analysis yields results strongly dependent on the location and amplitude of the initial perturbation. Thus, the only available tools to tackle this problem are experimental analysis and direct numerical simulation, both supported by dimensional analysis. However, and without a solid physical explanation of the results, the problem will remain open.

As mentioned in Sec.\ \ref{sec13}, \citet{U16} proposed a plausible physical picture of the above-mentioned problem when the breakup is dominated by capillarity and the jet issues from a circular orifice. The phenomenon is described in terms of Plateau-Rayleigh unstable waves emanating from the orifice, and a self-destabilizing loop associated with them. This could explain the collapse of data for capillary jets as a function of the Weber number. \citet{GCHWKDGHLCBM19} have recently shown that the natural breakup of those jets and those produced by gaseous flow focusing can be predicted by the single physical model (\ref{ble}), which involves the Weber and Capillary numbers and two universal fitting parameters.

Given that encouraging generality, it is an open problem for future research whether the same boldly simple explanation is extendable to other configurations, such as gravitational jets. It is also of interest to determine the potential role of bulk and interfacial rheology in this phenomenon. In particular, the increasing extensional viscosity of weakly viscoelastic liquids may dramatically affect both the rate at which unstable waves are damped upstream and the global energy balances along the jet, which determines the natural breakup length. In some cases, the experimentally observed development of long-term cyclical periods \citep{DBS17,MFMHC18} could be predicted theoretically.

\subsection{Steady tip streaming stability}

As explained in Sec.\ \ref{sec3.1.3}, an accurate stability analysis of the steady tip streaming regime generally requires the calculation of the eigenvalues characterizing the linear global modes of the system \citep{T11}, which constitutes a complex problem for a multiphase capillary system. In addition, global stability may fail to explain certain bifurcations of stability limits \citep{S07}. In this case, direct numerical simulations may be the only route to describe the problem theoretically. This scenario explains why the stability analysis of the steady jetting arising from tip streaming has been frequently carried out by splitting the fluid domain into the source and the emitted jet. This steady jetting is assumed to be stable if none of the mechanisms that destabilize each region separately comes into play. The source stability is commonly examined by rationalizing experimental results in terms of simple scaling laws, while the jet behavior is described in terms of the convective-to-absolute instability (Sec.\ \ref{sec4}).

Global stability of simple-jet tip streaming configurations \citep{GSC14,CHGM17,PRHGM18,BHGM19} has been systematically studied only until the recent development of numerical tools for this task \citep{HM16}. Current developments in simulation methods for flows in the presence of free surfaces promise a wealth of possibilities for applied research. The experimental stability limits of gaseous flow focusing have been rationalized in terms of three instability mechanisms: the jet absolute instability, the growth of the dominant global mode, and the superposition at short times of asymptotically stable global modes \citep{CHGM17}. The competition and relationship among these mechanisms should be elucidated for other tip streaming configurations, such as liquid-liquid flow focusing or electrospinning.

% Coaxial jets
\subsection{Coaxial capillary jets}

Coaxial jets can be driven by coflowing gas streams \citep{G98a}, electric fields \citep{LBGCMG02}, pressure gradients \citep{HMFG10}, etc. The number of parameters governing these flows considerably increases, which makes their systematic study a difficult task. The formation of coaxial jets involves the stability of the upstream flow from which they issue, which usually takes place in a compound capillary meniscus. The breakup of these jets is also strongly affected by the stability of the upstream flow, since the inner jet may break up and drip before the outer jet reaches its own breakup region. Again, the complexity associated with competing effects of different forces appearing in the inner and outer jets, together with the usual presence of bulk and interfacial rheology, keeps this field full of untrodden regions. Therefore, one may consider this area of study in its infancy. 

% Surfactants
\subsection{Surfactant-driven tip streaming in microfluidics}

The mechanisms triggered by the presence of surfactants in tip streaming have not been well elucidated yet. It is believed that the major effect is the reduction of the surface tension due to the accumulation of surfactant molecules in the tapering droplet/meniscus tip (the so-called soluto-capillarity effect). However, many questions have not been addressed so far. Do convection and interface expansion empty the tip of the tapering droplet/meniscus? Do Marangoni stress and surface viscosity play a relevant role in the process?

The theoretical analysis of tip streaming in the presence of soluble surfactants is a complicated problem. As happens in the absence of these molecules, there are numerical difficulties associated with the disparity between the time and spatial scales characterizing the flow in the tip and the rest of the fluid domain. In addition, the analysis requires not only solving the convective-diffusive transport of the surfactant across the bulks but also calculating the surfactant distribution over the interface, which involves modeling accurately the adsorption-desorption process. In many cases, the parameters necessary to feed the model are unknown. The analysis may be simplified in some particular cases. If diffusion over the interface ensures a quasi-uniform surfactant distribution, Marangoni stress can be neglected. If the adsorption-desorption characteristic time is much larger than the hydrodynamic time characterizing the tip streaming, the surfactant can be regarded as insoluble during the fluid ejection, and neither the surfactant transport across the bulk nor the adsorption/desorption process needs to be taken into account.

% Other geometrical and driving effects
\subsection{Other worth noting geometrical and driving effects in jetting}

A fertile ground for nonlinear phenomena and unexpected results appears when additional effects are introduced in capillary jetting. Some of those effects have been already considered in the literature. Some possible research lines are listed below. In all of them, bulk and interfacial rheology will enhance the complexity and interest of further studies.

\begin{itemize}

\item {\em Vibration and pulsation}. A wide variety of phenomena have been described depending on the relative weight of the energy introduced by vibration and pulsation compared to inertia, viscous and capillary forces. The influence of the ratio of vibration period over capillary time has also been examined \citep{MKG92,MH19}. The extreme cases in which the above-mentioned parameters take values of the order unity or larger still offer an ample field for research.

\item {\em Rotation}. The effect of rotation on the jet dynamics was reviewed by \citet{EV08}. Several works have described how non-axisymmetric instabilities with increasing azimuthal wavenumber may prevail for a single-phase liquid jet as the rotation factor is increased. One may consider rotated coaxial jets to favor specific features in microencapsulation. For this purpose, rotation can be combined with, for example, vibration or pulsation.

\item {\em Non-round discharge sections}. This is also a classical effect \citep{R79b} which can produce potentially unexpected results in combination with other factors.

\item {\em Impinging free jets}. This configuration has attracted the attention of some researchers for its intrinsic esthetic features \citep{HB02}. It has recently raised the interest of the Serial Femtosecond Crystallography community due to the extremely thin liquid sheets that can be produced \citep{Koralek2018}.
\end{itemize}

\vspace{1cm}
Partial support from the Ministerio de Econom\'{\i}a y Competitividad through Grant No. DPI2016-78887 and Junta de Extremadura (Spain) through Grant No. GR18175 is gratefully acknowledged.

%\bibliography{central}\bibliographystyle{unsrtnat}\end{document}
%\bibliographystyle{model2-names}%\biboptions{authoryear}

\end{document}